\documentclass[]{aa}
\usepackage{graphicx}
\usepackage{dirtytalk}
\usepackage{currvita}
\usepackage[T1]{fontenc}
\usepackage{txfonts}
\usepackage{hyperref}
\usepackage{xcolor}
\hypersetup{
colorlinks,
linkcolor={red!80!black},
citecolor={blue!80!black},
urlcolor={blue!80!black}
}

\usepackage{lscape}

\newenvironment{myfont}{\fontfamily{ppl}\selectfont}{\par}
\DeclareTextFontCommand{\textmyfont}{\myfont}

\newcommand{\code}[1]{\texttt{#1}}

\def\nifs{\iso{56}Ni}

\def\cm3{cm$^{-3}$}
\def\kms{km\,s$^{-1}$}

\def\msun{$M_{\odot}$}

\def\mic{\,$\mu$m}

\def\one{\ts {\,\sc i}}
\def\two{\ts {\,\sc ii}}
\def\three{\ts {\,\sc iii}}
\def\four{\ts {\,\sc iv}}
\def\five{\ts {\sc v}}

\def\beq{\begin{equation}}
\def\eeq{\end{equation}}

\def\lesssim{\mathrel{\hbox{\rlap{\hbox{\lower4pt\hbox{$\sim$}}}\hbox{$<$}}}}
\def\gtrsim{\mathrel{\hbox{\rlap{\hbox{\lower4pt\hbox{$\sim$}}}\hbox{$>$}}}}

\def\one{{\,\sc i}}
\def\two{{\,\sc ii}}
\def\three{{\,\sc iii}}
\def\four{{\,\sc iv}}
\def\five{{\sc v}}

\def\voned{{\code{V1D}}}

\def\mesa{{\code{MESA}}}
\def\cmfgen{{\code{CMFGEN}}}

\def\longpol{{\code{LONG\_POL}}}

\def\sn{SN\,2023ixf}

\def\ekin{$E_{\rm kin}$}
\def\mej{$M_{\rm ej}$}

\def\ergs{erg\,s$^{-1}$}

\def\esh{$\dot{e}_{\rm sh}$}
\def\vcds{$V_{\rm CDS}$}
\def\mcds{$M_{\rm CDS}$}

\def\ha{H$\alpha$}
\def\hb{H$\beta$}

\def\mgiiuv{Mg\two\,$\lambda\lambda$\,$2795,\,2802$}
\def\lya{Ly\,$\alpha$}
\def\ha{H$\alpha$}

\def\naid{Na\,{\sc i}\,D}
\def\pag{Pa\,$\gamma$}

\def\oitrip{O\one\,$\lambda\lambda$\,$7771-7775$}

\newcommand{\iso}[2]{\ensuremath{^{#1}\rm{#2}}}

\begin{document}

   \title{SN\,2023ixf: Radiative-transfer modeling of the photospheric phase evolution from the ultraviolet to the infrared}

   \titlerunning{Photospheric-phase modeling of SN\,2023ixf}

\author{
    Luc Dessart\inst{\ref{inst1},\ref{inst1a}}
\and
Wynn V. Jacobson-Gal\'an\inst{\ref{inst2},\ref{inst3}}
\and
K. Azalee Bostroem\inst{\ref{inst4},\ref{inst5}}
\and
Alexei V. Filippenko\inst{\ref{inst6}}
\and
WeiKang Zheng\inst{\ref{inst6}}
\and
Thomas G. Brink\inst{\ref{inst6}}
\and
Stefano Valenti\inst{\ref{inst7}}
 }

\institute{
    Institut d'Astrophysique de Paris, CNRS-Sorbonne Universit\'e, 98 bis boulevard Arago, F-75014 Paris, France\label{inst1}
    \and
    French-Chilean Laboratory for Astronomy, IRL 3386, CNRS and Instituto de Astrofísica,
       Pontificia Universidad Católica de Chile, Casilla 306, Santiago, Chile.\label{inst1a}
   \and
  Cahill Center for Astrophysics, California Institute of Technology, Pasadena, CA 91125, USA\label{inst2}
  \and
  NASA Hubble Fellow\label{inst3}
\and
Steward Observatory, University of Arizona, 933 North Cherry Avenue, Tucson, AZ 85721-0065, USA\label{inst4}
\and
LSST-DA Catalyst Fellow\label{inst5}
\and
Department of Astronomy, University of California, Berkeley, CA 94720-3411, USA\label{inst6}
\and
Department of Physics and Astronomy, University of California, Davis, 1 Shields Avenue, Davis, CA 95616-5270, USA\label{inst7}
}

   \date{}

  \abstract{
\sn, a Type II supernova (SN) showing early signs of interaction with circumstellar material (CSM), has been observed with unprecedented detail across the electromagnetic spectrum since shock breakout. Here, we present nonlocal thermodynamic equilibrium time-dependent radiative-transfer calculations of its photospheric-phase evolution (i.e., $\sim$\,20 to $\sim$\,120\,d), and for the first time encompassing from the ultraviolet (UV) to the infrared (IR). The explosion of a 15\,\msun\ progenitor star, evolved with enhanced mass loss during the red-supergiant phase, yielding an ejecta of 7--8\,\msun, a kinetic energy of $1.2 \times 10^{51}$\,erg, and a \nifs\ mass of 0.05\,\msun, yields a satisfactory match to the photospheric-phase duration and brightness. Prolonged interaction with a decreasing CSM density is required to match a number of salient features of \sn\ during the photospheric phase, including the persistent UV continuum and line fluxes, the optical brightness and line profiles (in particular \ha), as well as the IR flux (interaction boosts the free-free emission at long wavelengths). The presence of a cold dense shell (CDS), which is hard to infer at early times when the CDS and photosphere lie at similar velocities, becomes evident at later times and more so in the IR --- we find no evidence for material faster than the CDS at $\sim$\,8000\,\kms. Exploratory two-dimensional radiative-transfer calculations based on axially symmetric CSM or ejecta suggest that asymmetry can produce a diversity of profile shapes, with absorption troughs exhibiting a flat bottom or notches at any Doppler velocity. We emphasize the complexity of UV spectra influenced by complex metal-line blanketing at these phases. We document the sensitivity of model results to the adopted clumping in the CDS, though the largest offset is obtained here in the unlikely case of a smooth CDS.
}


   \maketitle



\section{Introduction}
\label{sect_intro}

\sn\ is a nearby, H-rich (Type II) supernova (SN) that showed early signs of interaction with circumstellar material (CSM). Its proximity at 6.85\,Mpc, and its discovery \citep{itagaki_23} as well as classification \citep{perley_23ixf_23} within hours of shock breakout, has made possible intense studies across the electromagnetic spectrum, from X-rays \citep{grefenstette_23ixf_23,chandra_23ixf_24,nayana_23ixf_25} to ultraviolet \citep[UV;][]{teja_23ixf_23,bostroem_23ixf_24,zimmerman_23ixf_24,bostroem_uv_25}, optical (see, e.g., \citealt{bostroem_23ixf_23}; \citealt{jacobson_galan_23ixf_23}; \citealt{zheng_23ixf_25}), near-infrared \citep[NIR;][]{park_23ixf_25,derkacy_23ixf_26}, mid-infrared \citep[MIR;][]{derkacy_23ixf_26}, to radio \citep{iwata_23ixf_25,nayana_23ixf_25} ranges, including high-resolution spectroscopy \citep{smith_23ixf_23,dickinson_23ixf_25} and spectropolarimetry \citep{vasylyev_23ixf_23,singh_23ixf_24,shrestha_23ixf_25,vasylyev_23ixf_26}. These data cover from the earliest times following first light through the photospheric and nebular phases, and continuing today as \sn\ evolves into a young SN remnant. \sn\ is an extraordinary transient in which a full coverage of the electromagnetic spectrum can be, and has been, obtained \citep{wynn_sed_25}.

Pre-explosion imaging has set some constraints on the progenitor star. There is a broad consensus that the progenitor is a luminous, dusty, red supergiant (RSG) with an abnormally large wind mass-loss rate and some modest level of photometric variability \citep{dong_23ixf_23,jencson_23ixf_23,kilpatrick_23ixf_23,niu_23ixf_23,soraisam_23ixf_23,neustadt_23ixf_24,qin_23ixf_24,ransome_23ixf_24,vandyk_23ixf_24,vandyk_23ixf_24b}. There is, however, much disparity in the inferred stellar luminosity and thus progenitor mass, with values covering most the available range for RSGs, from 11\,\msun\ \citep{kilpatrick_23ixf_23} to 20\,\msun\ \citep{soraisam_23ixf_23}.

Estimates of the progenitor mass of \sn\ have also resulted from the inference of the ejecta mass obtained with radiation-hydrodynamics simulations and comparison with the bolometric or multiband light curves, combined in some (though not all) cases with the important constraints provided by the evolving photospheric velocity. Here, too, there is much disparity in the results. To reproduce the shorter-than-standard photospheric-phase duration of \sn, some studies require a partially stripped progenitor \citep{fang_23ixf_25,forde_23ixf_25,hsu_23ixf_25,kozyreva_23ixf_25,zheng_23ixf_25}, but others invoke a standard RSG progenitor \citep{bersten_23ixf_24,moriya_23ixf_24,singh_23ixf_24,vinko_23ixf_25} though in some cases with a higher-than-standard explosion energy. This scatter may reflect differences in the methods (gray versus multigroup, flux-limited diffusion or solution of the moment equations, local thermodynamic equilibrium (LTE) versus nonLTE, opacity floors, etc.), but it is also deeply rooted in the fundamental degeneracies of SN light curves \citep{d19_sn2p,goldberg_sn2p_19}. Such degeneracies are further aggravated when the constraint from the photospheric velocity is ignored (i.e., when photometry is the only constraint used). Even if there were a consensus, inferences from light-curve modeling are compromised by the assumptions made regarding the exploding star, such as assuming a progenitor in hydrostatic equilibrium (e.g., \citealt{bronner_rsg_25,laplace_23ixf_26}), a spherically symmetric progenitor (e.g., \citealt{goldberg_3d_rsg_22,goldberg_sbo_22,ma_rsg_25}), or ignoring ejecta clumping \citep{d18_fcl,dessart_audit_rhd_3d_19}.

Archival nebular-phase models have been confronted with observations of \sn\ at 200--500\,d and suggest a 12--15\,\msun\ progenitor \citep{ferrari_23ixf_24,folatelli_23ixf_25,kumar_23ixf_25,wynn_iii_25,wynn_sed_25}. However, these studies relied on standard RSG progenitors and typically ignored the potential influence of interaction power or dust.

The present study is one of three that will examine in detail the full evolution of \sn\ from early to late times, focusing in turn on the different phenomena relevant to those main phases: the interaction phase lasting until about 20\,d (i.e., the SN~IIn and cold dense shell (CDS) phases in the nomenclature of \citealt{dessart_review_26}), the photospheric phase until the onset of the nebular phase (i.e., 20 to 115\,d; this study), and the nebular phase (i.e., 120 to 1000\,d). The methodology in each of these works will differ, in particular between the SN~IIn phase wherein the interaction is treated with radiation hydrodynamics and the nonmonotonic velocity solver in \cmfgen\ \citep{D15_2n} and the subsequent evolution (i.e., 20--1000\,d) wherein the nonLTE time-dependent radiative transfer is solved together with the interaction power in \cmfgen\ \citep{dessart_csm_22,dessart_late_23}.

This paper is organized as follows. In the next section, we summarize the observational dataset used in this work, covering the UV, optical, NIR, and MIR from 20 to 115\,d. Section~\ref{sect_setup} presents the numerical setup and describes in detail the procedure for the modeling of \sn. We confront in Section~\ref{sect_rt} the results of our modeling to the flux-calibrated multiwavelength multiepoch observations of \sn. A comparison of these models to the bolometric and $V$-band light curves of \sn\ is presented in Section~\ref{sect_lc}. We then discuss some specifics of \sn. Section~\ref{sect_kinks} explores the origin of profile kinks observed in optical and IR lines.\footnote{By kink, we mean any localized flux deficit or absorption appearing somewhere blueward of a line, typically within the trough and its outermost, blueward extent where the flux joins with the continuum or some other, overlapping line.} We explore in Section~\ref{sect_asym} the impact of asymmetry on the radiative properties, as would result from an asymmetric CSM or a combination of asymmetric ejecta and CSM. In Section~\ref{sect_uv} we discuss the UV properties of \sn\ and in particular the difficulty of extracting detailed information from the spectra in the UV, here obtained at two epochs with the Hubble Space Telescope \citep[HST;][]{bostroem_23ixf_24}. Section~\ref{sect_ir} presents our results in the IR, focusing on the JWST observations at 33.6\,d  \citep{derkacy_23ixf_26}. We discuss in Section~\ref{sect_fvol} the sensitivity of our results to the adopted clumping in the CDS. Section~\ref{sect_conc} summarizes our conclusions. Appendix~\ref{sect_more} includes additional figures that further document the results discussed in detail in the main text.


\section{Observations}
\label{sect_obs}

All the observations used here have been published in previous papers. Specifically, the optical spectra and photometry are from \citet{zheng_23ixf_25}, augmented at 71.5\,d with the data from \citet{park_23ixf_25}. The UV data at epochs of 24.7 and 66.5\,d are from \citet{bostroem_23ixf_24}. The NIR data at 37.5 and 67.5\,d are from \citet{park_23ixf_25}. The NIR and MIR observations obtained with JWST at 33.6\,d are from \citet{derkacy_23ixf_26}. Full details regarding these observations are to be found in the cited publications. A summary of the epochs used for the modeling and the sources of these data is given in Table~\ref{tab_obs}.

The optical spectra of \citet{zheng_23ixf_25} were flux-calibrated to match the optical photometry. When multiple datasets are used for the same epoch, some flux mismatches may occur. For example, at the epoch of 34.5\,d, the optical spectrum is well calibrated in flux with an offset of 0.04\,mag in $V$, 0.01\,mag in $R$, and 0.004\,mag in $I$. The $K$-band magnitude at that time is 10.68\,mag for the NIR data of \citet{park_23ixf_25} and the NIR/MIR of \citet{derkacy_23ixf_26}, and these two IR spectra overlap in the NIR. However, the NIR spectrum from \citet{park_23ixf_25} is offset from that of \citet{zheng_23ixf_25}. Part of the offset may arise from the different post-explosion times, which is a few days in most cases. We have not tried to resolve this issue since all the data we used have already been published. Simultaneous spectroscopy and photometry is not always available, in particular in the IR.

Throughout this work, we adopt for \sn\ a distance of 6.85\,Mpc \citep{riess_h0_22}, a redshift of $z = 0.000804$ \citep{perley_23ixf_23}, a total visual extinction $A_V=$\,0.127\,mag ($E(B-V)=$\,0.041\,mag; \citealt{jacobson_galan_23ixf_23}), and a time of first light at MJD\,$=$\,60082.788 \citep{li_23ixf_24}.

\begin{table}
\caption{Days since first light at which the observations of \sn\ were modeled.}
\label{tab_obs}
\begin{center}
\begin{tabular}{c|ccccc}
\hline
Epoch      &        UV     &      optical &       NIR      &      NIR/MIR    \\
\hline
Source     &       (a)     &      (b)     &        (c)     &         (d)     \\
\hline
22.5       &       24.7    &      22.5    &      $\cdots$  &      $\cdots$   \\
31.5       &     $\cdots$  &      31.5    &      $\cdots$  &      $\cdots$   \\
34.5       &     $\cdots$  &      34.5    &      $\cdots$  &      33.6       \\
38.5       &     $\cdots$  &      38.5    &        37.5    &      $\cdots$   \\
52.4       &     $\cdots$  &      52.4    &      $\cdots$  &      $\cdots$   \\
61.5       &       66.5    &      61.5    &        67.5    &      $\cdots$   \\
71.9       &     $\cdots$  &    71.9(e)  &      $\cdots$  &      $\cdots$   \\
84.5       &     $\cdots$  &      84.5    &      $\cdots$  &      $\cdots$   \\
97.4       &     $\cdots$  &      97.4    &      $\cdots$  &      $\cdots$   \\
\hline
\end{tabular}
\end{center}
{\bf Notes:} For each epoch labelled at left, we specify the corresponding time elapsed since first light for each spectral range when multiple datasets were used (the time corresponding to the optical dataset is used when defining the epoch label). The time offset is generally just a few days. The UV data are from \citet[(a)]{bostroem_23ixf_24}, the optical data from \citet[(b)]{zheng_23ixf_25}, the NIR data from \citet[(c)]{park_23ixf_25}, and the NIR/MIR data from \citet[(d)]{derkacy_23ixf_26}; also, (e), the optical data are from \citet{park_23ixf_25}.
\end{table}

\section{Numerical setup}
\label{sect_setup}

The progenitor and explosion models used as initial conditions in this work come from two distinct sources. The first is \citet{HD19}, in which models of a solar-metallicity 15\,\msun\ star were evolved with \mesa\ \citep{mesa1,mesa2} using different assumptions for the RSG mass-loss rate. That is, relative to the standard ``Dutch'' recipe used in \mesa\ at the time, the mass-loss rate was scaled during the RSG phase by a factor of 1.5 (model ``x1p5'') up to 10.0 (model ``x1e1''), yielding, as desired, a range of pre-SN progenitor masses, which with that choice and in that order went from 13.75 down to 4.96\,\msun. All the preSN progenitor models of \citet{HD19} were exploded with the radiation-hydrodynamics code \voned\ \citep{livne_93,dlw10a} to yield an ejecta kinetic energy of $1.2 \times 10^{51}$\,erg but with a range of \nifs\ masses left unconstrained and reflecting the differences in core structure and choice of mass cut. \citet{zheng_23ixf_25} showed that model x5p0 in that series yielded a photospheric-phase duration of 80--90\,d, which is comparable to that observed for \sn.  Because that model x5p0 had only 0.02\,\msun\ of \nifs\ and underestimated the nebular brightness of \sn, we took a slightly more stripped progenitor and thus reran the model x6p0 of \citet{HD19} by enforcing a higher \nifs\ mass of 0.05\,\msun\ (the greater \nifs\ mass extends the photospheric phase, and thus a model with a greater RSG mass loss, a lower progenitor mass, and correspondingly a lower ejecta mass, was needed). The progenitor H-rich envelope mass in models x5p0 and x6p0 is respectively 6.02 and 5.04\,\msun\ \citep{HD19}.

In this work, we used this modified model x6p0 from \citet{HD19} as our reference model for all nonLTE time-dependent radiative-transfer calculations with \cmfgen\ \citep{hm98,HD12,dessart_csm_22} --- we mean here the combined progenitor and explosion model x6p0. The preSN evolution of the models presented by \citet{HD19} were, however, computed in 2014, with a moderate resolution and a small nuclear network, such that the resulting composition was limited to the most abundant isotopes. Furthermore, this x6p0 model was found to have a peculiar composition, such as a small oxygen yield of 0.44\,\msun\ or a strong overabundance of magnesium (four times solar) in the H-rich envelope. Rather than exploring the origin of these peculiarities, and to ensure we used a detailed and robust composition structure for our model, we built a revised x6p0 ejecta model that inherited the original density (or mass), temperature, radius, and velocity structure at a given post-explosion time but took the composition from the s15.2 model of \citet{sukhbold_ccsn_16}. In practice, this was accomplished by remapping in Lagrangian-mass space the s15.2 model composition onto the x6p0 model. Since the x6p0 model has a lower ejecta mass than the s15.2 model (7.57\,\msun\ compared to 10.95\,\msun), this remapping was performed from the innermost ejecta layers outward. The \nifs\ mass was also reset to be 0.05\,\msun. In practice, because we used a simple linear interpolation for this remapping and the two model grids had different resolution, an offset of 10\,\% in \nifs\ mass was introduced, yielding slightly different brightnesses near the end of the photospheric phase and during the nebular phase. This is not a major concern for the present conceptual study, in which we will explore the many degeneracies and complications affecting the observed properties of \sn, including the relatively weak sensitivity of results to small abundance variations. At the 10\,\% level, variations in \nifs\ mass only introduce a global flux scaling but negligible relative-flux changes (i.e., line ratios; see \citealt{DH20_neb}).

Using this reference model x6p0 (together with its updated composition and \nifs\ mass), we then produced model counterparts that included the possibility of interaction with CSM. We thus introduced an outer dense shell (nicknamed hereafter CDS for cold dense shell) with a location at 8000\,\kms\ as required, for example, by the width of the \ha\ emission profile at late times. Many explorations were performed varying the CDS properties including its mass, width (in velocity space), or clumping. Similarly, numerous calculations were done to vary the power injected within the CDS, including location, width, and magnitude. This was done using the same formalism as by \citet{dessart_csm_22}. Here, only a subset of these calculations is shown, focusing on cases that yield the closest match to observations, or choices that highlight important dependencies (for example, with the adopted clumping level within the CDS). The selected models with interaction power (``x6p0/Pwr'' in Table~\ref{tab_init}) are named x6p0 + Pwr($t$) and x6p0 + Pwr1($t$), with specific powers given in Table~\ref{tab_pwr}. These powers are adjusted continuously along the time sequence starting at $\sim$\,20\,d for x6p0 + Pwr($t$), but adjusted only at a specific timestep for x6p0 + Pwr1($t$). These adjustments are made to track the observed changes in the SED as well as the observed evolution in line profile shapes and strengths. We also conducted simulations with a CDS but without interaction power, yet we found that the results were similar to those discussed by \citet{dessart_csm_22} so these simulations are not included here.

A summary of the ejecta model properties used in this work is given in Table~\ref{tab_init}. Slight differences between models arise from the linear interpolation used in remapping between codes or when resetting and smoothing the inner ejecta density to get rid of the density jump that inevitably forms when assuming spherical symmetry. These slight variations bear no relevance to the conclusions of this study.

\begin{figure}
\centering
\includegraphics[width=0.9\hsize]{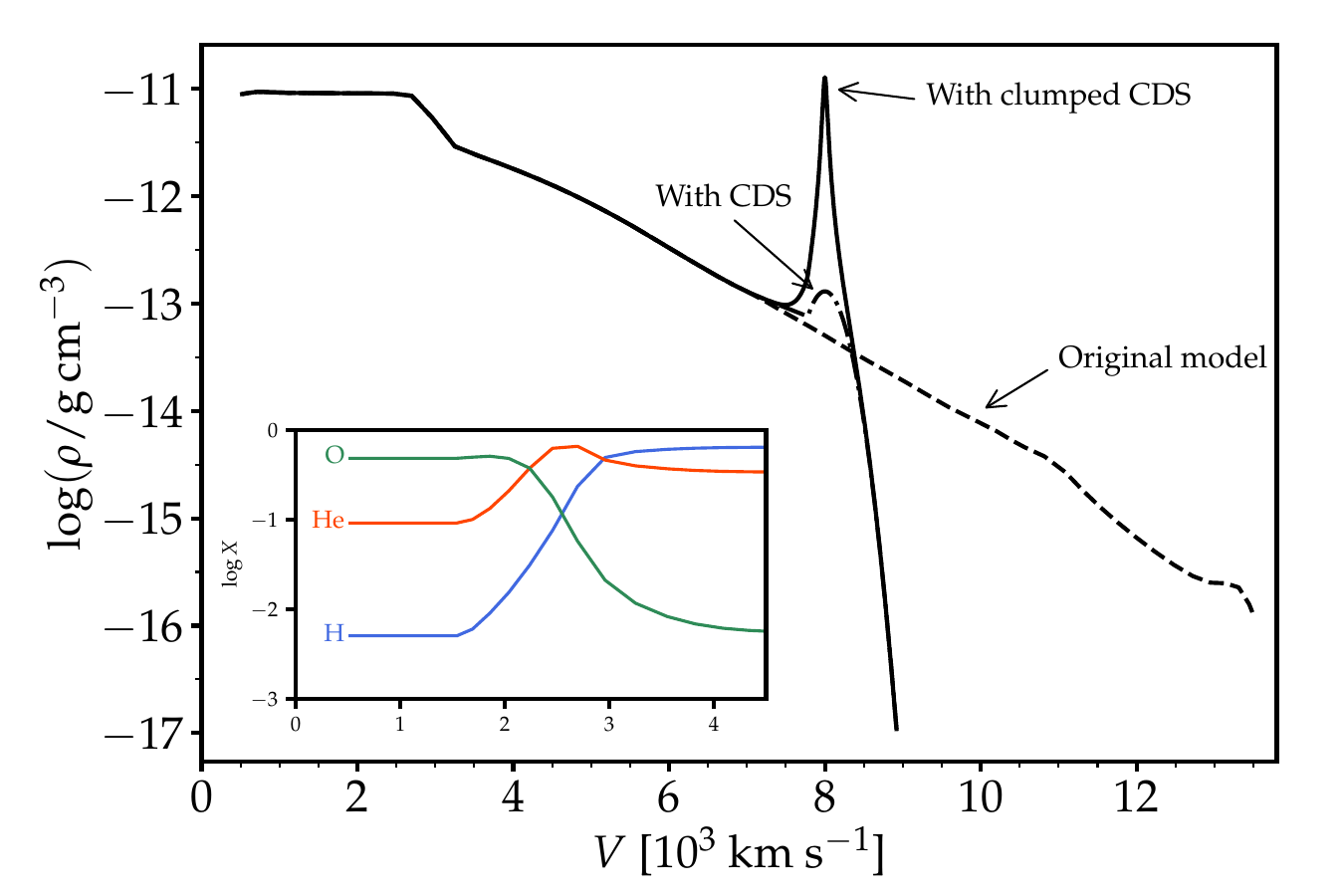}
\includegraphics[width=0.9\hsize]{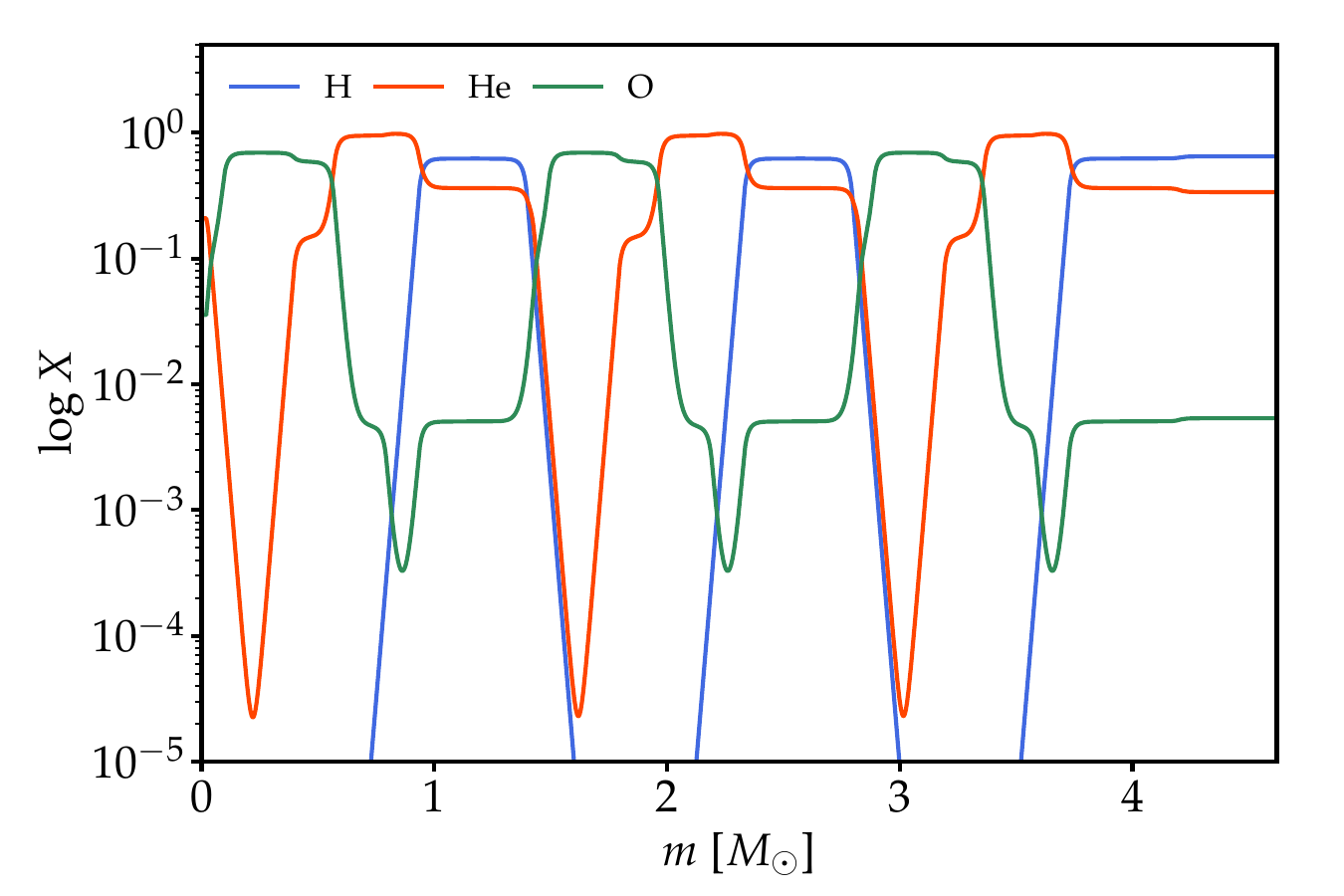}
\caption{{\it Top:} Density and composition structure at 22.3\,d for the ejecta models used in this work. This includes the original, reference model x6p0, and its variants with interaction power and a CDS (e.g., x6p0 + Pwr($t$)). {\it Bottom:} Shuffled-shell composition used in a subset of models at the end of the photospheric phase. Only H, He, and O are shown for better visibility.
\label{fig_init}
}
\end{figure}

We also show the density structure as well as the composition of the adopted models  in Fig.~\ref{fig_init}. In essentially all simulations, we assumed ejecta in which the composition was mixed using a standard boxcar algorithm because it can be resolved with fewer grid points (top panel of Fig.~\ref{fig_init}). At the end of the photospheric phase, when the inner metal-rich ejecta are revealed, we found that allowing for macroscopic mixing without microscopic mixing made a difference. We mimicked this chemical segregation with a shuffled-shell composition structure (bottom panel of Fig.~\ref{fig_init}; see details on the method in \citealt{dessart_shuffle_20} and its application to core-collapse SNe in \citet{dessart_sn2p_21} and \citet{dessart_snibc_21}). In those cases, we imported the shuffled-shell structure of the s15.2 model into a converged, microscopically mixed model at the same timestep, and reran \cmfgen\ for this timestep.

\begin{table}
\caption{Summary of model properties.}
\label{tab_init}
\begin{center}
\begin{tabular}{l@{\hspace{2mm}}c@{\hspace{2mm}}c@{\hspace{2mm}}c@{\hspace{2mm}}c@{\hspace{2mm}}c@{\hspace{2mm}}}
\hline
 Model     &    \mej\ &   \ekin\  &  $M$(\nifs)   & \mcds\      & \esh\            \\
           &  [\msun] &    [10$^{51}$\,erg]  &  [\msun]    & [\msun]     & [10$^{40}$\ergs] \\
\hline
x6p0o      &     7.57 &    1.20   &   0.045     &   $\cdots$  & $\cdots$         \\
x6p0       &     7.53 &    1.24   &   0.053     &   $\cdots$  & $\cdots$         \\
x6p0/Pwr   &     7.47 &    1.18   &   0.053     &   0.2       & 4--200           \\
\hline
\end{tabular}
\end{center}
{\bf Notes:} Model x6p0o refers to the original model x6p0 from \citet{HD19} but exploded to yield a \nifs\ mass roughly compatible with that inferred for \sn. Model x6p0 corresponds to the same progenitor model whose composition has been reset to that of model s15.2 from \citet{sukhbold_ccsn_16}. x6p0/Pwr is the same as x6p0 but with a CDS in which an interaction power \esh\ is injected, corresponding in this work to models x6p0 + Pwr($t$) and x6p0 + Pwr1($t$).
\end{table}

\begin{table}
\caption{Evolution of the powers injected in the CDS for models x6p0 + Pwr($t$) and x6p0 + Pwr1($t$).}
\label{tab_pwr}
\begin{center}
\begin{tabular}{|c|c|c||c|c|c|}
\hline
Age    &  Pwr($t$)   & Pwr1($t$)                 & Age    &  Pwr($t$)   & Pwr1($t$)                 \\
\hline
\multicolumn{1}{|c|}{[d]}     &  \multicolumn{2}{c||}{[10$^{40}$\,\ergs]} & \multicolumn{1}{|c|}{[d]}    &  \multicolumn{2}{c|}{[10$^{40}$\,\ergs]} \\
\hline
27.0   &  178.9      &      &  63.6   &  60.0       & 8.0  \\
29.7   &  155.6      &      &  70.0   &  45.0       &      \\
32.7   &  130.0      &      &  77.0   &  28.6       & 5.0  \\
35.9   &  120.2      & 40.0 &  82.0   &  16.8       &      \\
39.5   &  109.4      &      &  87.0   &  5.0        &      \\
43.5   &  97.5       &      &  92.0   &  4.8        &      \\
47.8   &  84.4       &      &  97.0   &  4.6        &      \\
52.6   &  70.0       & 10.0 &  102.0  &  4.4        &      \\
57.8   &  65.2       &      &  112.0  &  4.4        &      \\
\hline
\end{tabular}
\end{center}
\end{table}

As in \citet{dessart_csm_22}, we introduce clumping in the CDS in essentially all cases.\footnote{The mass per grid zone, or spherical shell in 1D, is $f_{\rm vol}\,\rho\,dV$, where $f_{\rm vol}$ is the volume filling factor, $\rho$ is the density, and $dV$ is the volume of the cell. When introducing clumping (i.e., a radial compression in 1D), $\rho$ is boosted by some factor but the material occupies a volume reduced by the same factor, so that the zone mass is unchanged. Similarly, for constant ionization, such a clumping does not change the optical depth (assuming all opacity sources were only linearly dependent on density, which they are not).}  This choice was initially made to mimic the strong compression that takes place in ejecta interacting with CSM as well as to favor recombination within the CDS and limit a surge in temperature or ionization. In reality, the CDS would have a different structure in three dimensions (3D), being both compressed, extended, and broken into a shrapnel structure (see, e.g., \citealt{chevalier_snr_92,chevalier_blondin_95}). In this work, any surge in ionization will be more limited than in \citet{dessart_csm_22} because we adopted a much larger model atom and in particular treated additional ionization stages. That is, we covered stages \one\ to \four\ for C, N, O, Ne , Mg, Si, and S (except Ne\one). In addition, we included a huge model atom for Fe, Co, and Ni, with a total of 1000--3000 levels for each ionization stage. With these additional ions and expanded number of levels, there is less risk of running out of coolants at any grid location, which is otherwise the primary reason for the temperature to rise unphysically. Observations of \sn\ indicate that the ionization is rather low and standard for an SN~II during the post-SN~IIn, photospheric phase, despite the clear influence of ejecta interaction with CSM.

More specifically, the model atoms used in all calculations of this study are based on the atomic data reported by \citet{blondin_21aefx_23} and included (in order of increasing atomic weight and then ionization) H\one\ (26,36), He\one\ (40,51), He\two\ (13,30), C\one\ (14,26), C\two\ (14,26), C\three\ (62,112), C\four\ (59,64), N\one\ (44,104), N\two\ (23,41), N\three\ (25,53), N\four\ (56,92), O\one\ (21,51), O\two\ (54,123), O\three\ (44,86), O\four\ (76,152), Ne\two\ (22,91), Ne\three\ (32,80), Ne\four\ (30,67), Na\one\ (22,71), Mg\one\ (39,122), Mg\two\ (31,80), Mg\three\ (31,99), Mg\four\ (18,100), Al\two\ (26,44), Al\three\ (27,60), Si\one\ (100,187), Si\two\ (31,59), Si\three\ (33,61), Si\four\ (37,48), S\one\ (106,322), S\two\ (56,324), S\three\ (48,98), S\four\ (27,67), Ar\one\ (56,110), Ar\two\ (134,415), Ar\three\ (32,346), K\one\ (25,44), Ca\two\ (21,77), Ca\three\ (16,40), Ca\four\ (18,69), Sc\two\ (38,85), Sc\three\ (25,45), Ti\two\ (37,152), Ti\three\ (33,206), Cr\two\ (28,196), Cr\three\ (30,145), Cr\four\ (29,234), Fe\one\ (413,1142), Fe\two\ (228,2698), Fe\three\ (96,1001), Fe\four\ (100,1000), Fe\five\ (139,1000), Co\two\ (112,1005), Co\three\ (88,1075),  Co\four\ (56,1000), Ni\one\ (56,301), Ni\two\ (59,1000), Ni\three\ (47,1000), and Ni\four\ (64,1000); here, the numbers in parentheses correspond to the number of super levels and full levels (for additional information on super levels, see \citealt{hm98}).

In all simulations of this work, we accounted for five two-step decay chains whose parent isotopes are \nifs, \iso{57}Ni, \iso{52}Fe, \iso{48}Cr, and \iso{44}Ti, with a proper description of all isotopes associated with a given species. Finally, a careful gridding was employed to finely resolve the CDS, generally assumed to be clumped, by allocating about 70 points for the CDS and as many for the rest of the ejecta. In the special case of shuffled-shell models, a fine resolution was also needed to capture the rapid abundance variations with depth, requiring a total of about 300 grid points --- this situation is analogous to the late-time simulations presented by \citet{dessart_late_23}.

\section{Radiative-transfer models for the spectroscopic evolution of \sn}
\label{sect_rt}

In this work, we developed tailored models for \sn\ using multiple constraints from photometry and spectroscopy at multiple epochs. A bad way to proceed would be to take some random progenitor, explode it, and iterate on preSN evolution and explosion parameters until a good match to \sn\ was found at 20\,d. A valid model must first match the photospheric-phase duration, early- and late-time brightness, ejecta expansion rate, etc. That is, one must first have a global model that matches these main properties before focusing on secondary aspects concerning the CDS and other factors since \sn\ was no longer an interacting SN after 20\,d.

Hence, the model x6p0 described above (originally from \citealt{HD19}) was first modeled throughout the photospheric phase to verify that it matched the photospheric-phase duration, the evolution of various spectral line widths (e.g., Fe\two\ lines, \ha), as well as the nebular-phase brightness. This model was then augmented with a CDS and by various time-dependent interaction powers, and subsequently compared with observations from about 20\,d until the onset of the nebular phase. In this section, we compare these model spectra to observations, primarily in the optical, but also including UV, NIR, and even MIR data whenever available. We discuss in detail just a few epochs, and show the results for additional epochs in Appendix~\ref{sect_more}.

We focused primarily on modeling flux-calibrated spectra, both relatively (i.e., the accurate distribution of the flux vs. wavelength) and absolutely (all optical spectra were scaled to match the observed $V$-band brightness of \sn\ at the corresponding epoch). The present models are thus constrained by a broad spectral energy distribution (SED) and thus bypass the need to infer a bolometric luminosity, which is the main constraint typically used in light-curve modeling. The radiative-transfer computations being done at several 10$^5$ frequencies and in full nonLTE, the relative distribution of the flux is also accurately computed, in contrast to radiation-hydrodynamics codes in which the gas is treated in LTE and blackbody arguments are made. A final reason for focusing on (flux-calibrated) spectra first, is that the interaction power affecting \sn\ cannot be constrained from photometry. It is through the careful analysis of line profiles, and the UV spectra when available, that this interaction power may be constrained. Thus, in this work, light-curve comparisons to \sn\ will be presented in the next section, after the model confrontation to spectra.

Overall, we found that interaction power in \sn\ at epochs 20 to 120\,d affected in a nonobvious manner the SED, for example by channeling a few percent of the total flux into the UV range. The optical typically contains 70--80\,\% of the model escaping radiation, with the rest falling mostly in the NIR (only about 5\,\% of the SN flux falls beyond 2.0\mic). Hence, to assess the level of interaction, we relied mostly on the \ha\ profile, and specifically on the emission arising from the CDS in which the ejecta interaction with CSM injects power. This also emphasizes the importance of \ha\ for diagnosing interaction when UV data are absent.


\begin{figure}
\centering
\includegraphics[width=0.9\hsize]{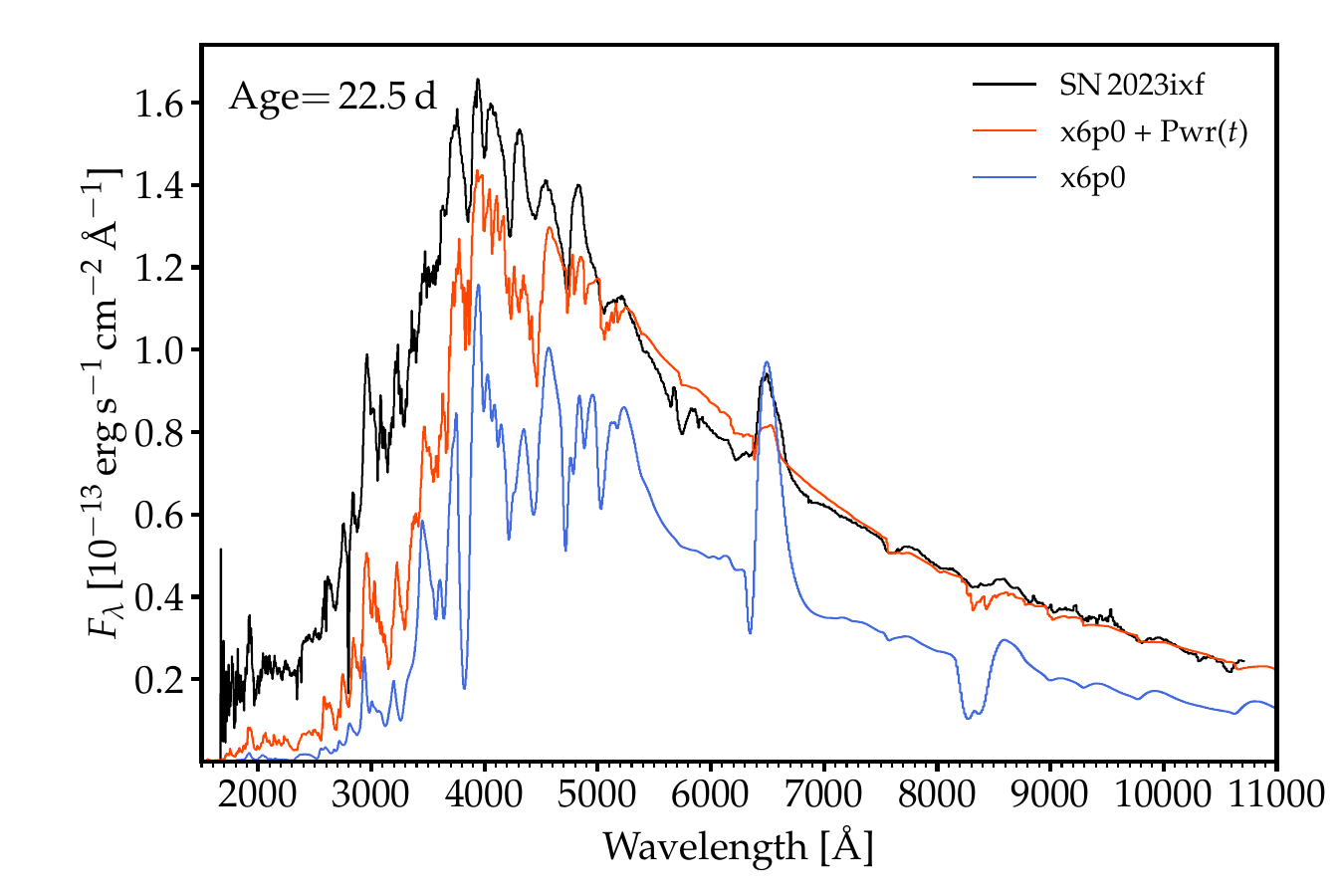}
\caption{Comparison between the UV and optical observations of \sn\ at an epoch of 22.5\,d (black) and model x6p0 with an interaction power of $2 \times 10^{42}$\,\ergs\ (red) and without (blue).
\label{fig_spec_22p5d}
}
\end{figure}

\subsection{Comparison with observations at 22.5\,d}
\label{sect_22p5d}


Figure~\ref{fig_spec_22p5d} shows a comparison between the observations of \sn\ at an epoch of 22.5\,d (see Table~\ref{tab_obs} for the exact times since first light for each spectral range) and model x6p0 at 24.5\,d influenced by an interaction power of $2 \times 10^{42}$\,\ergs\ as well as with the reference model that exploded in a vacuum (no CDS, no injection power; see Fig.~\ref{fig_init} for the corresponding ejecta structures). The model with interaction power yields a better match to the overall SED, with a $\sim$\,50\,\% boost to the flux throughout the UV and optical ranges and a widespread impact on all line profiles. Specifically, whereas the reference model x6p0 exhibits line profiles with both strong absorption and emission, the model with interaction power exhibits line profiles with weak emission and absorption relative to the continuum --- the same lines are, however, present in both models and thus irrespective of the interaction. The origin of these line-profile changes is primarily the change in density structure since in the x6p0 model with interaction, the spectrum forms exclusively in the CDS (which is marginally optically thick at that time; the electron-scattering photosphere is at 8020\,\kms\ and thus lies in the CDS at 24.5\,d). The steep density fall-off outside of the CDS reduces the volume of emission and leads to a more featureless spectrum. These profile characteristics are clearly apparent in \sn, with weak emission and absorption in Ca\two\,H\&K, \ha, or the Ca\two\ NIR triplet. The choice of a CDS velocity of 8000\,\kms\ is also confirmed by the good match to the width of the line profiles.

Interaction power and the presence of a CDS have therefore a strong impact on the full spectrum (including brightness and thus photometry) even though there is no direct signature of interaction as commonly defined (e.g., presence of lines with electron-scattering wings). This influence is of two distinct natures. First, the CDS is largely inherited from previously swept-up CSM, and its structure (mass, density, velocity) will remain essentially unchanged for weeks and months (it can only grow in mass and recede in velocity as more CSM is swept up, but this is a slow process). Second, the interaction power, which we inject within the CDS, reflects the ongoing interaction of the ejecta with CSM. This power is nonzero only if some ``fresh'' CSM is continuously made available, and is instrumental to boost the SN luminosity. This impact on line profiles was already reported by \citet{HD19}, in particular relative to the absorption-to-emission flux ratio in lines like H$\alpha$ (for the observational counterpart at the time, see \citealt{gutierrez_ha_14}). However, in the simulations of \citet{HD19}, only the CDS was accounted for and was present as a fossil of interaction with CSM at the time of and immediately after shock breakout --- all the CSM had been swept up before the start of the simulations at 10--15\,d.

Using the model for line identification, we find that Fe\two\ below 3000\,\AA\ and Ti\two\ beyond 3000\,\AA\ dominate the metal-line blanketing in the UV and in the blue part of the optical (Ni and Cr would contribute if Fe were taken out but it seems that solar-metallicity Fe swamps most contributions by other metals; \citealt{bostroem_23ixf_24}). Ti\two\ is the origin of the broad flux depression between 4000 and 4500\,\AA. Fe\two\ also causes isolated absorptions in the optical, the main ones being located at 4923, 5018, and 5169\,\AA. Ca\two\ produces essentially two features with Ca\two\,H\&K and the NIR triplet. The Si\two\ doublet at 6355\,\AA\ causes the notch blueward of \ha\ --- this notch is present in both models but stronger in the model with interaction (the excess absorption arises from the CDS). \mgiiuv\ is the only predicted line from Mg\two, although its contribution in the UV is swamped by Fe\two\ and Ti\two\ blanketing. The absorption around 5900\,\AA\ is due to \naid\ rather than He\one\,5875\,\AA. O\one\,7774\,\AA\ is also present. Finally, the models predict the full set of H\one\ Balmer lines, but only \ha\ is clearly visible. \hb\ is affected by Fe\two\ blanketing and the higher transitions in the series nearer the Balmer edge are affected by Ti\two\ blanketing. The model also predicts a strong Balmer jump in the continuum flux that is erased by metal-line blanketing.

Overall, model x6p0 with interaction power captures the salient features of the observations of \sn\ at 22.5\,d. Model deficiencies include an underestimate of the UV flux, which may arise in part from the fact that the model has not yet fully relaxed to the injected power (i.e., we inject the power in the first step of the time sequence at 22.3\,d, whereas in reality interaction in \sn\ has been ongoing at all times since shock breakout). Later in this study, we will explore how the adopted level of clumping in the CDS (see also \citealt{bostroem_23ixf_24}) as well as asymmetry can also modulate the UV flux. In the optical, the main deficiency is the overestimate of the blanketing in the blue part, which may arise from a too low ionization or possibly a metallicity effect \citep{d13_sn2p,li_23ixf_25}. The other mismatch is the emission strength of \ha, which is too weak in the model, although the absorption and the filling-in effect from the CDS emission is well matched. Asymmetry and clumping are two possible origins of this mismatch.


\begin{figure}[h]
\centering
\includegraphics[width=0.9\hsize]{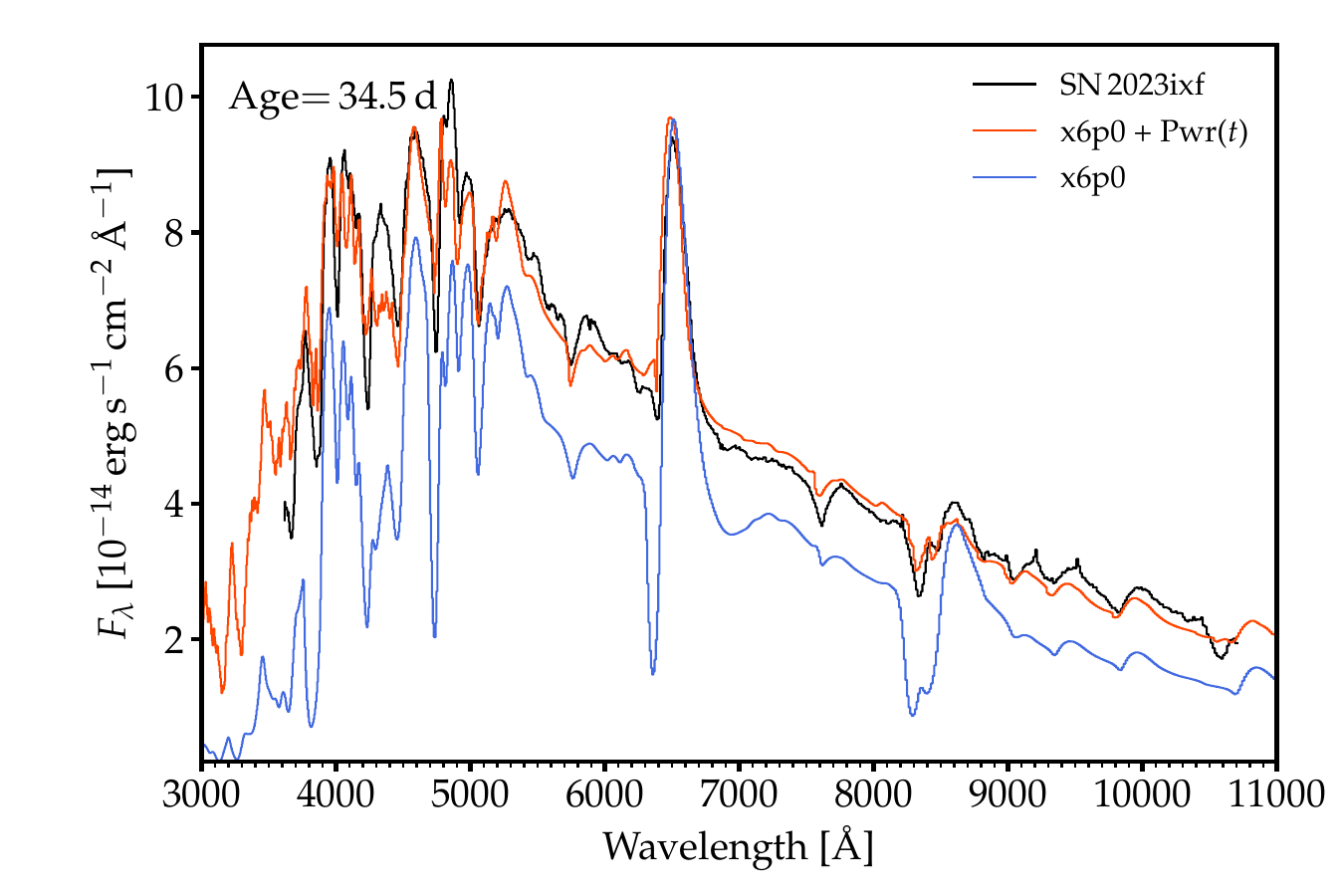}
\includegraphics[width=0.9\hsize]{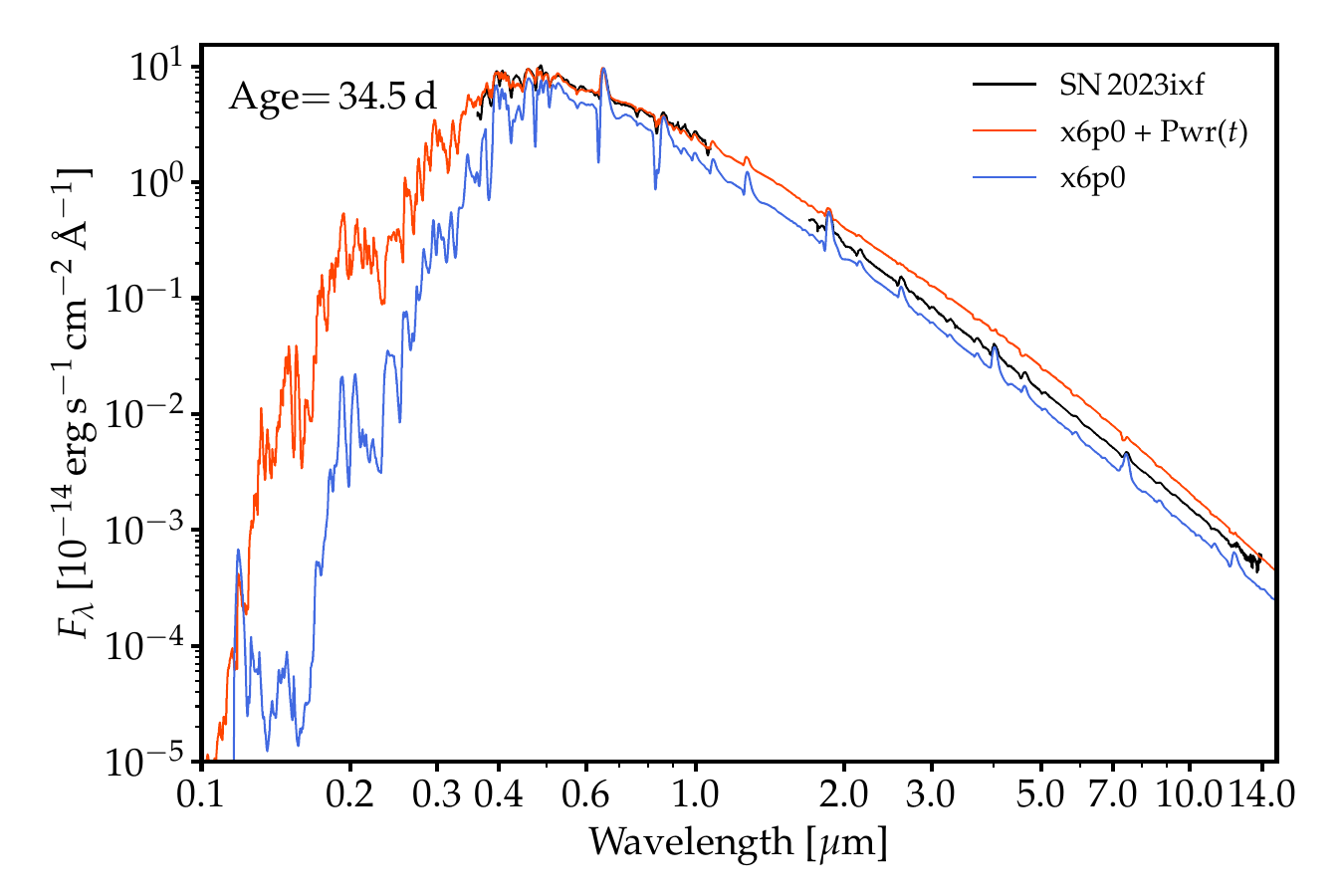}
\caption{Comparison between the optical (top) and optical and IR (bottom) observations of \sn\ at an epoch of 34.5\,d (black) and model x6p0 with an interaction power of $1.2 \times 10^{42}$\,\ergs\ (red) and without (blue).
\label{fig_spec_34p5d}
}
\end{figure}

\subsection{Comparison with observations at 34.5\,d}
\label{sect_34d}


Figure~\ref{fig_spec_34p5d} shows a comparison of the optical (top panel) and optical/IR (bottom) observations of \sn\ at 34.5\,d with model x6p0 at 35.9\,d influenced by an interaction power of $1.2 \times 10^{42}$\,\ergs\ as well as with the reference model that exploded in a vacuum. Model x6p0 with interaction power yields a good match in the optical, is slightly overluminous in the blue part of the optical and thus perhaps in the UV too, and is somewhat too luminous in the IR. Interaction tends to inhibit recombination and limits in this way the strength of metal-line blanketing in the optical. Evidently, the model with interaction yields a much better match to all line profiles,  with line identifications remaining largely unchanged from the previous epoch. Notably, the \ha\ profile is better matched in strength by the interaction model, with contributions blueward of line center due to a combination of Si\two\ and Fe\two\ lines. In the red part of the optical, beyond the Ca\two\ NIR triplet, H\one\ transitions in the Paschen series are observed and roughly reproduced, together with the blue edge of He\one\,10,830\,\AA.

The bottom panel of Fig.~\ref{fig_spec_34p5d} illustrates the full SED of both models, from the UV to the IR, relative to the observations of \sn\ in the optical and IR. Interaction power, which is known to boost the UV flux \citep{chevalier_fransson_94,dessart_csm_22}, is also found here to boost the IR flux. This arises from the presence of the CDS, which is partially ionized and optically thick owing to bound-free and free-free opacities in the IR. With interaction power, the level of ionization and the temperature in the CDS rise and enhance the IR flux. However, we find that model x6p0 with an interaction power of $1.2 \times 10^{42}$\,\ergs, despite a satisfactory match to the optical range, overestimates the IR flux. Further discussion on this topic is provided in Section~\ref{sect_ir}.


\begin{figure}[h]
\centering
\includegraphics[width=0.9\hsize]{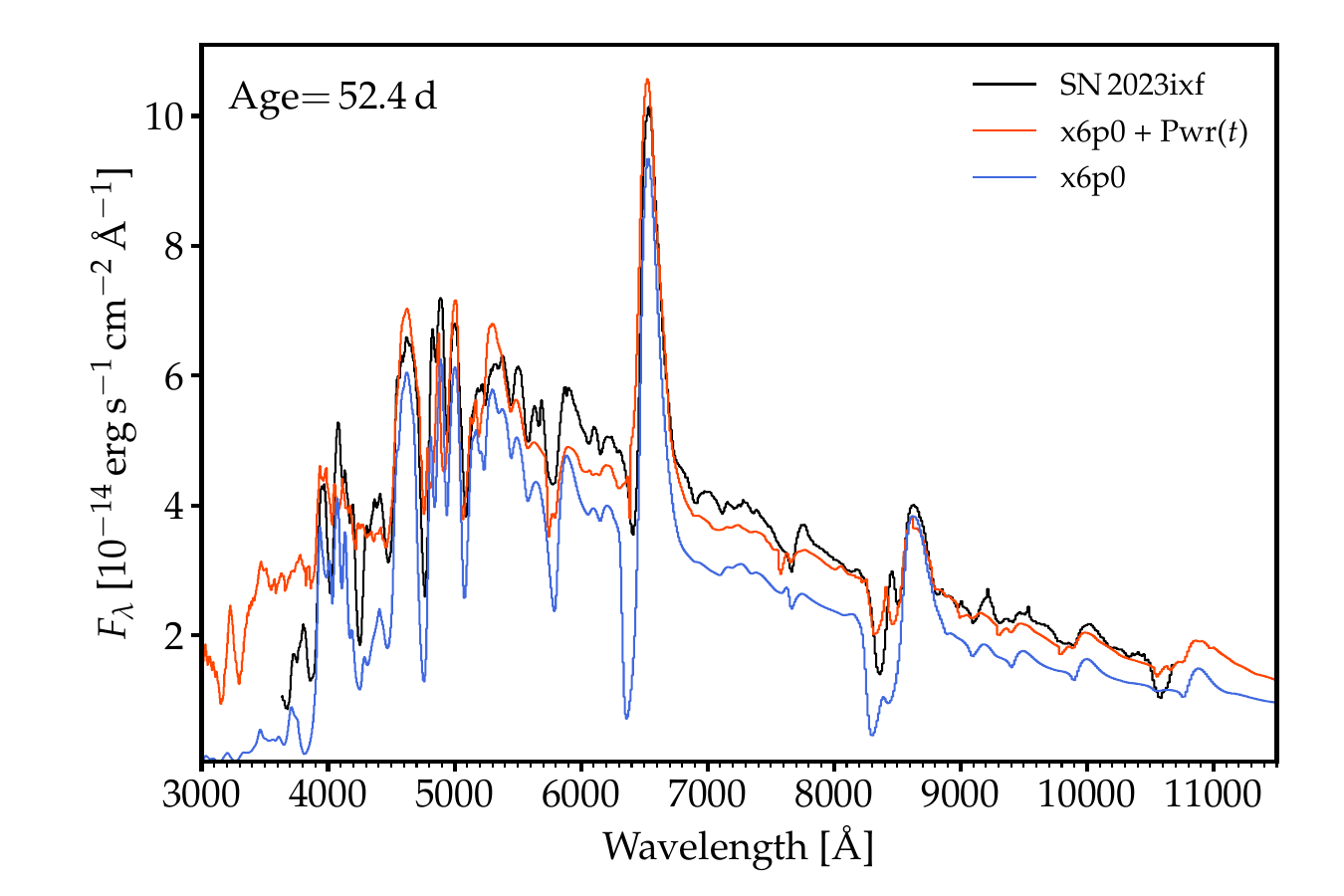}
\caption{Comparison between the optical observations of \sn\ at an epoch of 52.4\,d (black) and model x6p0 at 52.6\,d with an interaction power of $7.0 \times 10^{41}$\,\ergs\ (red) and without (blue).
\label{fig_spec_52p4d}
}
\end{figure}

\subsection{Comparison with observations at 52.4\,d}
\label{sect_52p4d}

%
%

Figure~\ref{fig_spec_52p4d} shows a comparison of the optical observations of \sn\ at 52.4\,d with model x6p0 at 52.6\,d influenced by an interaction power of $7.0 \times 10^{41}$\,\ergs\ as well as with the reference model that exploded in a vacuum. The model with interaction yields a good match to the observations, both for the overall flux level and line profiles. \sn\ exhibits a typical behavior observed in all Type II SNe at the recombination epoch with a narrowing of line profiles testifying for the recession of the spectrum formation region as well as a strengthening of metal-line blanketing by Fe or Ti and of strong lines like \naid\ or the Ca\two\ NIR triplet. Compared to the previous epoch, the \ha\ profile has also strengthened relative to the continuum and even exhibits an absorption trough on its blue side. The model with interaction yields a much better match to all line profiles compared to the model without interaction, which predicts strong absorption troughs in many lines in conflict with observations. We also find that the model with interaction produces double-absorption features in strong optical lines like \naid, O\one\,7774\,\AA, or the H\one\ lines from the Paschen series around 1\,$\mu$m. In observations, these kinks are absent in some lines (likely because of a combination of, for example, optical-depth effects, line overlap, poor signal, or poor resolution) but are clearly present in the P-Cygni profiles of transitions in the red part of the optical such as H\one\ lines at 9546.0, 10,049.4, or 10,938.1\,\AA\ (the last of these is blended with He\one\,10,830.2\,\AA).

In SNe with past or ongoing signatures of interaction, a pair of absorptions associated with the outer CDS and the deeper photosphere are naturally expected \citep{chugai_hv_07,dessart_csm_22}. The lack of such double kinks in line profiles is in no way a proof that there is no CDS (i.e., no past or ongoing interaction). Asymmetry can explain this peculiarity (Section~\ref{sect_asym}), but it may also arise from a temporal or an observational bias affecting the optical range at early times since these double absorptions are observed in \sn\ at later times in the NIR (see Sections~\ref{sect_61p5d} and \ref{sect_kinks}).


\begin{figure}
\centering
\includegraphics[width=0.9\hsize]{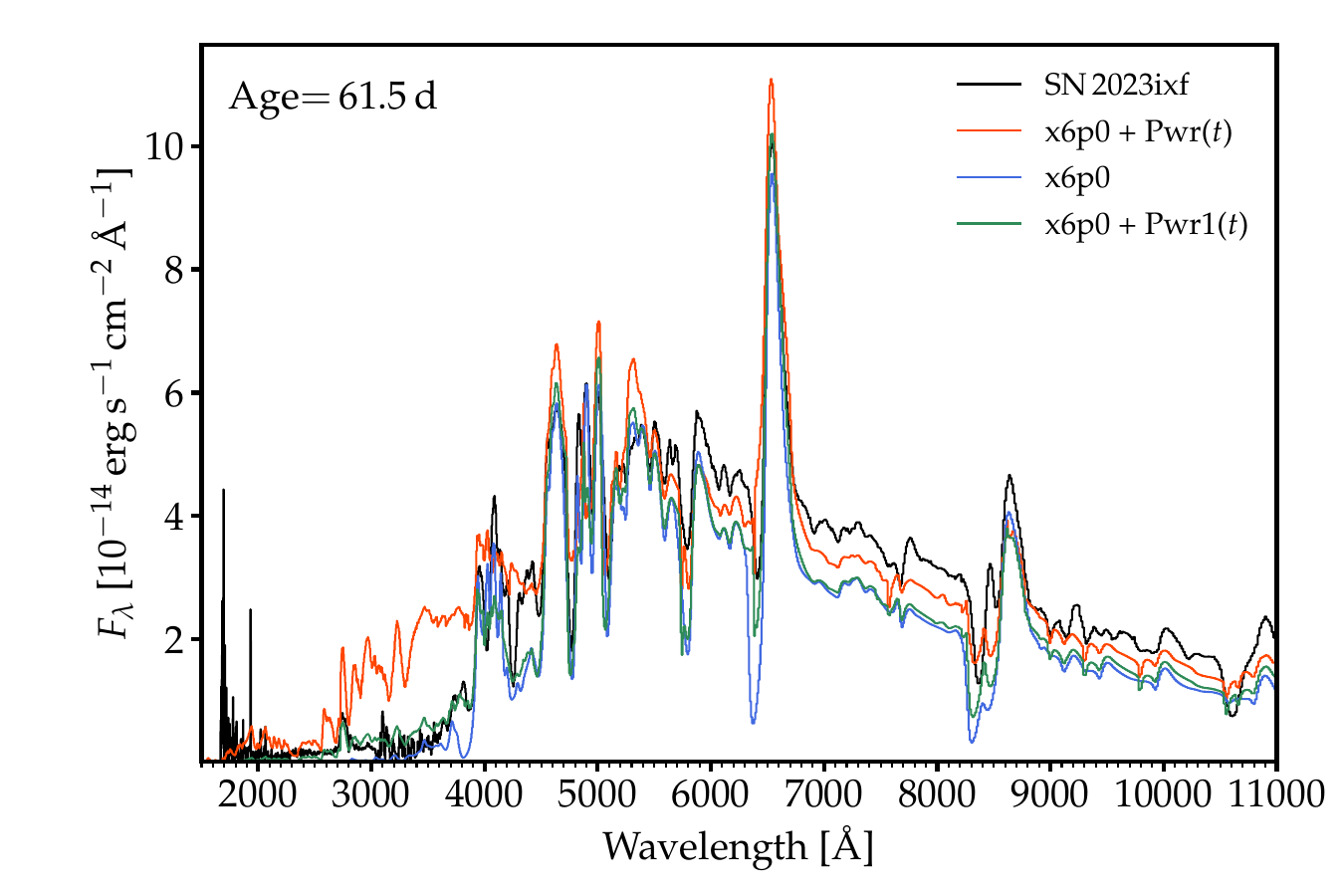}
\includegraphics[width=0.9\hsize]{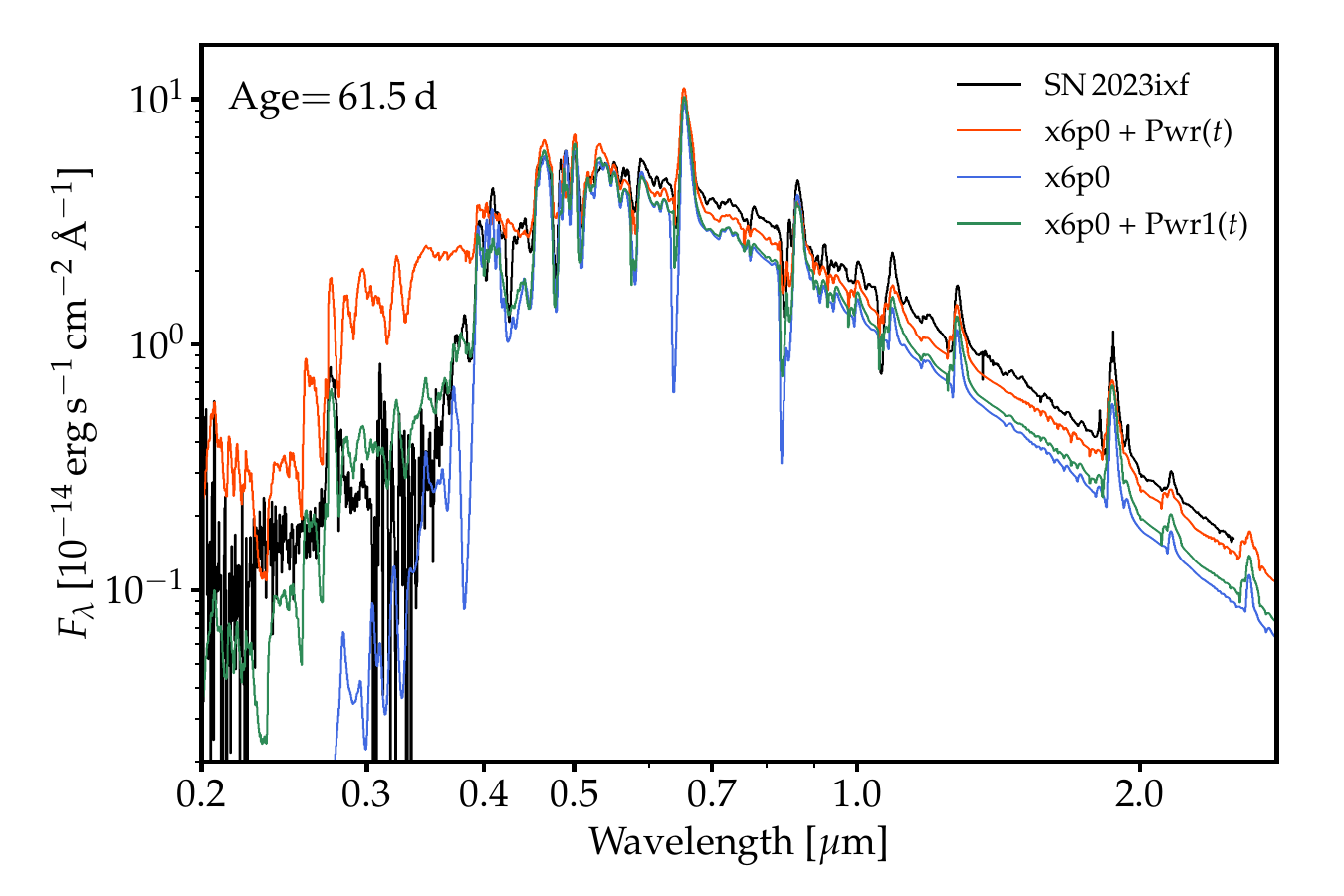}
\includegraphics[width=0.9\hsize]{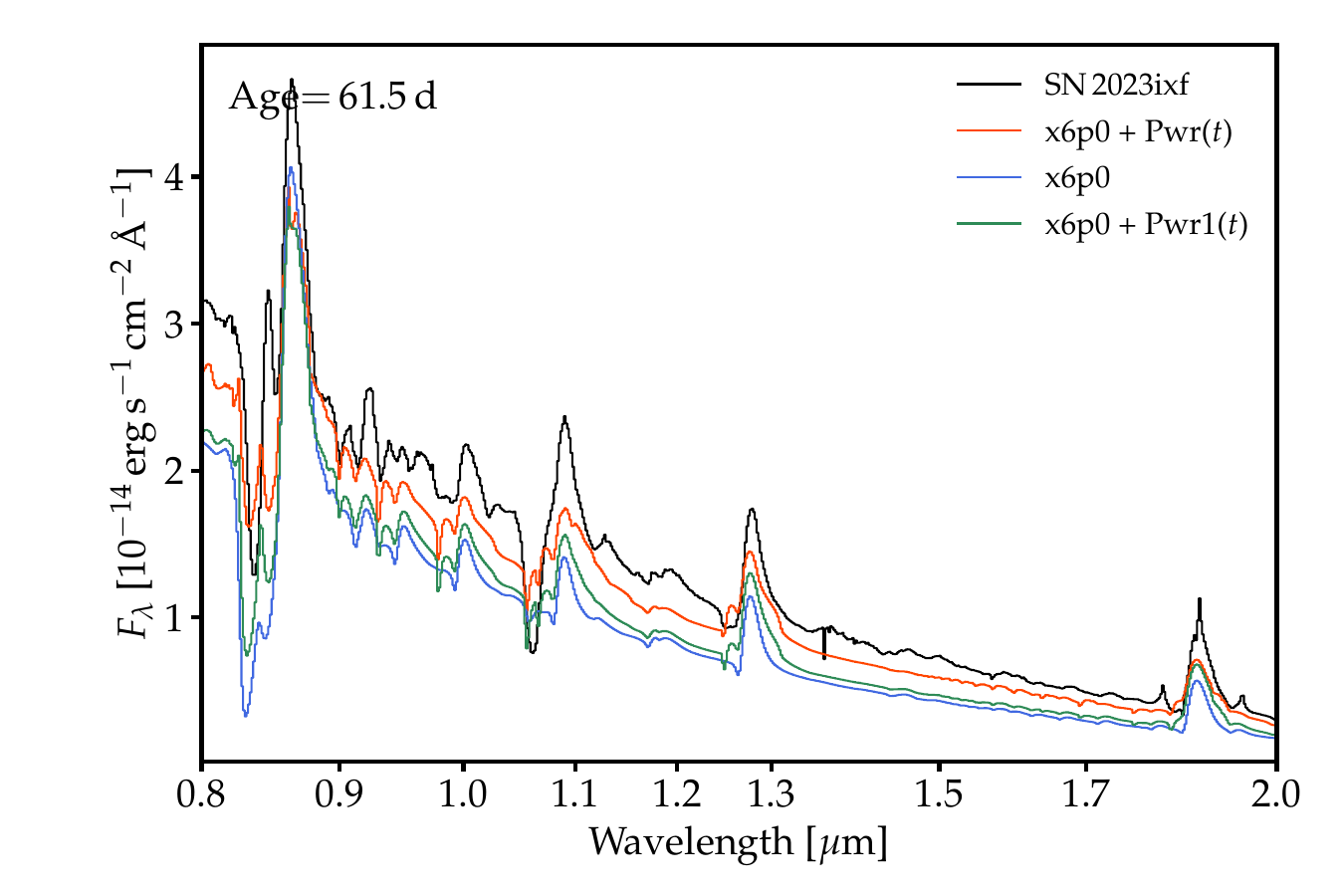}
\caption{Comparison between the UV, optical, and NIR observations of \sn\ at an epoch of 61.5\,d (black; different spectral ranges were observed at distinct times; see Table~\ref{tab_obs} and Section~\ref{sect_61p5d}) and model x6p0 at 63.6\,d with an interaction power of $6.0 \times 10^{41}$\,\ergs\ (red), of $8.0 \times 10^{40}$\,\ergs\ (green), and without (blue).
\label{fig_spec_61p5d}
}
\end{figure}

\subsection{Comparison with observations at 61.5\,d}
\label{sect_61p5d}


Figure~\ref{fig_spec_61p5d} shows a comparison of the UV to NIR observations of \sn\ at about 61.5\,d with the three different incarnations of model x6p0 at 63.6\,d.\footnote{The epochs of observations in the UV, optical, and NIR are, in this order, 66.5, 61.5, and 67.5\,d. The NIR spectrum was scaled by a factor of 1.25 to make it align with the optical, flux-calibrated data from Lick Observatory. See Table~\ref{tab_obs} and Section~\ref{sect_obs} for details.} We show the predictions for model x6p0 with an interaction power of $6.0 \times 10^{41}$\,\ergs\ (red), x6p0 with an interaction power of $8.0 \times 10^{40}$\,\ergs\ (green), as well as the reference model x6p0 that exploded in a vacuum (blue).

At this epoch, \sn\ is well into the recombination phase, with a UV flux that is strongly reduced due to the low temperature in the outer ejecta (and CDS) and around the photosphere, and the strong metal-line blanketing in the UV and the blue part of the optical range, but with a clear contribution from interaction without which the flux is predicted to be essentially zero below about 3000\,\AA. Another indication of interaction is the clear presence of \mgiiuv, which is absent in the x6p0 model, but present in both models with interaction power, with a good match by the model with a moderate power of $8.0 \times 10^{40}$\,\ergs\ (see also \citealt{bostroem_23ixf_24}). The higher-power model overestimates the UV flux but yields a better match of the optical photometry. Obtaining a good match throughout is a challenge and suggests some interaction power is required at $\sim 10^{41}$\,\ergs. One key constraint in our modeling of \sn\ is the morphology of \ha, with the strength and width of the emission (i.e., the appearance of a ledge in the red-most part of the emission) as well as the magnitude of the filling-in of the \ha\ trough. Evidently, the model without interaction fails entirely at matching the \ha\ profile, but the model with a power of $8.0 \times 10^{40}$\,\ergs captures the presence of a moderate absorption or trough (emission from the CDS only partially fills-in the trough). The model with $6.0 \times 10^{41}$\,\ergs\ predicts essentially no trough in \ha\ (i.e., there is only a narrow notch at $-$\vcds) and thus overpredicts the \ha\ emission from the CDS.

When considering the full UV to NIR SED, all models tend to underestimate the NIR flux, regardless of the interaction-power strength (provided the NIR flux is accurately calibrated). This suggests the model is lacking a fundamental component. Increasing the interaction power raises the NIR flux but enhances the discrepancy in the UV. The intermediate interaction-power case, which matches well the UV flux of \sn, is nearly as discrepant in the NIR as the reference model x6p0. The discrepancy might be related to our simplistic handling of interaction power or the CDS properties. We explored the impact of clumping on the IR flux and found a complex behavior with an opposite effect between NIR and MIR (see Section~\ref{sect_fvol} for details). Invoking a larger progenitor radius would boost the luminosity, although at such late times in the photospheric phase, this effect can be mitigated by asymmetry or mixing of material from the metal-rich inner ejecta (increasing $R_\ast$ would boost the luminosity but would have an adverse effect on line-profile widths).

Finally, the NIR range (bottom panel of Fig.~\ref{fig_spec_61p5d}) shows clear evidence now of the presence of the CDS, which formed already at the time of shock breakout. Indeed, the lower transitions of the Paschen series exhibit peculiar troughs in their associated P-Cygni profiles. This trough is extended with a blue edge at about 8000\,\kms\ from the line rest wavelength (e.g., H\one\,12,818.1\,\AA). In some cases, a double absorption is clearly present (e.g., H\one\,10,049.4\,\AA). These features are tentatively present in H\one\,18,751.0\,\AA\ (this transition is affected by atmospheric absorption) or H\one\,10,938.1\,\AA\ (line overlap with He\one\,10,830.2\,\AA\ compromises the analysis). The two models with interaction (and a CDS) predict double absorptions too in H\one\,9546.0\,\AA\ and H\one\,9229.0\,\AA, although admittedly the close proximity of these transitions challenges this inference from the observations. Further discussion about these kinks and how they evolve in time in both the optical and NIR ranges is presented in Section~\ref{sect_kinks}.



\begin{figure}
\centering
\includegraphics[width=0.9\hsize]{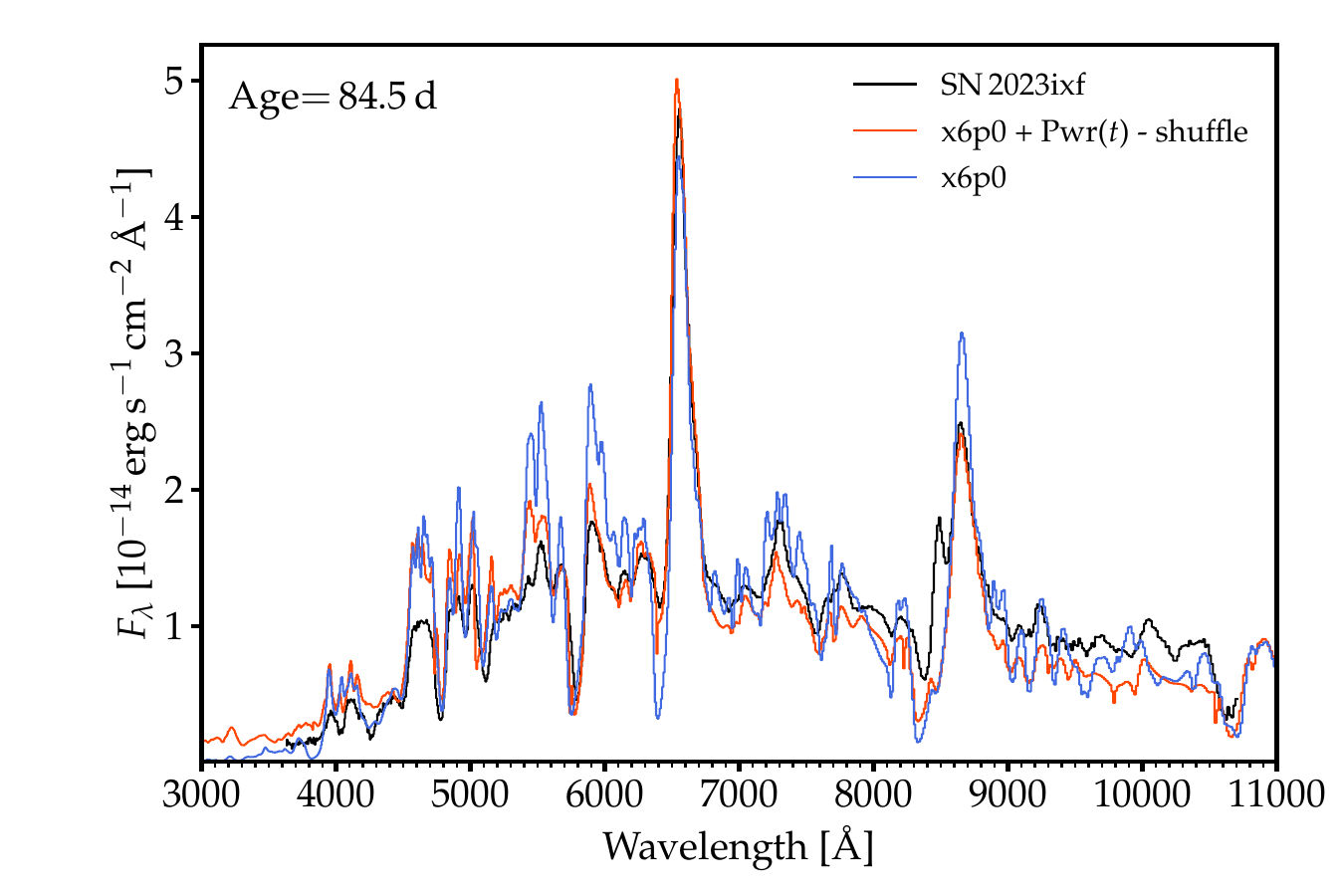}
\caption{Comparison between the optical observations of \sn\ at an epoch of 84.5\,d (black) and model x6p0 at 87.0\,d with an interaction power of $5.0 \times 10^{40}$\,\ergs\ (red; this model also has a shuffled-shell structure) and without interaction power (nor CDS; blue).
\label{fig_spec_84p5d}
}
\end{figure}

\subsection{Comparison with observations at 84.5\,d}
\label{sect_84p5d}

Figure~\ref{fig_spec_84p5d} shows a comparison of the optical observations of \sn\ at about 84.5\,d with model x6p0 at 87.0\,d influenced by an interaction power of $5.0 \times 10^{40}$\,\ergs\ (the \nifs\ mass in this model is 0.047\,\msun). Because this time is within the transition to the nebular phase, the inner metal-rich ejecta start to contribute directly to the emergent spectrum and we therefore adopt a shuffled-shell structure for the ejecta (see bottom panel of Fig.~\ref{fig_init}) in which macroscopic mixing is allowed without introducing any microscopic mixing.  As before and for comparison, we show the reference model x6p0 (with a standard chemical mixing) at the same epoch of 87.0\,d (the \nifs\ mass in this model is 0.054\,\msun).

Both models yield a satisfactory match to the SED but the shuffled-shell model with interaction achieves better in a number of sectors. First, the strength of metal lines is reduced and more compatible with the observations. This may in part arise from the $\sim$\,10\,\% reduction in \nifs\ mass, but it more likely results from the fact that the metal abundances are truly solar in the H-rich material rather than artificially boosted above solar by the boxcar (microscopic) mixing --- this affects all abundances including those of Na, Ca, or Fe, etc.

Secondly, the \ha\ profile shape clearly requires a contribution at high velocity from the CDS, which acts to extend the emission in the red part of the profile and to fill-in the trough on the blue side. With the reduced interaction power from earlier epochs (this adopted interaction power has been continuously decreasing from  $2.0 \times 10^{42}$\,\ergs\ down to $5.0 \times 10^{40}$\,\ergs\ at 84.5\,d and continues thereafter), this filling-in of the trough is partial and in good agreement with observations (i.e., a trough is clearly present but not as strong as in the reference model x6p0).


\begin{figure}
\includegraphics[width=0.9\hsize]{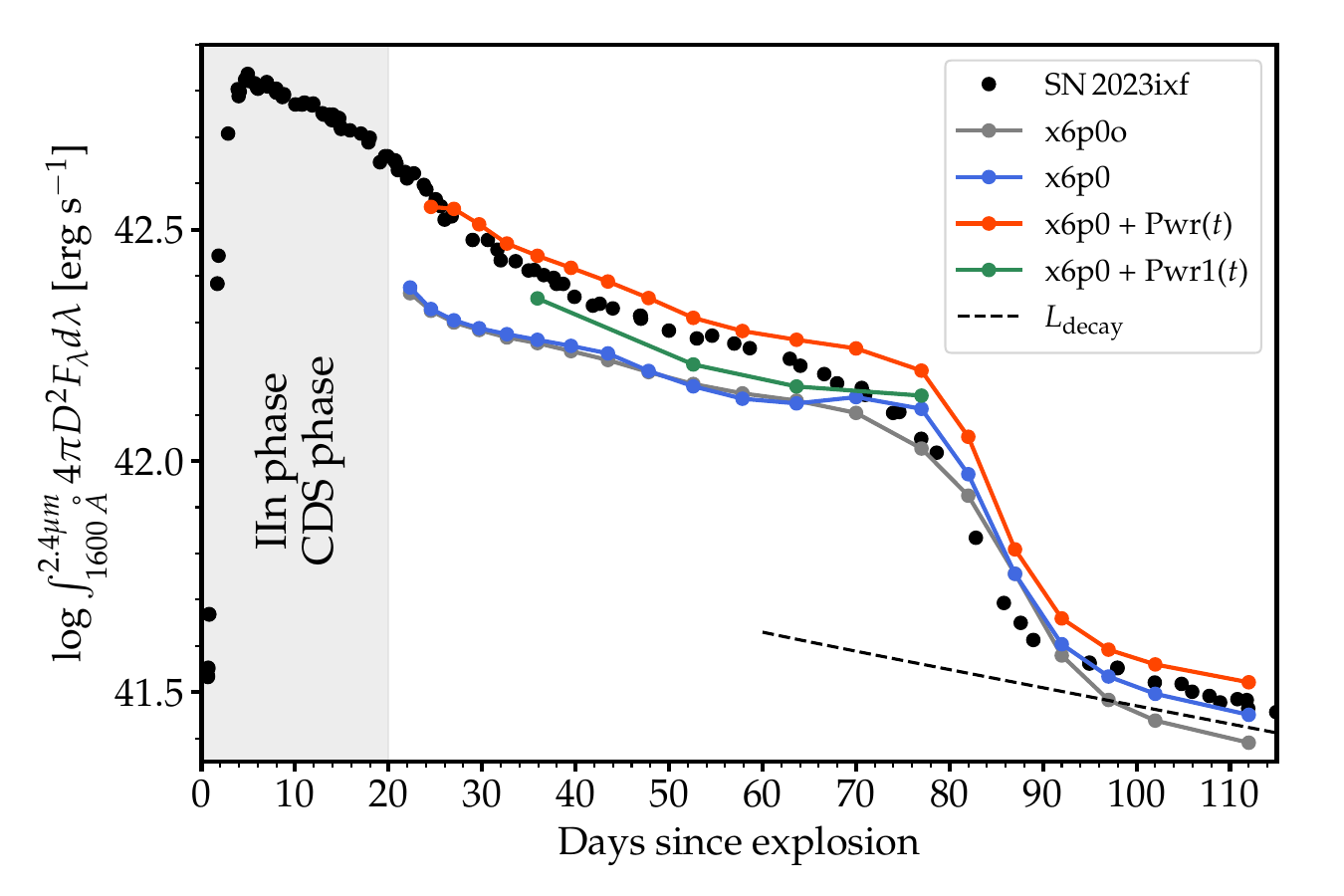}
\includegraphics[width=0.9\hsize]{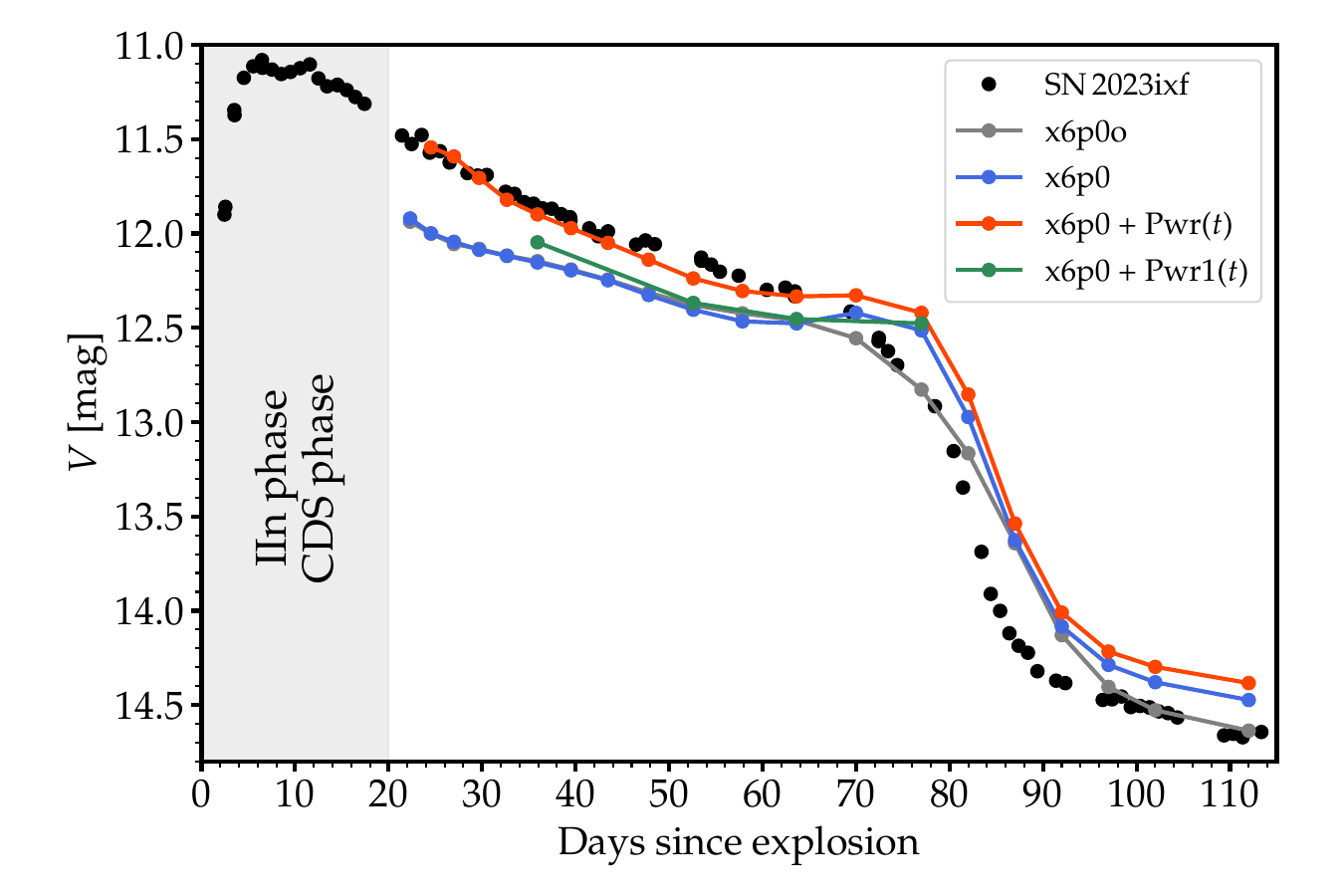}
\caption{Comparison between the inferred bolometric light curve (left) and observed $V$-band light curve of \sn\ (right), and models x6p0o (gray), x6p0 (blue), and x6p0 augmented by various amounts of interaction power (red and green curves; see Table~\ref{tab_pwr}). Models were redshifted, distance scaled, and reddened according to \sn\ characteristics. Models were computed until the onset of the nebular phase but skip the early SN~IIn phase (gray shading), whereas the full photospheric-phase evolution is shown for \sn. The black dashed line gives the instantaneous decay power from 0.05\,\msun\ of \nifs.
\label{fig_lc}
}
\end{figure}

\section{Model comparison with light curves}
\label{sect_lc}

Figure~\ref{fig_lc} shows the inferred quasibolometric luminosity as well as the $V$-band light curve of \sn\ in the initial 115\,d of evolution following first light. Overplotted are various incarnations of model x6p0, as discussed in the preceding section and summarized in Table~\ref{tab_init}. For \sn, the luminosity is based on the photometry covering from the UV ($\sim$\,1600\,\AA) out to the NIR ($\sim$\,2.4\mic), together with the characteristics (distance and reddening) of \sn. For the models, we directly integrated the spectra between these two wavelength bounds. For the magnitude plot, the model results are produced by scaling the spectra to the distance of \sn\ and adding the $V$-band extinction.

The various models shown in Fig.~\ref{fig_lc} are first the original model x6p0o from \citet{HD19} but with a \nifs\ mass of 0.045\,\msun. Second, there is the reference model x6p0 with 0.053\,\msun\ of \nifs\ and an updated composition based on model s15.2 of \citet{sukhbold_ccsn_16}. Third, there are the two model counterparts with low and high interaction power (see Table~\ref{tab_pwr}) that aim to bracket the actual interaction power released by the shock and absorbed by the ejecta (i.e., in practice the CDS).

Overall, all models are in rough agreement with the observations. There is a common offset in photospheric-phase duration by $\sim$\,5\,d, which would be cured by slightly reducing \mej\ or increasing \ekin. Models without interaction power underestimate the luminosity at all times prior to the end of the photospheric phase. With interaction power, the luminosity is boosted such that the offset is modulated and possibly cured. A moderate interaction power reduces the underestimate in the luminosity (model x6p0 + Pwr1($t$)), with only a modest change in $V$. Raising the interaction power further (model x6p0 + Pwr($t$)) slightly overestimates the luminosity but yields a good match to $V$. At the end of the photospheric phase, the various models exhibit different shapes. The old model x6p0o has a soft transition because of the stronger chemical mixing (including that for \nifs). In other models, the luminosity and $V$-band light curve exhibit a bump, which results from weaker chemical mixing (as well as a slightly larger \nifs\ mass). This phase is typically poorly handled in spherical symmetry because both density and composition profiles are affected by the Rayleigh-Taylor instability \citep{utrobin_sn2p_17}. At nebular times, the various models lie close to \sn\ with an offset of at most $\sim$\,0.2\,mag.

What this exercise shows is that estimating the interaction power from the optical spectrum and the \ha\ profile leads to rough agreement with the bolometric or $V$-band light curve, which is not so surprising since the photometry is an integral quantity derived from the information contained in spectra. All models here were computed in a time-dependent manner and thus rest on a similar approach as done with radiation hydrodynamics, the significant difference being that hydrodynamics is ignored. However, as pointed out by \citet{dessart_csm_22}, dynamical effects are essentially limited to the power injection into the CDS.

Overall, our work confirms the previous studies that concluded for the need of a reduced ejecta mass
(see, e.g., \citealt{fang_23ixf_25}; \citealt{forde_23ixf_25}; \citealt{hsu_23ixf_25}; \citealt{kozyreva_23ixf_25}) relative to standard Type II SNe in which the photospheric phase is longer and typically around 100\,d. Consequently, our results are in tension with other studies that suggest a standard ejecta mass \citep{bersten_23ixf_24,vinko_23ixf_25,moriya_23ixf_24}.


\begin{figure}
\includegraphics[width=0.9\hsize]{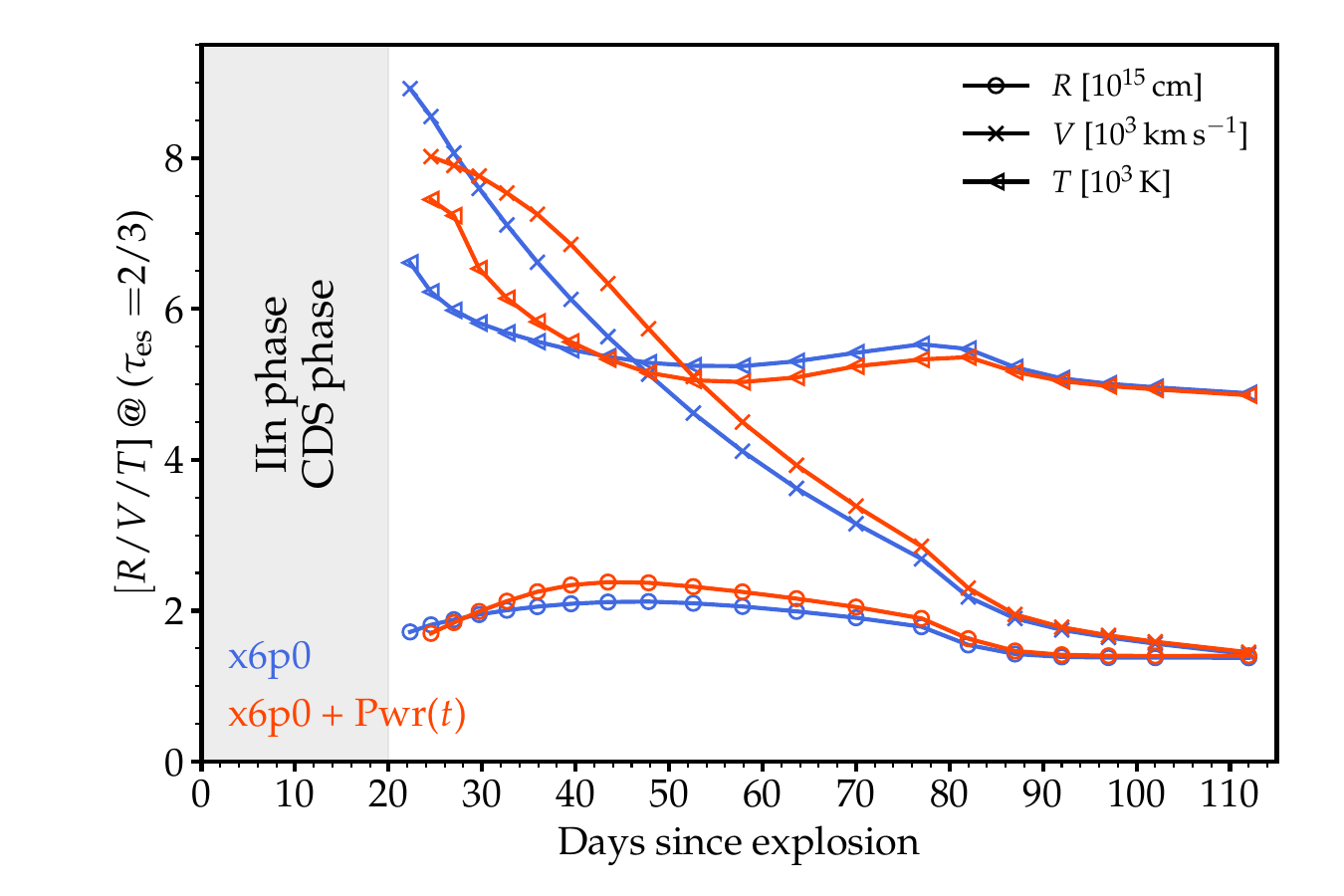}
\caption{Evolution of photospheric properties in model x6p0 and its counterpart with interaction power.
\label{fig_phot_prop}
}
\end{figure}

\section{Evolution of ejecta properties with and without interaction or CDS}

Figure~\ref{fig_phot_prop} illustrates the evolution of the photospheric properties (radius, density, and temperature) for the reference model x6p0 and its counterpart with interaction power. With interaction power, additional energy is made available in the outer ejecta (and CDS), causing a larger temperature, ionization, and thus optical depth, so that the photosphere resides in the CDS at 25\,d in the interacting model whereas it rapidly recedes in the reference model.

After about 50\,d, the difference in those properties between the two models is 10\,\% or less, indicating that the role of interaction is really separated from the underlying properties of the photosphere. That is, the interaction adds an extra power source within the CDS, located in the optically thin, outermost regions of the ejecta. This also emphasizes the suitability of the method used here and introduced by \citet{dessart_csm_22}, with a full nonLTE treatment that couples gas and radiation rather than performing superfluous radiation hydrodynamics and assuming (inadequately) that the gas is in LTE. This is also important for estimating the UV flux since UV photons decouple from the gas at much lower density than optical or IR photons owing to the much larger opacities in this wavelength range.

By the end of the time range shown in Fig.~\ref{fig_phot_prop}, the total electron-scattering optical depth of the ejecta is about 1.2, even though the light curve has already started its evolution in the nebular phase. Evidently, the drop in optical depth is progressive and will take many weeks to drop below 2/3 or reach 0.1 (when the conditions are really starting to be optically thin). Obviously, this applies to the innermost ejecta regions since the layers above may have been optically thin for days or weeks.

Although not relevant for \sn, if the interaction power were much larger, it could change this situation by inhibiting recombination and making the ejecta (and possibly the CDS too) optically thick for longer. The impact of the interaction is obviously dependent on the actual power injected (for the potential impact of larger interaction powers, see \citealt{dessart_csm_22}).


\begin{figure*}
\centering
\includegraphics[width=0.245\hsize]{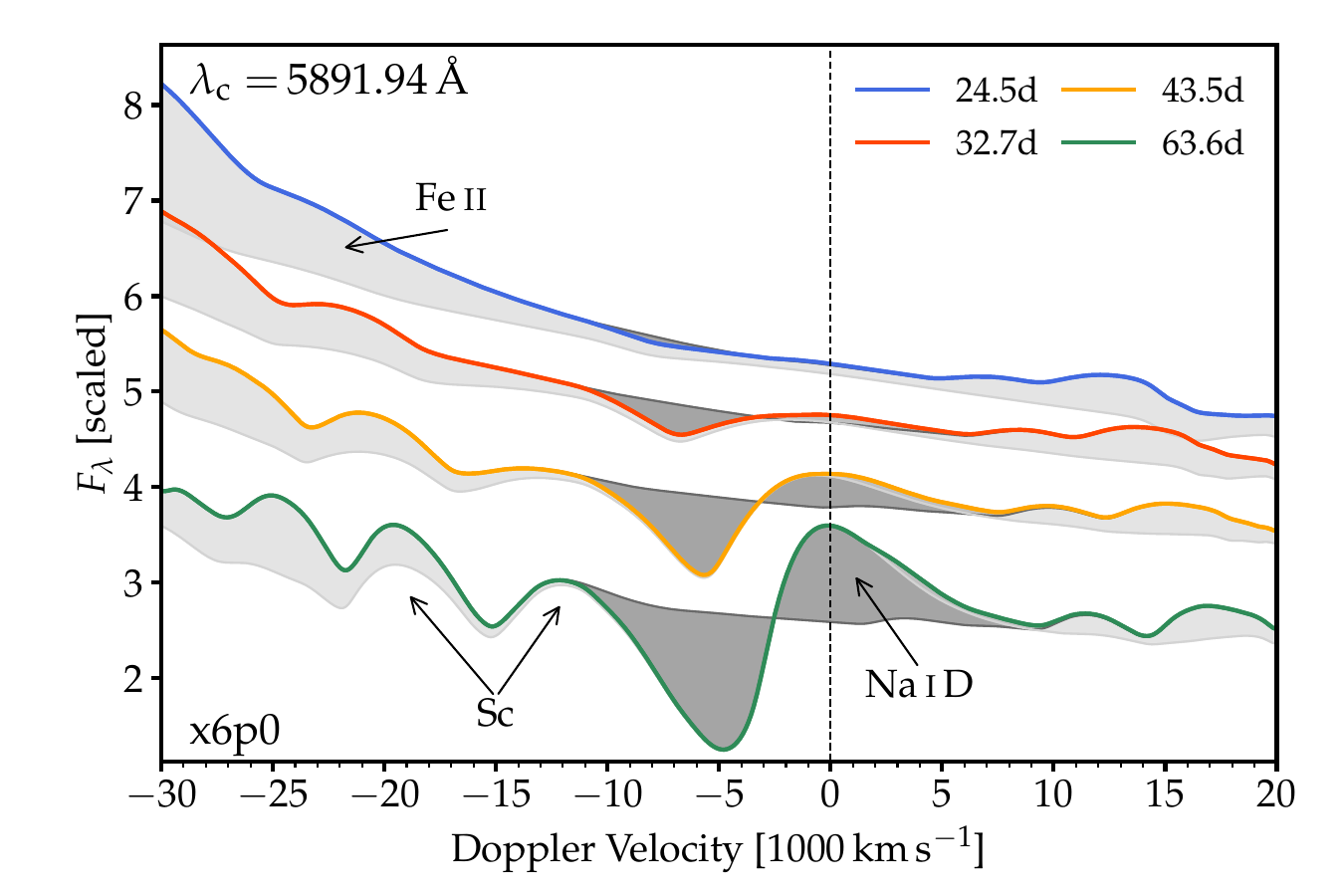}
\includegraphics[width=0.245\hsize]{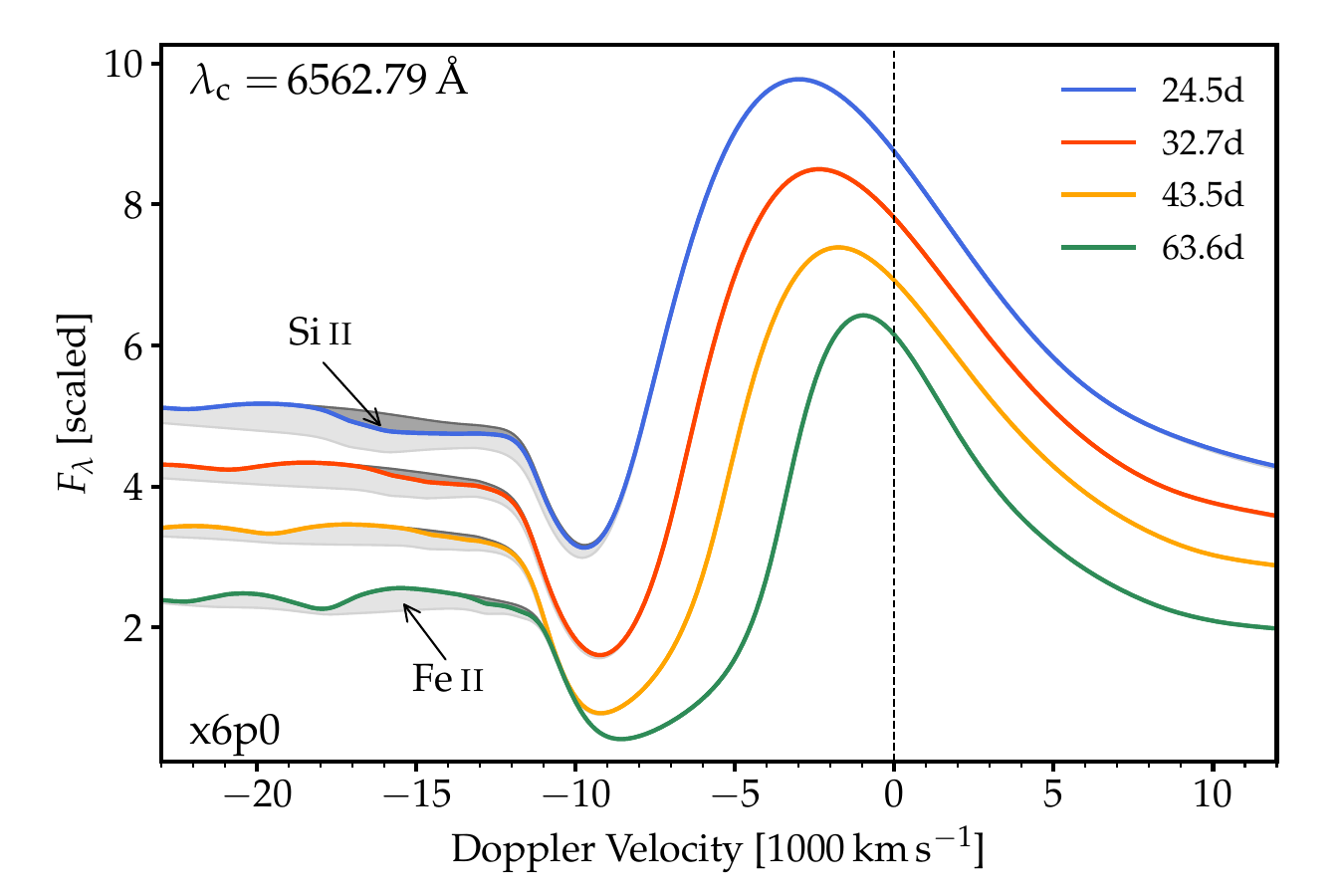}
\includegraphics[width=0.245\hsize]{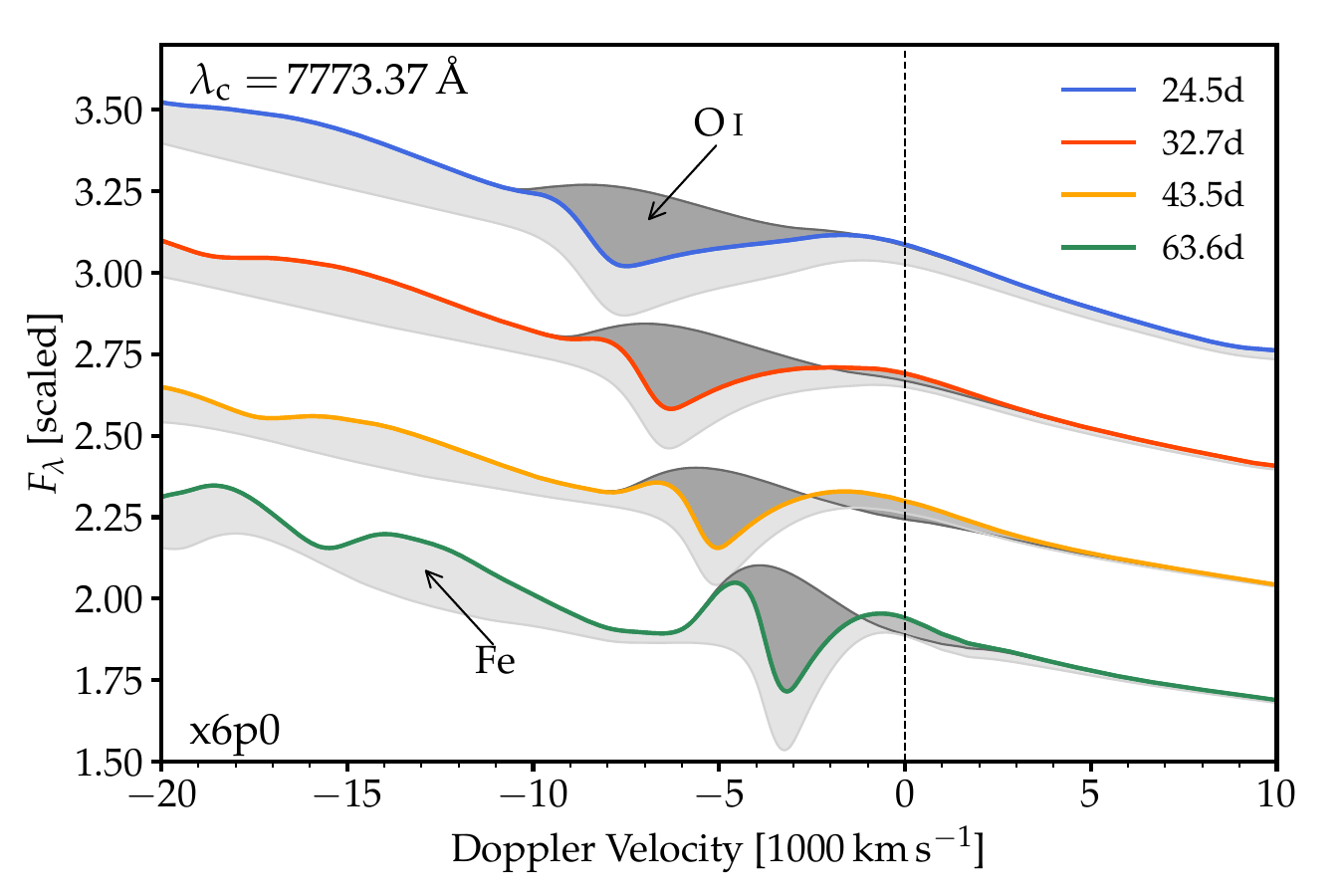}
\includegraphics[width=0.245\hsize]{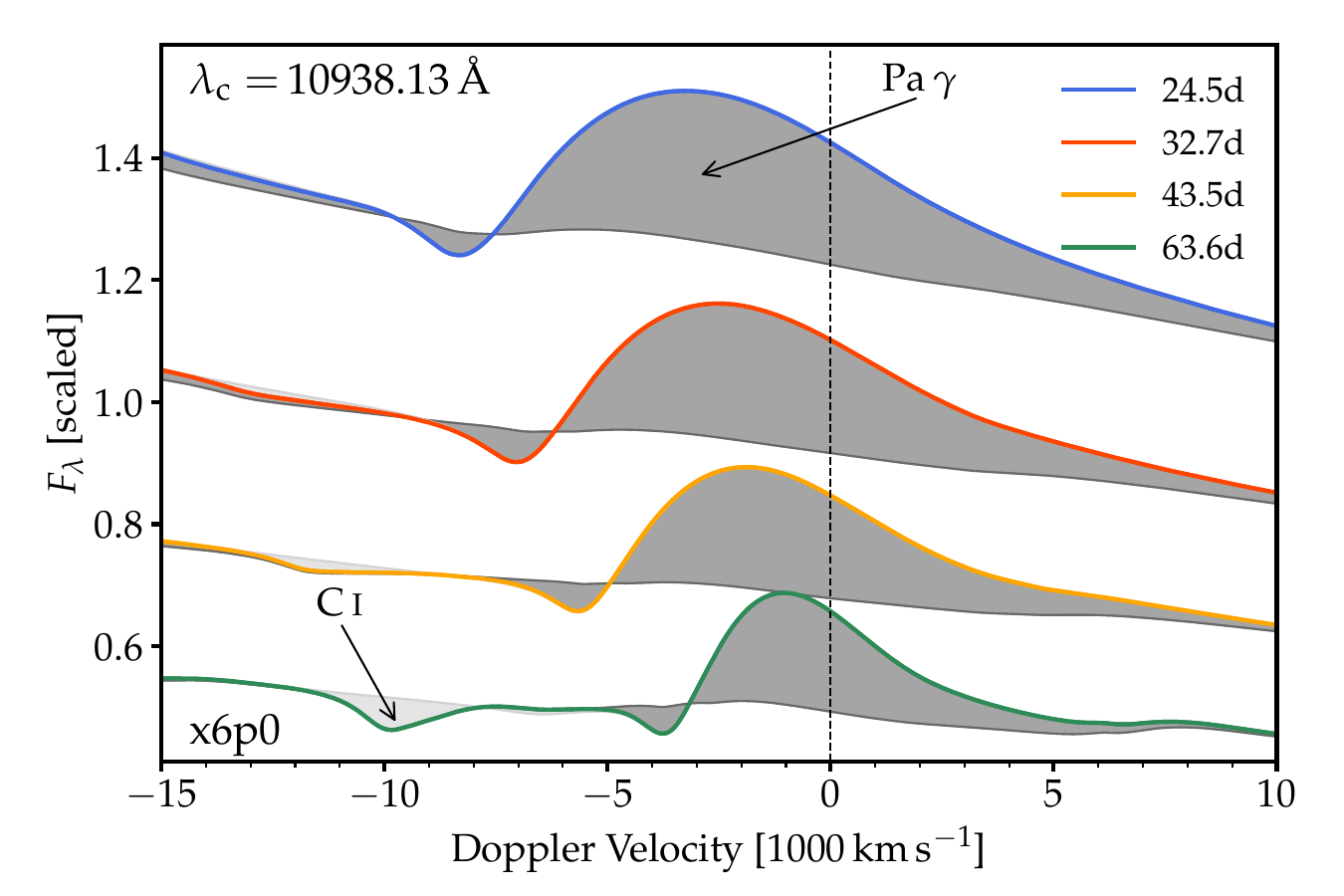}
\includegraphics[width=0.245\hsize]{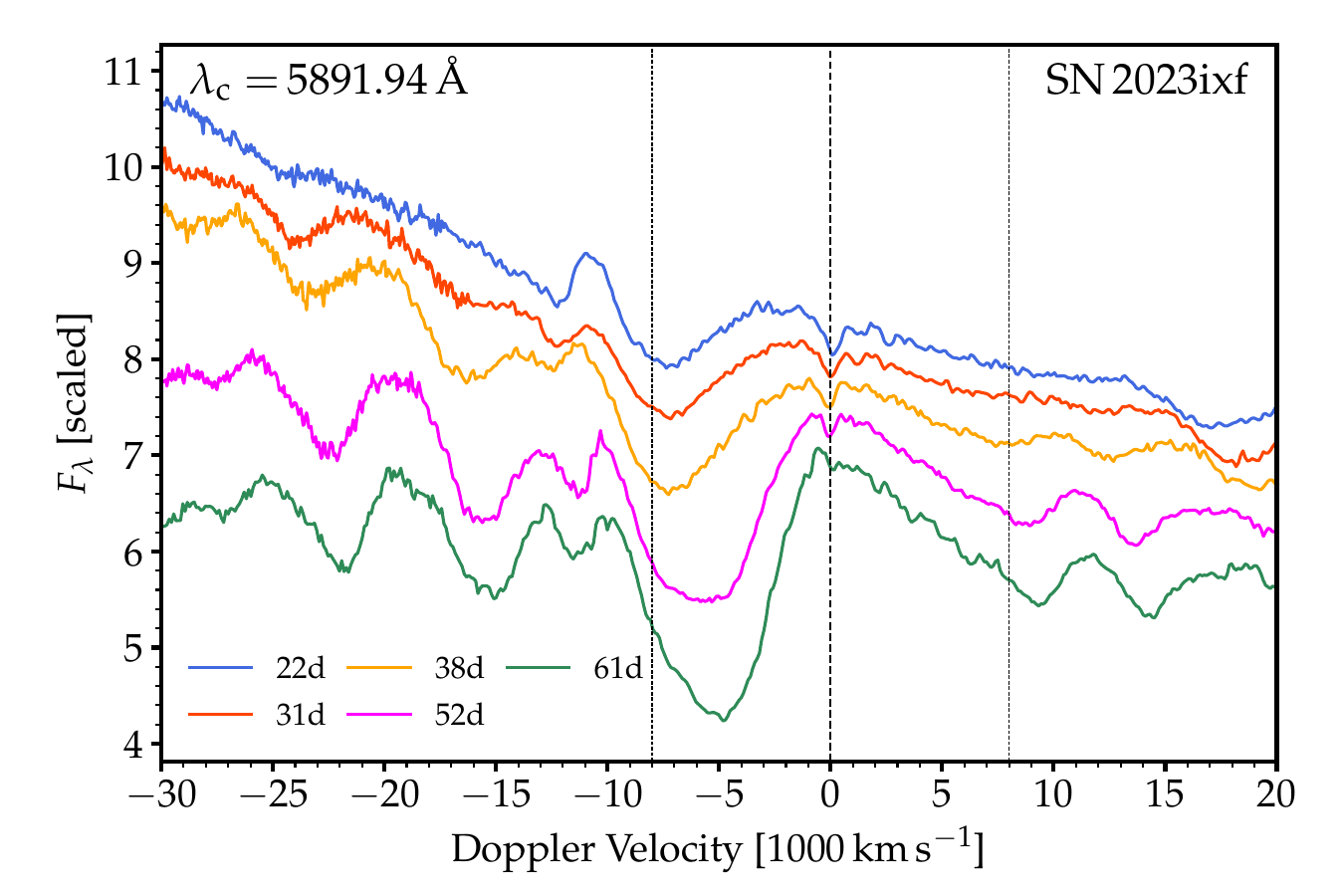}
\includegraphics[width=0.245\hsize]{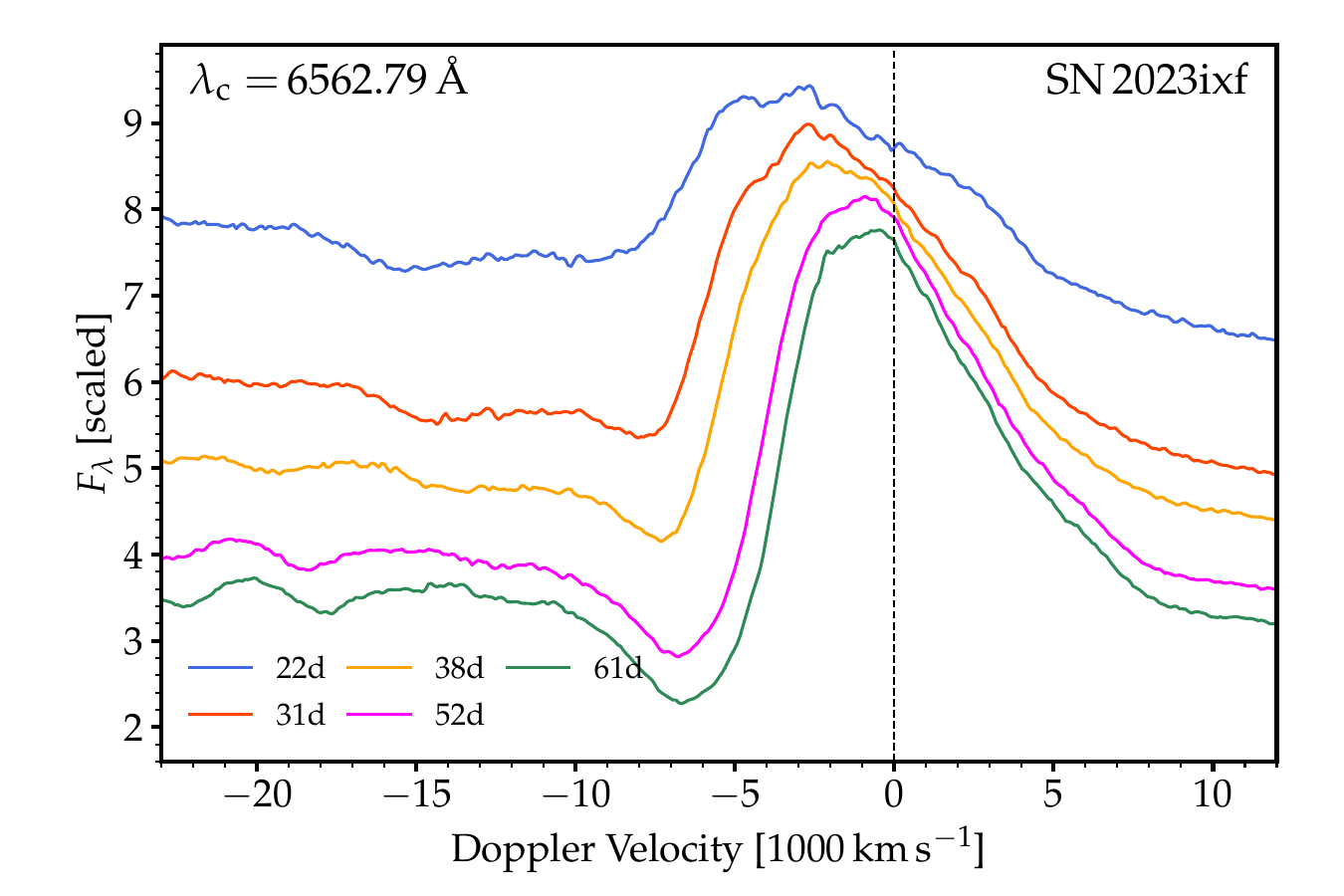}
\includegraphics[width=0.245\hsize]{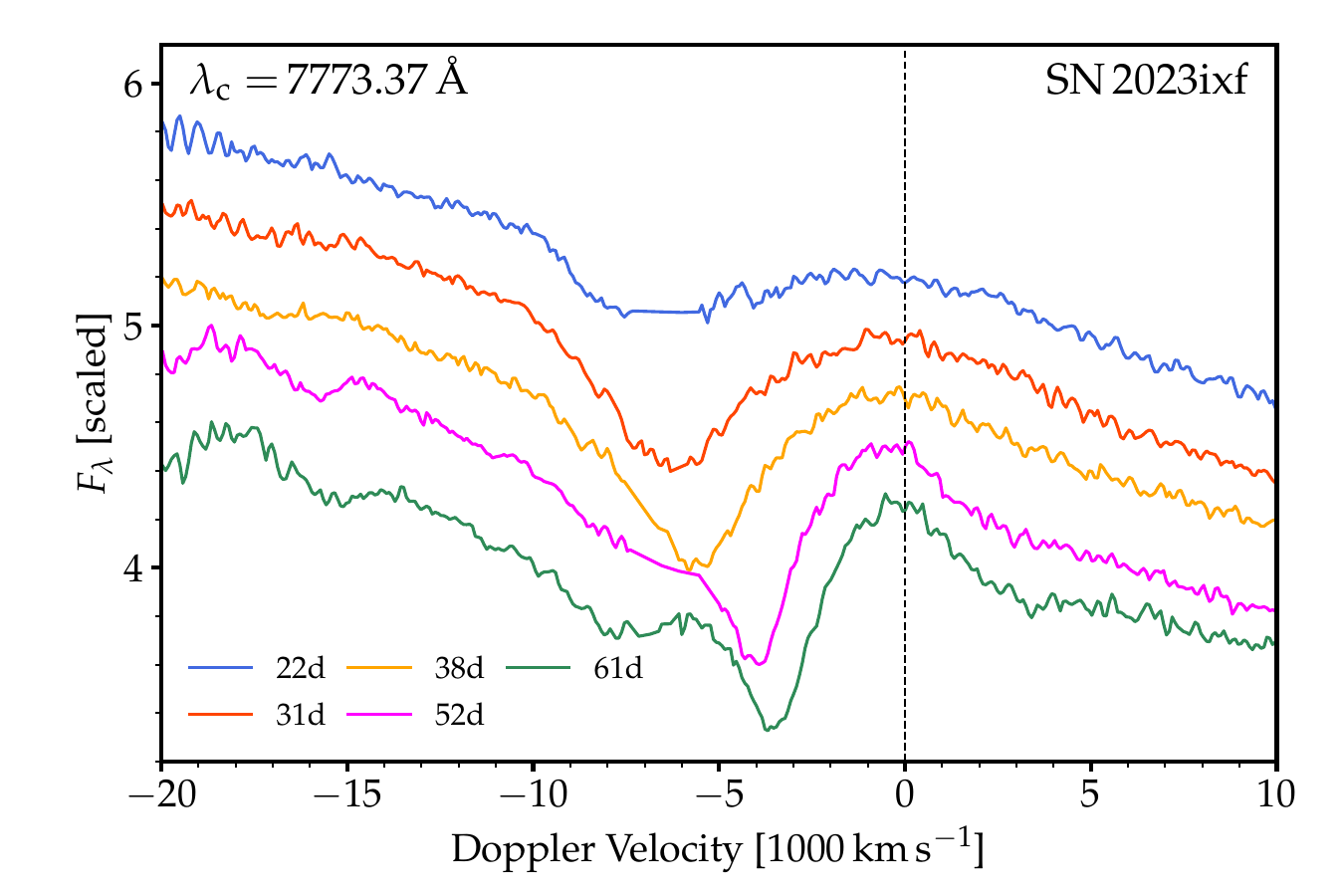}
\includegraphics[width=0.245\hsize]{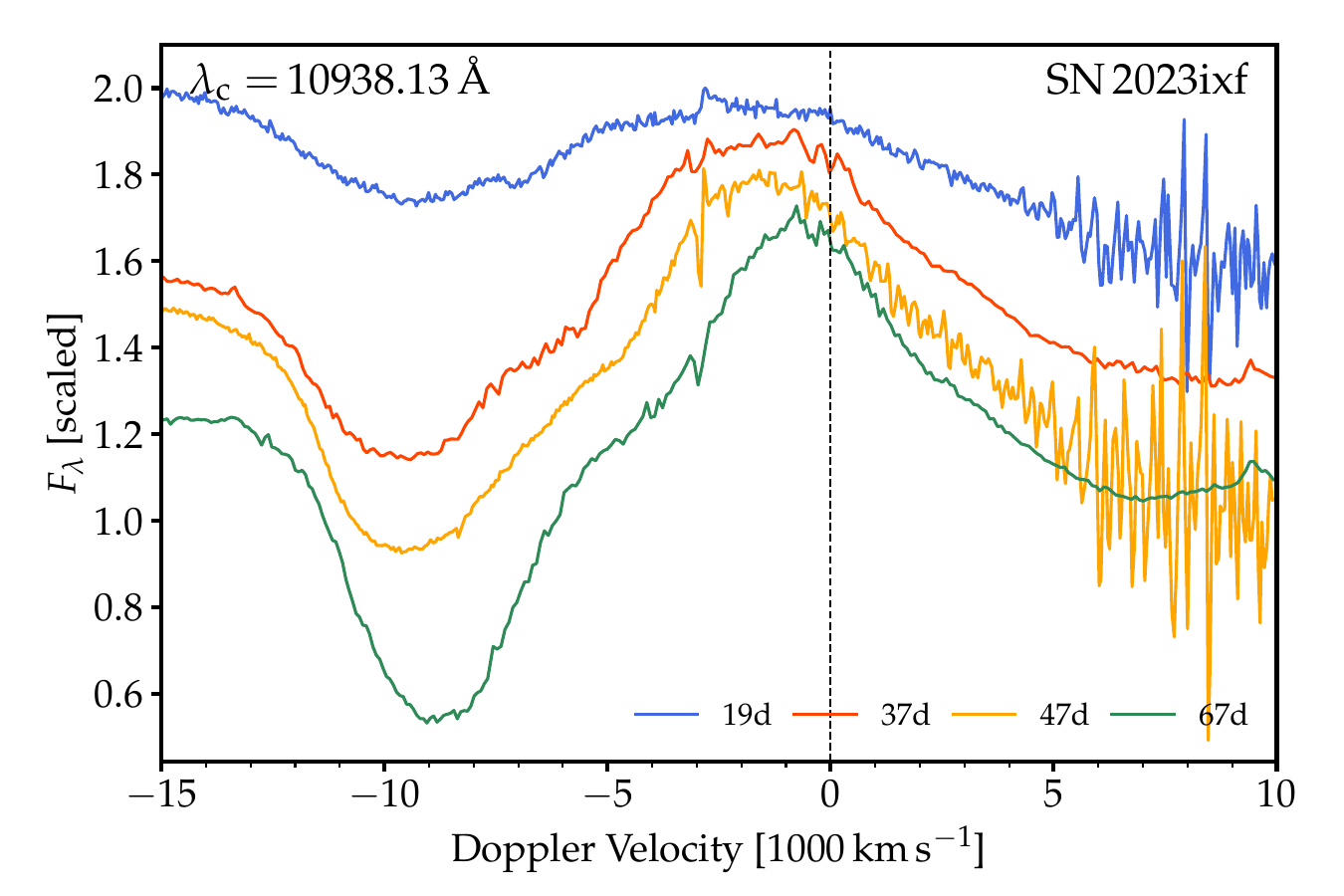}
\includegraphics[width=0.245\hsize]{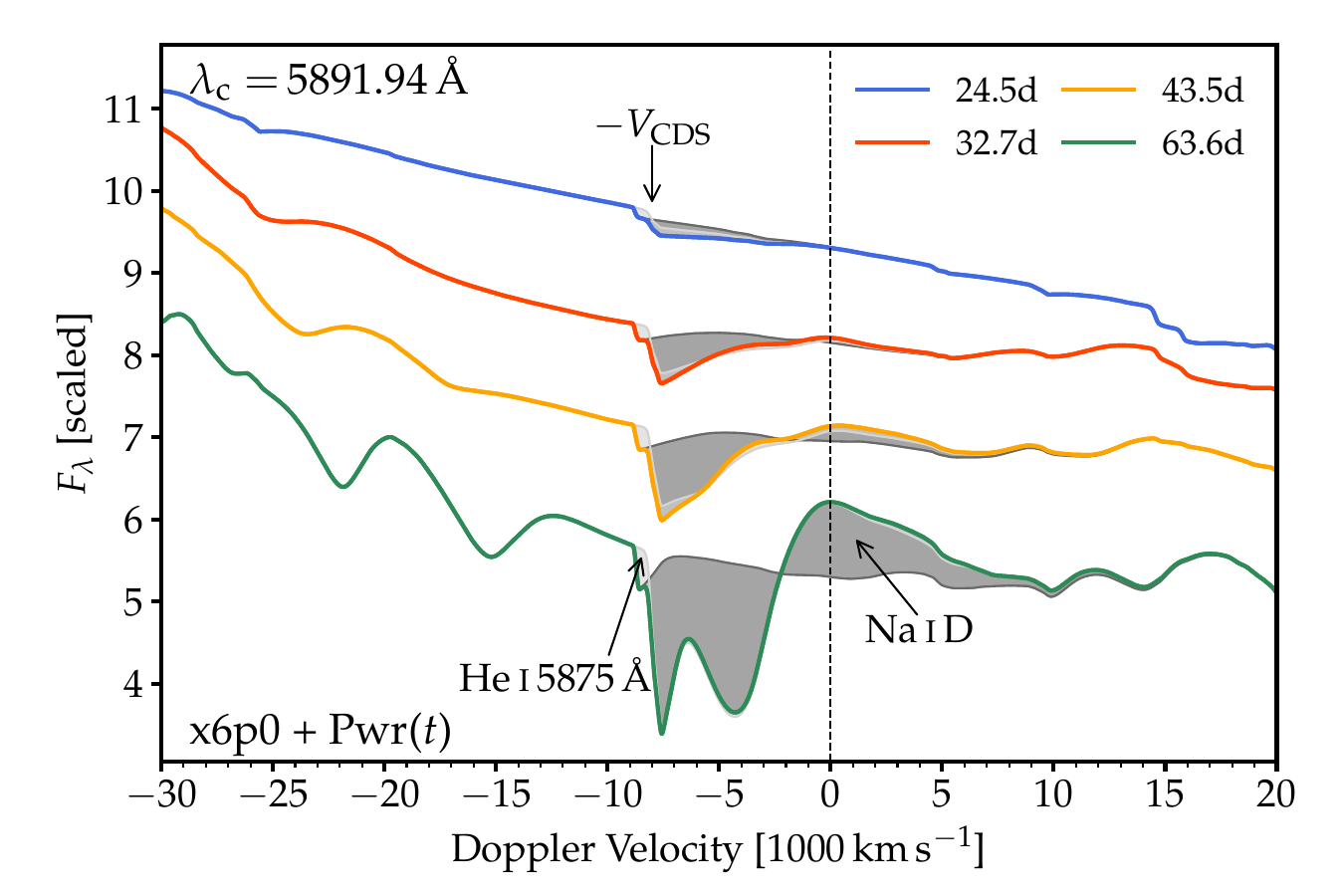}
\includegraphics[width=0.245\hsize]{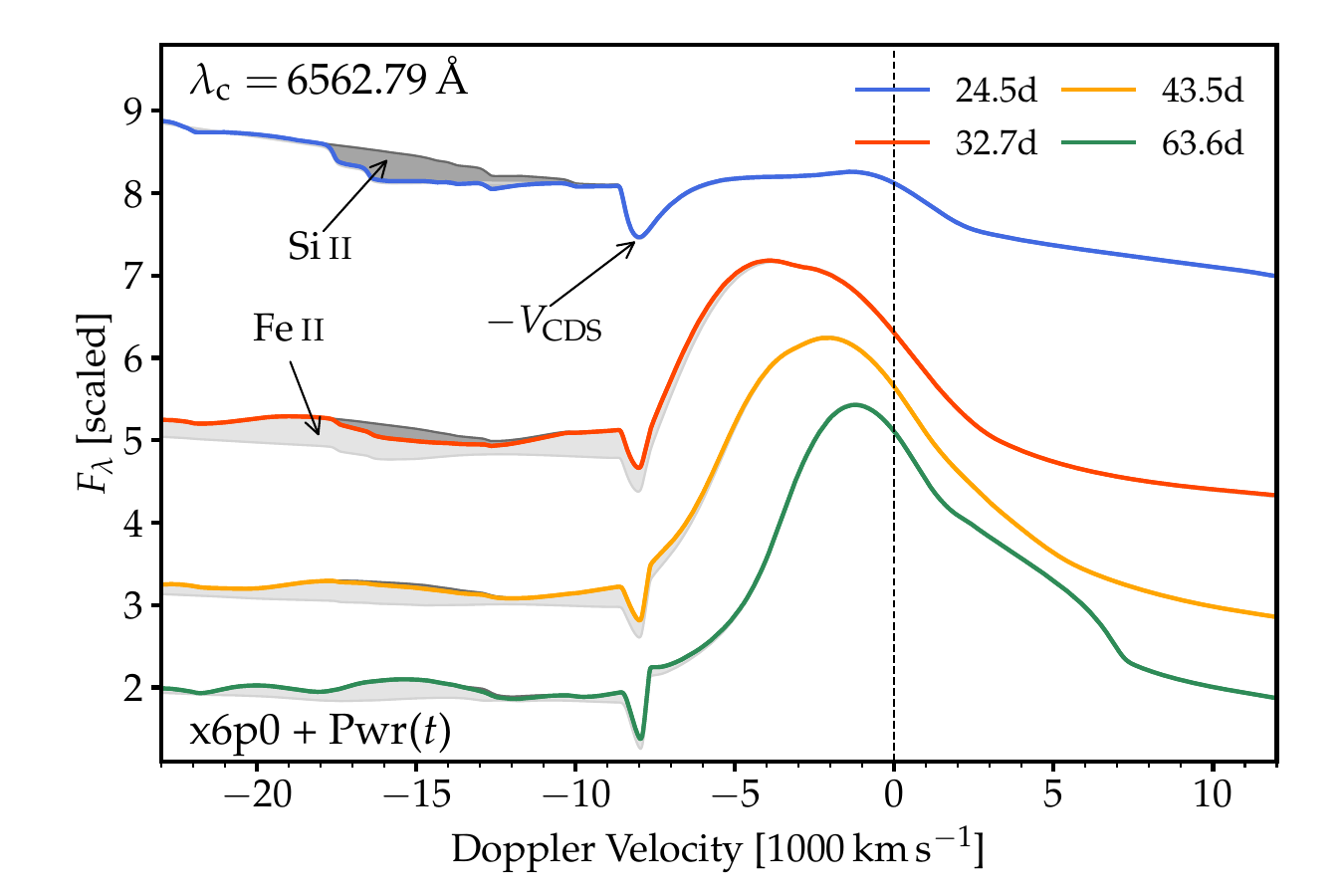}
\includegraphics[width=0.245\hsize]{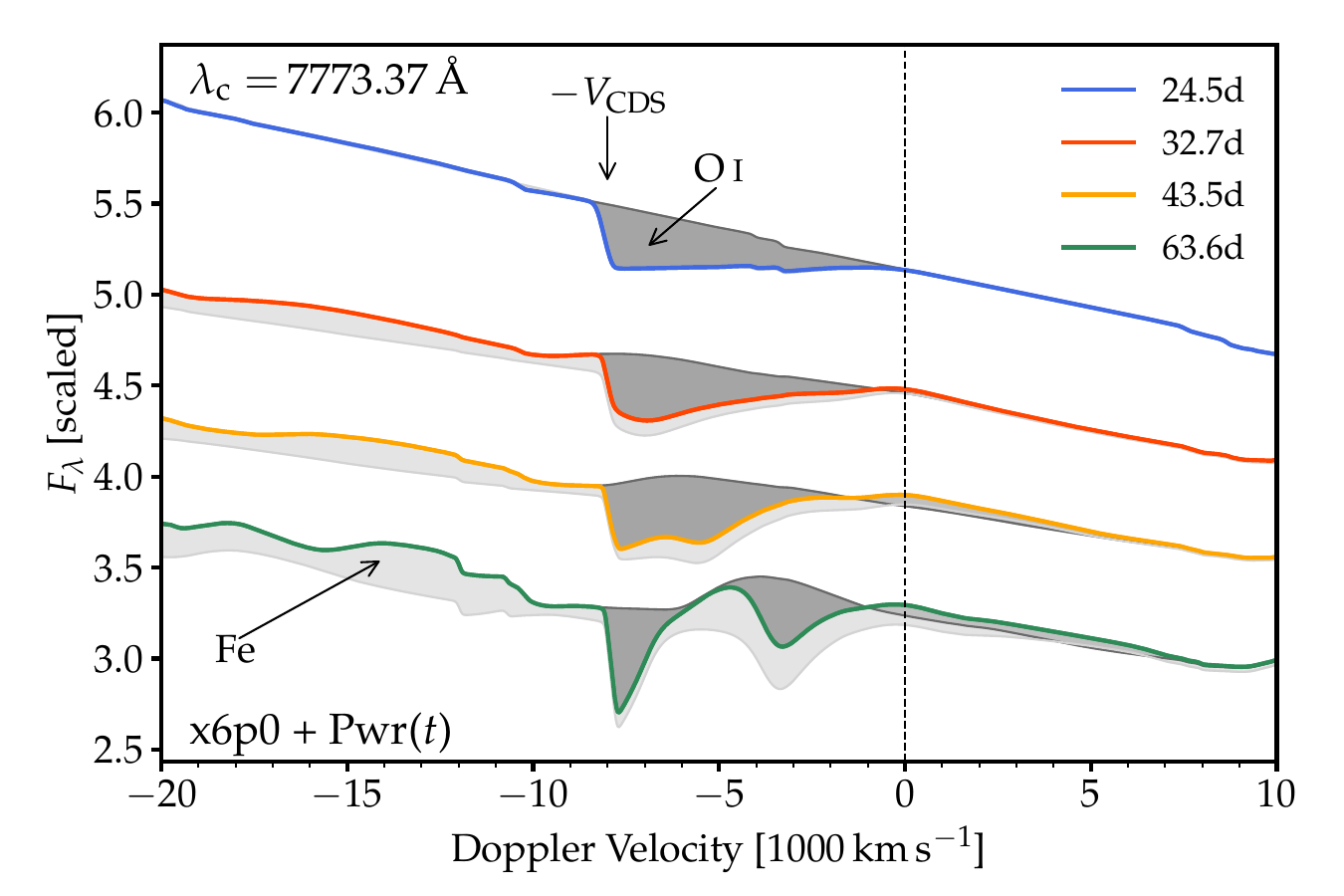}
\includegraphics[width=0.245\hsize]{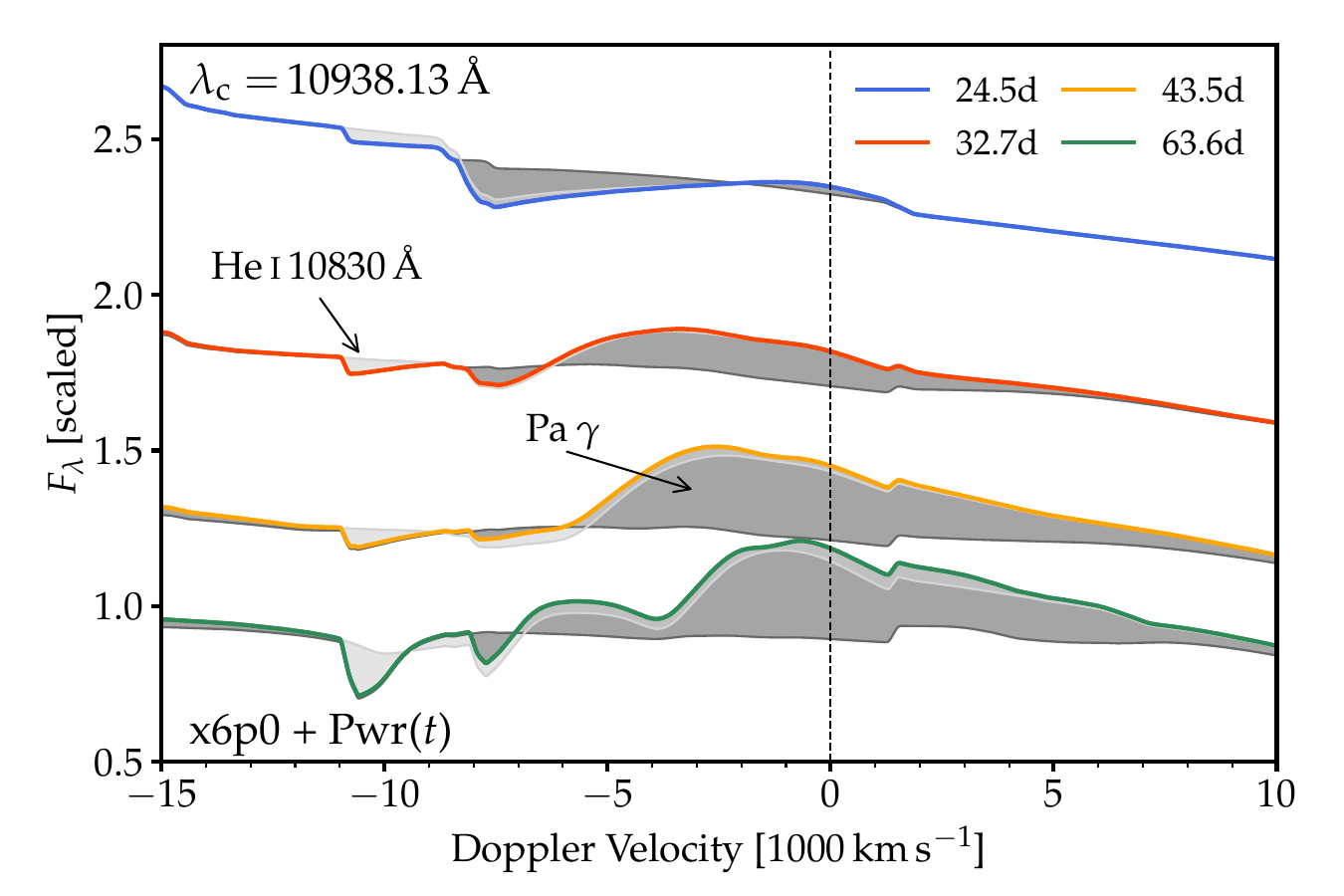}
\caption{Illustration of the photospheric-phase evolution of spectral regions centered on the rest wavelengths of \naid, \ha, \oitrip, and \pag\ for models x6p0 and x6p0/Pwr. Shading is used to highlight the contribution from selected species.
}
\label{fig_prof_kinks}
\end{figure*}

\section{Profile kinks}
\label{sect_kinks}

In this section, we explore the formation of profile kinks arising from the presence of an interaction, or more specifically the presence of a CDS. Indeed, as shown by \citet{dessart_csm_22}, ejecta surrounded by a spherical dense shell from previously swept-up material produce, for a distant observer, spectra in which strong lines exhibit a double absorption, the main one typically associated with the broad absorption occurring over a range of depths within the ejecta (and generally associated with the photospheric region), and a narrow component associated with the dense shell located at the outer edge of the ejecta (see also \citealt{chugai_hv_07}). Obviously, at sufficiently early times, the photosphere will be located in the outer ejecta or even in the CDS, so one expects only a single absorption component at such times (this is likely the reason why double-absorption features were not observed in the optical and IR observations analyzed by \citealt{derkacy_23ixf_26}). But as time progresses, the photosphere inevitably recedes in mass space, allowing for the appearance of two extrema in the projected line optical depth along the line of sight.
The present analysis complements the earlier work of \citet{park_23ixf_25}, but here guided by detailed radiative-transfer simulations.

Interaction power plays no role in the making of these double kinks since, if anything, it tends to erase these absorption components by filling them
in with extra emission --- the main reason for these double kinks is the presence of a CDS. Such kinks are of a different nature than those that may
arise from the overlap of distinct lines over a narrow wavelength range, as may occur with Si\two\,6355\,\AA\ and \ha\ in early-time spectra of SNe~II,
and particularly visible in low-energy explosions (e.g., SN\,2005cs; \citealt{pastorello_etal_05cs2,dessart_05cs_06bp}).

Figure~\ref{fig_prof_kinks} illustrates the spectral evolution from $\sim$\,20 to $\sim$\,65\,d for both \sn\ and models in the regions centered on 5891.94\,\AA\ (blue component of \naid; left column), \ha\ (second column from left), 7773.37\,\AA\ (rest wavelength of the O\one\ multiplet; third column from left), and 10,938.13\,\AA\ (i.e., rest wavelength of Pa\,$\gamma$; rightmost column). The observations are displayed in the middle row, whereas the top row shows the results from the reference model x6p0 and the bottom row model x6p0 + Pwr($t$) (see Table~\ref{tab_pwr} for the powers used at the corresponding epochs). In all panels, we shade with light and dark gray the line-flux contributions from various atoms and ions such as H\one, He\one, O\one, Na\one, Fe\two\, Sc\two, and Si\two\ (the species name is shown when there are multiple ionization stages contributing, e.g., Fe\one\ and Fe\two).

Focusing first on the observations illustrated in the middle row, all profiles tend to shrink in velocity space (i.e., they narrow from both the red emission part and the blue absorption part), strengthen in emission, and strengthen in absorption, with a deepening of the trough on the blue side (both relative to the adjacent continuum). For \naid, the blue edge of the trough extends to about $-9000$\,\kms\ and remains unchanged in time, whereas the absorption maximum recedes from about $-8000$ down to $-4000$\,\kms. The properties are similar in the O\one\ line except that at late times, a double absorption has formed rather than a single broad trough. In both \ha\ and Pa\,$\gamma$ panels, the troughs are much more extended at all times. For \ha\ at early times, some absorption extends out to $-18,000$\,\kms\ but with a double-kink structure (as discussed below, this extended absorption is due primarily to Si\two). Both absorptions recede as time passes, and the absorption at $-12,000$\,\kms\ is hardly noticeable at the latest epoch of 61\,d. For Pa\,$\gamma$, the trough extends out to about $-12,000$\,\kms\ at all times, with a strong and broad trough, and with the development of a bump in the profile nearer line center. To reconcile these distinct behaviors and understand what they imply, we now consider the behavior observed in the models first without interaction power nor a CDS and then with interaction power as well as a CDS.

\begin{figure*}
\centering
\includegraphics[width=0.9\hsize]{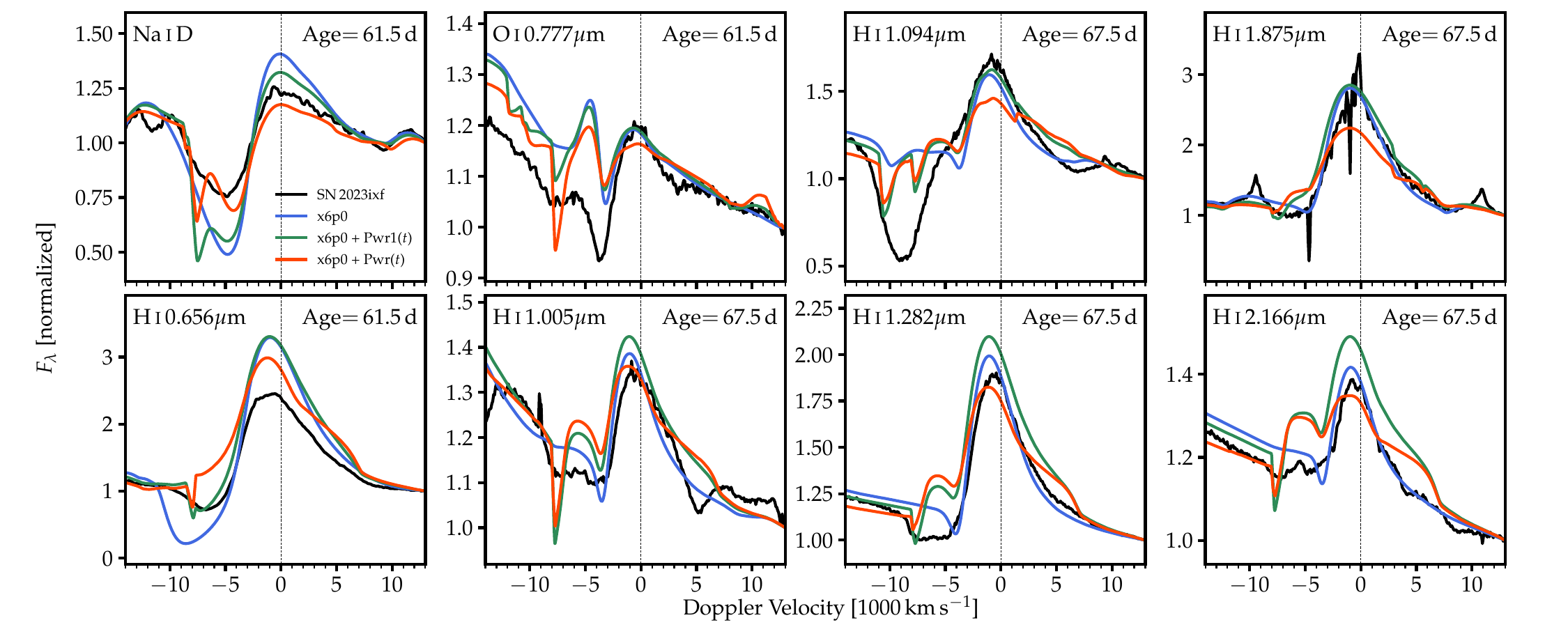}
\caption{Counterpart of Fig.~\ref{fig_prof_kinks} but comparing optical (age of 61.5\,d) and NIR (age of 67.5\,d) observations with models at 63.6\,d. Observations are corrected for redshift and reddening. Observed and model line profiles are normalized to unity at the blue edge of the depicted range. Models include the reference model x6p0 and two counterparts with interaction power (see Table~\ref{tab_pwr}).
\label{fig_NIR_66d}
}
\end{figure*}

In the reference model without interaction nor CDS (maximum ejecta velocity of 13,500\,\kms; top row of Fig.~\ref{fig_prof_kinks}), all four lines exhibit a similar qualitative behavior with a recession and a strengthening relative to the continuum. Some kinks are present and located at large velocities blueward of the rest wavelength of each line and in all cases attributed to overlap with one or multiple lines. For \naid\ (which forms a little later in this model relative to observations), all lines in the blue are due to Fe (Fe\two\ and to some smaller extent Fe\one) and Sc\two\ lines. Interestingly, \naid\ is such a strong transition that the width of the trough hardly changes with passing time and thus despite the drop in density. For \ha, we recover the usual double absorption due to Si\two\ at large negative Doppler velocities and \ha\ causing the absorption between $-12,000$\,\kms\ and line center --- there is additionally the typical strong blueshift of the peak emission \citep{DH05a}, also present in the observations. At later times, some bumps associated with Fe\two\ appear on the blue side. The O\one\ line exhibits a similar behavior to \naid, with a comparable contamination by Fe lines at later epochs. The evolution of Pa\,$\gamma$ is comparable to that of \ha\ but the overlap is with C\one\,10,685.4\,\AA\ ($\sim$\,7090\,\kms\ blueward of Pa\,$\gamma$) and only at late times (whereas Si\two\ is present at earlier times, C\one\ forms later under conditions of low ionization and temperature).

In the model with interaction, the results share some similarities with the reference model x6p0 but differ in two fundamental ways. First, the presence of a (spherical) dense shell at 8000\,\kms\ leads to a localized enhanced absorption in the outer ejecta (whereas model x6p0 has a smooth density structure extending out to 13,500\,\kms), with no absorption beyond (i.e., the density steeply declines beyond the CDS; see Fig.~\ref{fig_init}). The second distinction is that the injected power alters the ionization in and around the CDS, allowing for the formation of lines absent in the reference model x6p0. The bottom row illustrates these features. In all four lines shown, once the photosphere has receded to low velocities, the CDS leaves a clear narrow absorption at a Doppler velocity of $-$\vcds, distinct from the location of maximum absorption. This ``high-velocity'' notch is also present in the Si\two\,6355\,\AA\ profile. In addition, the boost in He\one\ excitation leads to the formation of He\one\ lines in the \naid\ and Pa\,$\gamma$ region. Finally, there is excess emission on the red side of \ha\ (i.e., the redward ledge) and additional kinks in the Pa\,$\gamma$ profile, both reminiscent of what is observed in \sn\ (though with some clear offsets that suggest some model deficiencies).

Overall, we find that the model with interaction (and a CDS) matches closely the observed lines widths of \sn\ shown in Fig.~\ref{fig_prof_kinks}. In particular, Si\two\,6355\,\AA\ (and later on Fe\two) naturally explains the high-velocity kink blueward of the rest wavelength of \ha. The peculiar morphology of the Pa\,$\gamma$ region likely arises from the simultaneous contamination of C\one\ and He\one\ lines, with emission and absorption from both ejecta and the CDS. We thus find no evidence for material present at larger velocities than about 8000\,\kms. The lack of a clear high-velocity notch in \naid\ or \ha\ indicates, however, that the CDS is either more spread out in velocity space than adopted here or that the CDS is asymmetric (we discuss this aspect in Section~\ref{sect_asym}). This notch may be present in the O\one\ line but the feature is uncertain since it coincides with a region affected by absorption by Earth's atmosphere. The broad and extended absorption seen in Pa\,$\gamma$ is likely due, in part, to He\one\,10,830.2\,\AA, which the model underestimates.
One feature unexplained by the models is the bump at about $-11,000$\,\kms\ from \naid\ present in the Lick observations throughout the photospheric phase (but potentially absent in those of \citealt{derkacy_23ixf_26}). This may be due to N\two\ (3p-3s multiplet around 5676\,\AA\ and not predicted by the model), or some instrumental error.

Figure~\ref{fig_NIR_66d} illustrates further the impact of interaction power and the CDS on model lines profiles and how these compare to \sn. Here, the reference model is complemented with two models having interaction power (so-called Pwr($t$) and Pwr1($t$); see Table~\ref{tab_pwr}), all at a time of 63.6\,d post explosion, and compared with the optical observations at 61.5\,d and the NIR observations at 67.5\,d of \sn. The comparison includes \naid, O\one\,0.777\,\mic, and a sample of H\one\ lines from the Balmer, Paschen, and Brackett series. Evidently, line profiles in the reference model exhibit only one absorption blueward of the rest wavelength, with the notable exceptions of O\one\,0.777\,\mic\ (high-velocity absorption is due to K\one\,0.768\,\mic) and Pa\,$\gamma$ (high-velocity absorption due to C\one\,1.069\,\mic). Both models with interaction exhibit a notch at $-8000$\,\kms, which coincides with the blue edge of the trough of NIR H\one\ lines. These exhibit either an extended trough with a flat bottom or with a double kink, reminiscent of what is seen in the model with interaction and in stark contrast with the reference model in which only one absorption is present closer to the line rest wavelength. Such double absorptions are absent at early times (see Fig.~\ref{fig_prof_kinks}) because the photosphere lies within or close to the CDS (Fig.~\ref{fig_phot_prop}). Near the end of the photospheric phase is the time when the photosphere is farthest away from the CDS and when such double kinks are best witnessed, confirming the presence of a CDS, even if not perfectly symmetric. This detection is facilitated in the NIR because the continuum photosphere is larger than in the optical owing to the enhanced continuum opacity (i.e., bound-free and free-free; see Section~\ref{sect_ir}).

We have here analyzed a similar dataset as \citet{park_23ixf_25} but we arrive at different conclusions. Our analysis suggests that there is essentially no material at velocities larger than about 8000\,\kms, which corresponds to the velocity of the CDS. By extension, this should also apply to UV lines, although in this spectral range, the widespread influence of metal-line blanketing is a complication. Notches at higher velocities are due to other lines (e.g., Si\two\,6355\,\AA\ in H$\alpha$). We also find that the overlap between transitions of the Paschen and Brackett series can give misleading signatures of high-velocity features. This is the case with Br\,$\epsilon$ at 18,174.12\,\AA, which is located about 10,000\,\kms\ blueward of Pa\,$\alpha$ at 18,751.00\,\AA.


\begin{figure}
\centering
\includegraphics[width=0.9\hsize]{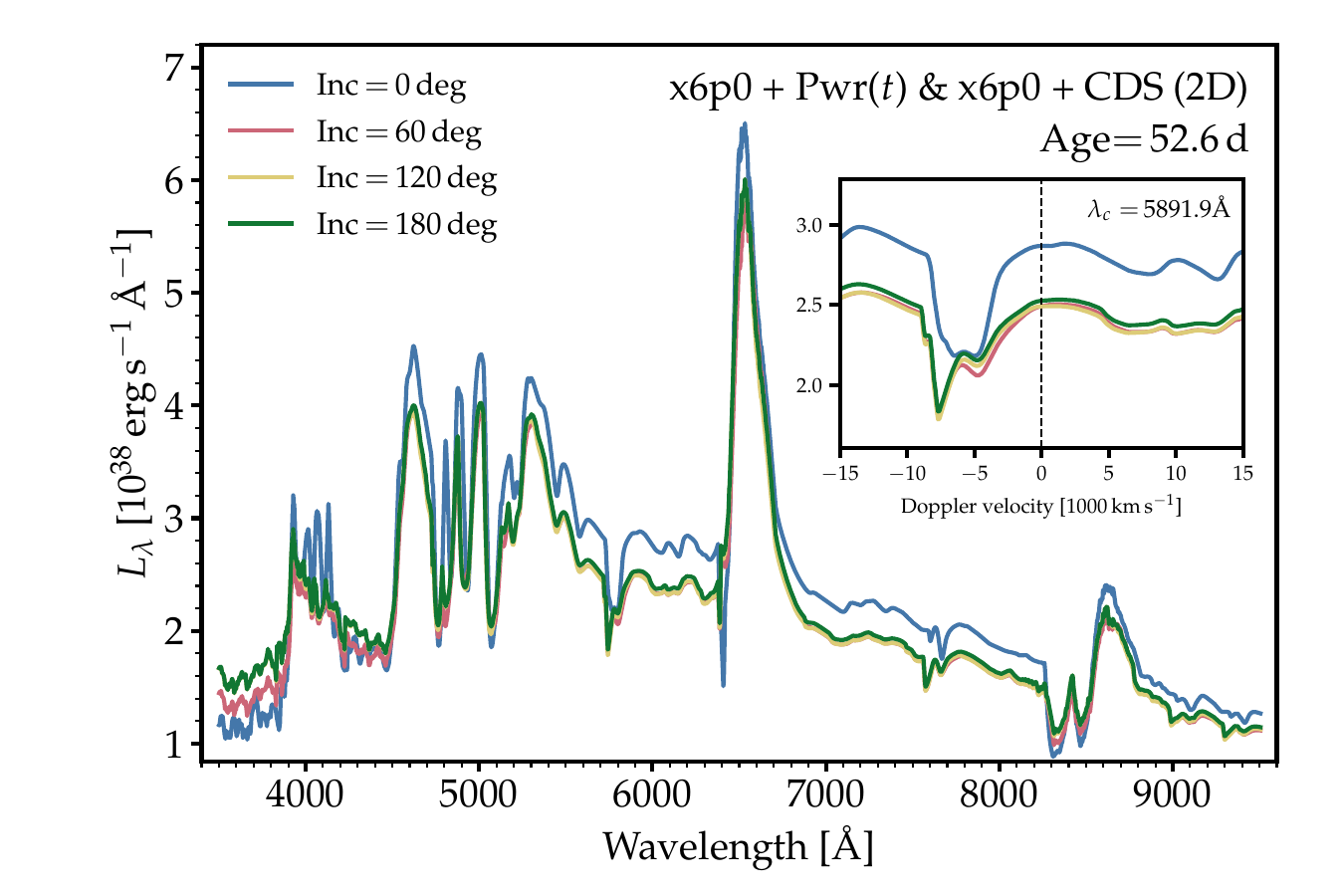}
\includegraphics[width=0.9\hsize]{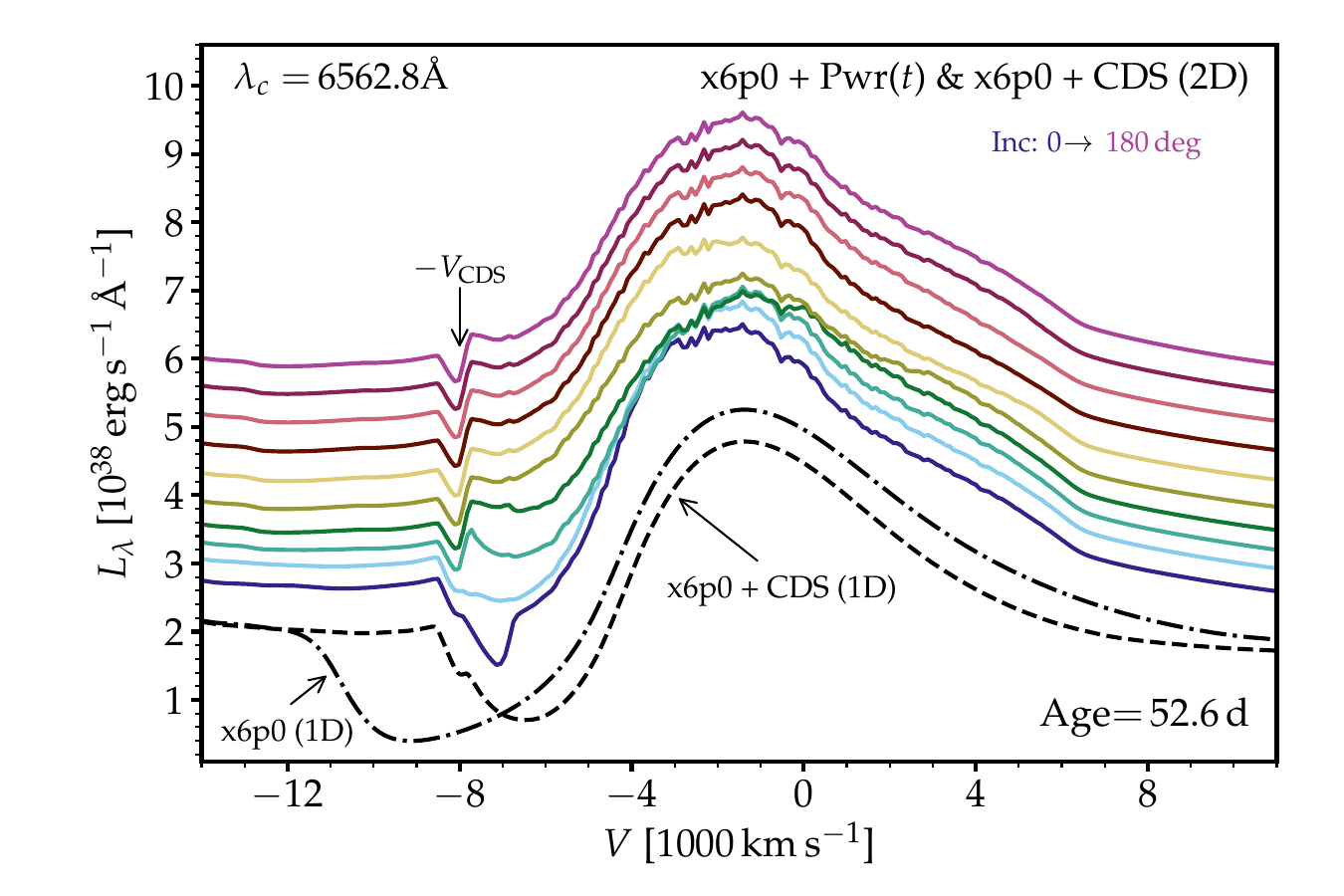}
\includegraphics[width=0.9\hsize]{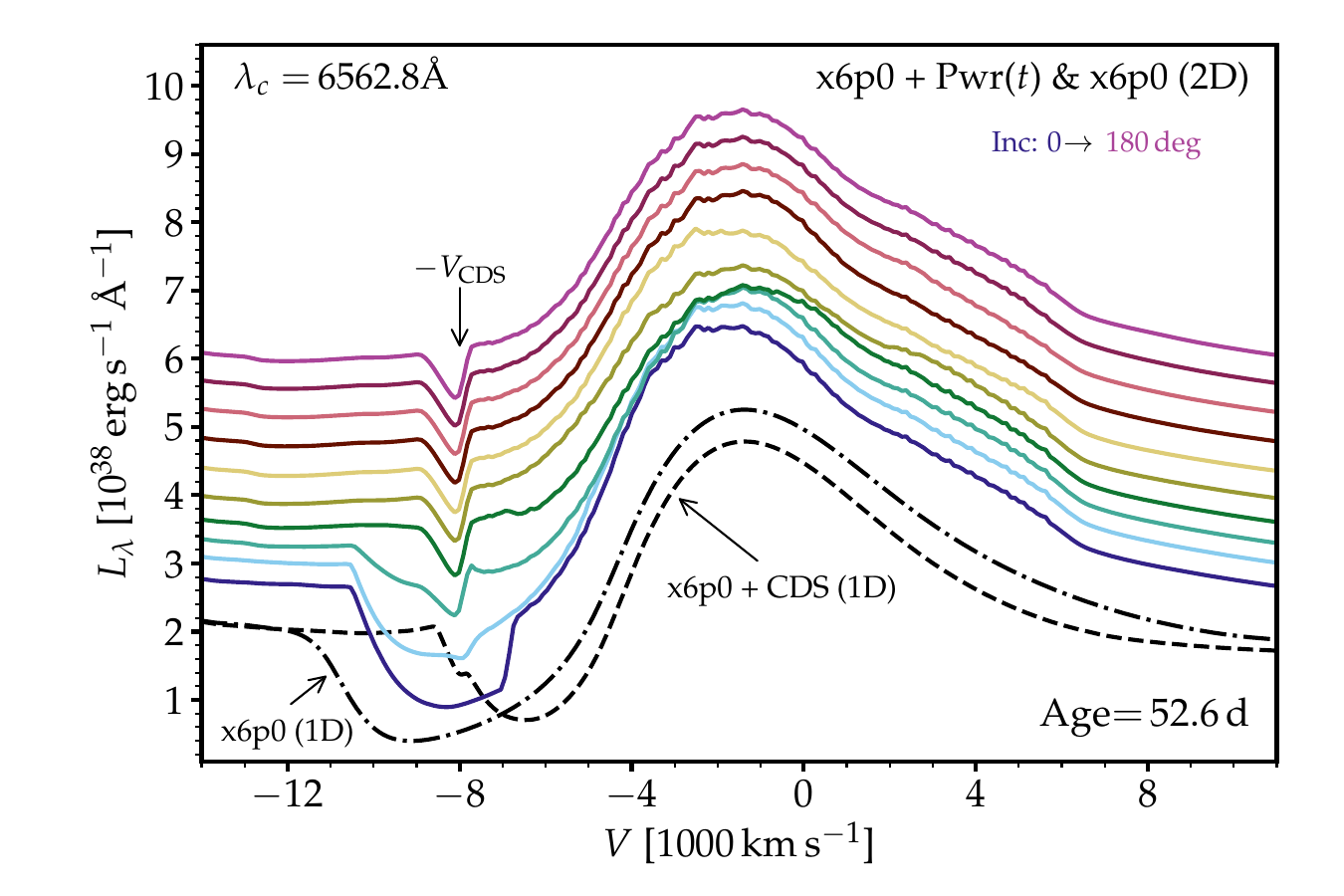}
\caption{Variation of the optical spectrum (top) or \ha\ region (bottom panels) with inclination for a 2D, axisymmetric radiative-transfer model based on model x6p0 with interaction along all directions except within a cone of 60\,$\deg$ opening angle (i.e., within 30\,$\deg$ of the polar axis). An inclination of 0\,$\deg$ corresponds to viewing down along the center of that cone. The material within 30\,$\deg$ of the polar axis is either a model with a CDS at 8000\,\kms\ but no interaction power (top two panels), or a model without interaction power nor a CDS (bottom panel). The colored lines indicate the 2D results for different viewing angles uniformly spread every 20\,$\deg$ from zero (black) to 180\,$\deg$ (magenta). The black broken curves in the middle and bottom panels correspond to the results for the spherical (1D) models with a CDS but no interaction power (dashed) or without interaction power nor a CDS (dash-dotted).
\label{fig_spec_2D_52p6d}
}
\end{figure}

\section{Impact of asymmetry}
\label{sect_asym}



Nonzero and time-dependent intrinsic polarization detected in \sn\ suggests that both the ejecta and CSM were asymmetric \citep{vasylyev_23ixf_23,singh_23ixf_24,shrestha_23ixf_25,vasylyev_23ixf_26}. While polarization is a compelling diagnostic of asymmetry in SNe \citep{shapiro_sutherland_82,wang_wheeler_rev_08}, it is hard to detect and interpret. So, finding evidence for asymmetry by other means would be a desirable asset.

Here, we explored 2D, axisymmetric ejecta configurations by combining the interaction model x6p0 + Pwr($t$) with either the reference model x6p0 (no CDS, no interaction power, maximum radial velocity of 13,500\,\kms) or model x6p0 augmented with a CDS but without interacting power (maximum radial velocity of $\sim$\,9000\,\kms). The approach is the same as that used for polarization studies conducted with \longpol\ \citep{hillier_94,hillier_96,DH11_pol} and applied to multiepoch polarization of SNe~II (e.g., SN\,2012aw; \citealt{dessart_12aw_21}). In practice, we assigned the properties of model x6p0 + Pwr($t$) to all polar angles between 40 and 180\,deg, and the properties of the other model (i.e., x6p0 or x6p0 + CDS) to polar angles below 30\,deg, interpolating between the two for polar angles between 30 and 40\,deg. We focused on just one epoch of 52.6\,d post explosion. To decrease the computation cost, we limited the \longpol\ calculation to the optical spectrum between 3500 and 9500\,\AA\ (the viewing angle dependence would also apply in the UV, including \lya, but this discussion is left to future work). The output from \longpol\ provides a spectrum for 19 inclinations spaced uniformly every 10\,deg between 0\,deg (pole-on) and 180\,deg (for details, see \citealt{dessart_12aw_21}).

The top panel of Fig.~\ref{fig_spec_2D_52p6d} shows the optical spectrum for the 2D model composed of model x6p0 + Pwr($t$) (polar angles 40 to 180\,deg) and model x6p0 + CDS (polar angles below 30\,deg), and for inclinations of 0, 60, 120, and 180\,deg. The inset zooms-in on the \naid\ region for the same inclinations. The middle panel of Fig.~\ref{fig_spec_2D_52p6d} is a zoom-in on the \ha\ region for the same 2D model but shown for inclinations uniformly spaced every 20\,deg from 0 and 180\,deg. With the breaking of spherical symmetry, the optical color becomes inclination dependent. Along the axis of symmetry, one sees down a column of material that is cooler and more recombined so that the electron-scattering optical depth is lower (the electron-scattering photosphere is located deeper) but the metal-line blanketing is greater (e.g., more opacity from Fe\two\ and Ti\two). This allows for more photons to escape at long wavelengths but fewer at short wavelength. Beyond an inclination of 40\,deg, the emergent spectrum does not vary much. The second impact of asymmetry is that the trough of lines like \naid\ and \ha\ becomes inclination dependent, with the high-velocity notch associated with the CDS potentially disappearing, primarily due to the variation in the strength of emission from the CDS filling-in that trough.

The bottom panel of Fig.~\ref{fig_spec_2D_52p6d} is the counterpart of the middle panel in which the 2D model is a composite of model x6p0 + Pwr($t$) and the reference (smooth, no CDS, no interaction) model x6p0. This 2D configuration mimics the presence of a ``hole'' in the CSM for a viewing angle of 0\,$\deg$ to the SN. We can see a much greater variation in H$\alpha$ than in the previous case because the fastest material in the x6p0 model is 13,500\,\kms. So, for an inclination of 0\,deg, the bluest parts of the \ha\ trough is like in the model x6p0, with the CDS emission appearing for Doppler velocities between $-V_{\rm CDS} \cos(35\,\deg)\approx -6500$\,\kms\ and \vcds\ (modulo optical-depth effects occulting the back side of the ejecta). As the inclination is increased, the \ha\ trough first adopts an extended flat bottom, before recovering the essentially pure-emission profile with a notch at $-$\vcds.

Obviously, a great variety of asymmetries may occur in nature, with a CDS broken not just in one location but in multiple directions. The CDS emission may also be modulated by varying the CSM density at a given time --- it is the current power that affects the CDS emission at any given time because the CDS is optically thin (i.e., here when considering a post-explosion age of 52.6\,d; see also Fig.~\ref{fig_phot_prop}). Any radial direction where the CSM was less dense would have material stretching to larger velocities (as in the reference model x6p0), allowing line absorption to occur potentially at Doppler velocities larger than the average radial CDS velocity. With asymmetry, profile kinks may also occur within the blueward  and redward bounds of the line profiles. Overall, these results demonstrate that ejecta asymmetry, combined with ongoing or past interaction with CSM, may explain the observed diversity of line profiles in SNe~II, as documented for example for \ha\ \citep{gutierrez_ha_14}. In the context of \sn, it explains how some profiles can have an extended trough with a flat bottom or with internal flux modulations or kinks (see, e.g., Fig.~\ref{fig_NIR_66d}).


\begin{figure*}
\centering
\includegraphics[width=0.9\hsize]{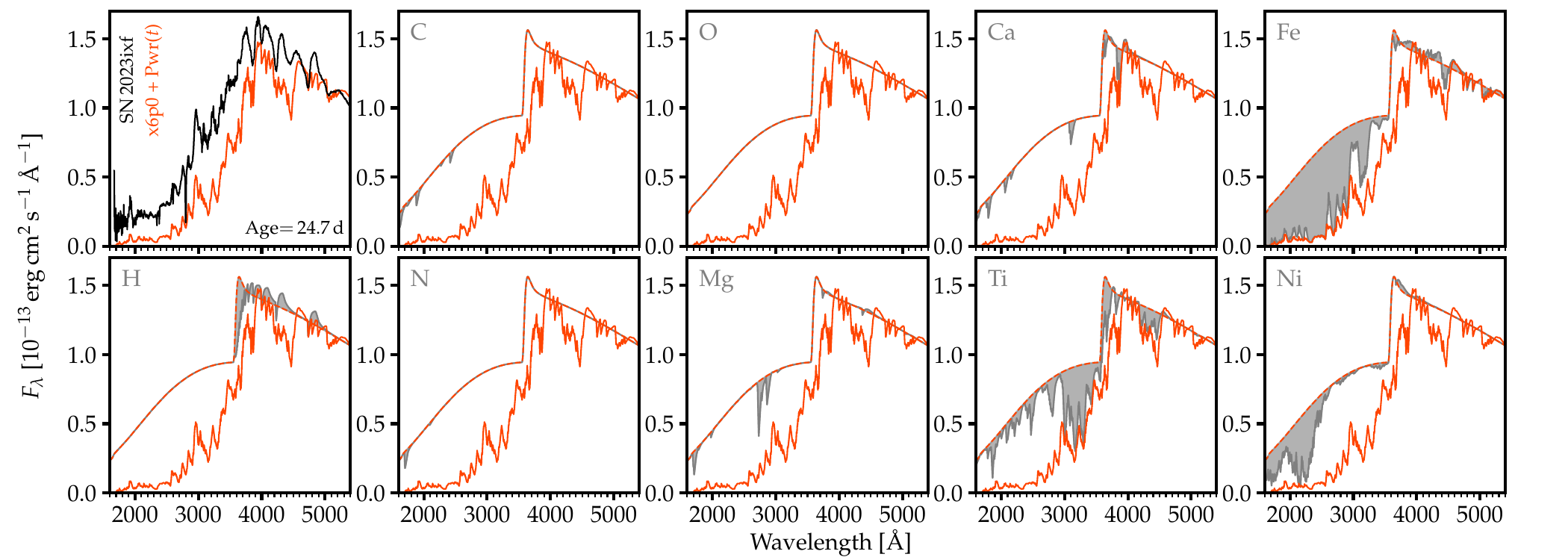}
\caption{Spectral properties blueward of 5500\,\AA\ for \sn\ and the x6p0 model with interaction power at 24.7\,d. The top-left panel compares observations and model, whereas the other panels show the total flux (red), the continuum flux (dashed red curve) and the contribution from selected species (gray shading; see label). A similar illustration with similar properties is shown for the epoch of 67.5\,d in Fig.~\ref{fig_UV_66d}.
\label{fig_UV_24d}
}
\end{figure*}

\section{Ultraviolet properties: comparison with HST observations at 24.7 and 66.5\,d}
\label{sect_uv}

Photometric observations of \sn\ were conducted in the UV using Swift \citep{jacobson_galan_23ixf_23,teja_23ixf_23,singh_23ixf_24,zimmerman_23ixf_24}. Spectra were obtained at early times (during the SN~IIn and CDS phases) in the far-UV by \citet{teja_23ixf_23} at 6.9, 11.9, 23.4\,d, and in the near-UV by \cite{zimmerman_23ixf_24} at 3.6, 4.7, 5.6, 8.7, and 11.7\,d. Modeling of these data will be presented in a separated study. Spectroscopic observations were obtained in the UV at later epochs during the photospheric phase at 24.7 and 66.5\,d \citep{bostroem_23ixf_24}. We have already shown model comparisons to these data in preceding sections but we now explore in greater depth the information contained in those UV spectra.

At the two epochs of interest, the SED peaks in the optical, typical of a cool photosphere roughly at the H recombination temperature (Fig.~\ref{fig_phot_prop}). Under such conditions, the UV flux is modest because of this low temperature, and further reduced by the effects of metal-line blanketing. Assessing the origin of the emission ``bumps'' in the UV range is difficult. When analyzing the UV spectrum of SN\,1999em obtained at one week after explosion \citep{baron_99em_00}, \citet{DH06} emphasized how these bumps coincided essentially with holes or gaps in the blanketing --- that is, spectral regions coinciding with opacity depressions. The present situation is similar (see also the UV properties of the noninteracting Type II SN\,2022acko where similar conclusions were reached; \citealt{bostroem_22acko_23}).

To complement the analysis presented by \citet{bostroem_23ixf_24}, we investigate the UV properties in a different way. Specifically, we explore how a given species affects the continuum distribution and drives it toward the full spectrum that results from the combined influence of line and continuum processes. This approach is instructive because in the absence of any metals, as would obtain at zero metallicity, the only lines present in a Type II SN spectrum during the photospheric phase would likely be limited to H\one\ and possibly He\one, with perhaps some small contribution from a few metal lines arising from metals dredged up from the metal-rich core. Thus, at zero metallicity, the UV flux would be the continuum flux with essentially only Ly$\alpha$ in the far-UV. Instead, at solar or near-solar metallicity, primordial metals present in the H-rich material (i.e., the progenitor H-rich envelope) carve into or sculpt this continuum flux over specific spectral windows. This effect generally leads to a reduction of the emergent flux in the UV, but a more mitigated effect with reductions and enhancements in flux in the optical and beyond.

Figure~\ref{fig_UV_24d} illustrates the impact on the flux due to a selection of species (i.e., those with the greatest impact) between H and Ni for the epochs of 24.7\,d (a similar figure is presented in the Appendix for the observations and model at 67.5\,d; see Fig.~\ref{fig_UV_66d}). The top-left panel shows a comparison of the total flux to the observations, and the other panels show a comparison of the total model flux, the continuum flux, and the corresponding flux (gray shaded area) arising from bound-bound transitions of a given species (see label).

Beyond about 1500\,\AA, the model predicts the presence of specific transitions (often multiplets) from specific ions including H\one\ with lines present only in the optical range longward of the Balmer jump, N\two\ lines at 1676 and 1740\,\AA, an N\three\ line at 1750\,\AA, C\two\ lines at 1761 and 2325\,\AA, C\three\ lines at 1909 and 2297\,\AA, a Mg\two\, line at 1738\,\AA\ together with \mgiiuv, and finally Ca\two\ lines at 1840, 2112, 3179, 3737\,\AA\ together with Ca\two\,H\,\&\,K. But these individual contributions are swamped by the metal-line blanketing due to Fe\two\ and Ti\two, whose opacity operates over overlapping regions but with Ti more present at longer wavelengths, especially with the strong blanketed region around 4000--5000\,\AA. Figure~\ref{fig_UV_24d} clearly shows how the variations in opacity from Fe and Ti carve the total-flux spectrum, such that the distribution of the latter (with bumps and dips) correspond to reductions and enhancements in the blanketing caused by Ti and Fe. At 24.7\,d, the influence of lines in the UV is primarily absorptive. At the second epoch (Fig.~\ref{fig_UV_66d}), the CDS is less optically thick in the UV and optically thin in the optical, and some emission is then visible (i.e., on top of the continuum flux) from \mgiiuv, Ca\two\,H\,\&\,K, as well as some emission from Ti\two\ and Fe\two\ lines in the optical range.

Overall, the contribution of the interaction to emission in the UV is modest in \sn. The UV flux remains faint and not far above the UV flux in the reference model without interaction. Irrespective of interaction, the outer ejecta optical depth (whether there is a CDS or not) remains significant in the UV and strong metal-line blanketing results. Relative to the total flux, the UV flux will be boosted much later when the ejecta and CDS density as well as the optical depth continue to drop while the interaction power remains significant \citep{bostroem_uv_25,wynn_sed_25}, leading also to an enhancement in the outer ejecta temperature and thus the UV emissivity \citep{dessart_csm_22,dessart_late_23}.


\begin{figure*}
\centering
\includegraphics[width=0.9\hsize]{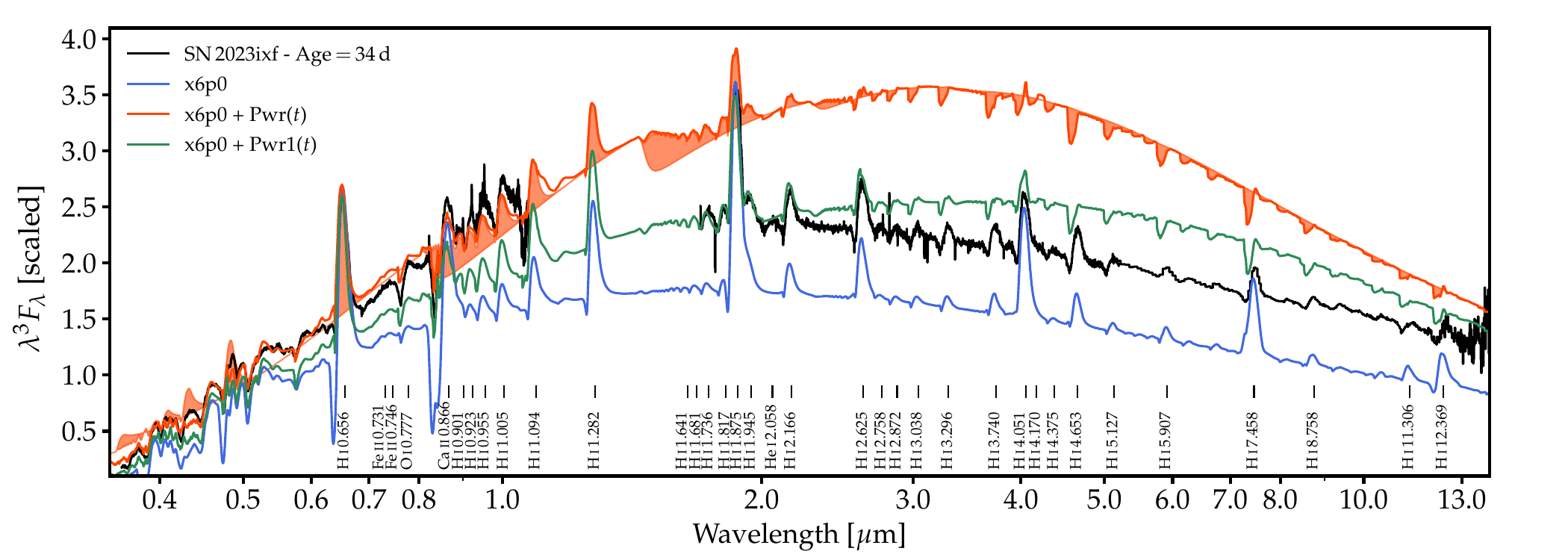}
\caption{Comparison between optical and IR observations of \sn\ with various incarnations of the x6p0 model with and without interaction power. The shaded red area represents the contribution from bound-bound transitions of H\one. Labels indicate the main contributor to the spectral feature at the corresponding wavelength. Observations have been corrected for redshift and reddening. Models have been scaled to the distance of \sn. For better visibility, all fluxes have been scaled by $\lambda^3$.
\label{fig_IR_34d}
}
\end{figure*}

\section{Infrared properties; comparison with JWST observations at 33.6\,d}
\label{sect_ir}

Figure~\ref{fig_IR_34d} shows the NIR and MIR observations of \sn\ at 33.6\,d post explosion \citep{derkacy_23ixf_26} together with the optical observations of \sn\ at 34.5\,d \citep{zheng_23ixf_25} and compared to the reference model x6p0 and the two counterparts with interaction power at 35.9\,d. Being halfway through the photospheric phase, the SED exhibits a strong continuum flux on top of which appear numerous lines. These lines, which are essentially all due to H\one\ beyond $\sim$\,1\,$\mu$m, tend to show a P-Cygni profile morphology with a strong emission component and a weak absorption component. All three models predict the same set of lines as observed, but each model presents a different continuum flux level and different profile morphologies with a distinct absorption-to-emission ratio. The reference model x6p0 shows the weakest continuum and most lines are in emission, whereas with increasing interaction power, lines progressively turn to being pure (blueshifted) absorption.

\begin{figure}
\centering
\includegraphics[width=0.9\hsize]{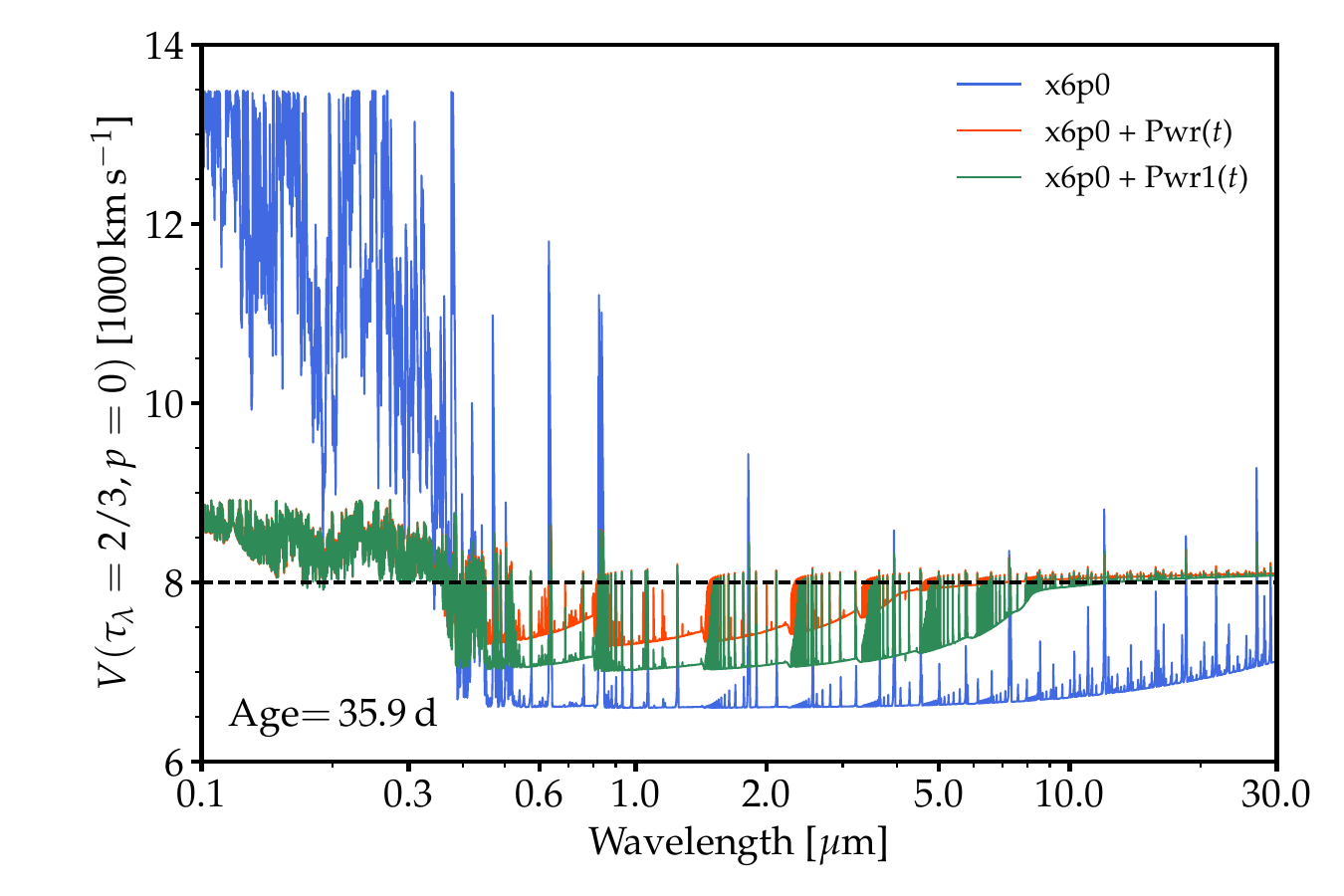}
\caption{Wavelength dependence of the ejecta velocity $V(\tau_\lambda=2/3)$ at which the radially (i.e., impact parameter $p=$\,0) inward integrated total optical depth is 2/3 for the reference model x6p0 and two counterparts with shock power at 35.9\,d after explosion. The near-constant velocity minimum in each model corresponds to the velocity of the electron-scattering photosphere --- this velocity increases for greater interaction power. The dashed-black line indicates the CDS velocity, to which $V(\tau_\lambda=2/3)$ converges at long wavelengths in models with interaction power (and a CDS). \label{fig_vtau_34d}
}
\end{figure}

The origin of these model differences stems from the different optical depth of their ejecta (together with the CDS when present) versus velocity and wavelength. To illustrate these differences, we show in Fig.~\ref{fig_vtau_34d} the wavelength-dependent photospheric velocity $V(\tau_\lambda=2/3)$ along a radial ray striking the SN center in each of the three models --- the photospheric radius is a less convenient quantity since it changes with SN age. In the UV, $V(\tau_\lambda=2/3)$ reaches the outermost grid location, set at 13,500\,\kms\ in model x6p0 and 8900\,\kms\ in models with interaction power. That is, all three models are optically thick in the UV all the way to the outermost grid point. In the reference model x6p0, $V(\tau_\lambda=2/3)$ is essentially constant all the way to 10\,\mic\ and rises beyond with the increase in free-free opacity. In the two model counterparts with interaction, $V(\tau_\lambda=2/3)$ is systematically greater, and rises in the IR at shorter wavelength, and the more so for increasing power. This rise in both bound-free and free-free opacity likely arises from the greater electron density in the outer ejecta. On top of these continuum processes, narrow spikes in  $V(\tau_\lambda=2/3)$ appear for every bound-bound transition, here mostly due to H\one.

Figure~\ref{fig_vtau_34d} thus explains the different profile morphology in the IR between models. In the reference model x6p0, the lines form at both optical and IR wavelengths over a large volume above the photosphere, leading to strong emission relative to the absorption part. With interaction power, the photosphere shifts closer to the CDS, reducing the available volume for emission. In the optical the reduction is moderate, but in the IR both lines and continua form within the CDS at that time, and only an absorption component can result (model x6p0 + Pwr($t$)). The larger radiating surface in those models with interaction also lead to a greater continuum flux in the IR.

The three models shown in Fig.~\ref{fig_vtau_34d} bracket the observed flux. The reference model x6p0 is too faint throughout but captures well the emission strength observed in H\one\ lines in the IR. Model x6p0 + Pwr1($t$) is too faint in the optical but matches roughly the IR (the actual distribution is skewed). Model x6p0 + Pwr($t$) matches the optical but strongly overestimates the IR flux, also predicting lines that are systematically in absorption and conflicting with observations. While interaction seems required to explain the flux level and the morphology of line profiles in the IR, the properties within the CDS are probably more complex than currently assumed in the modeling. One possibility not considered so far is that the IR data may have flux-calibration issues, either in an absolute or relative sense --- the optical data used here match the photometry of \sn\ to within a few 0.01\,mag.

Our simulations confirm the findings of \citet{derkacy_23ixf_26} who proposed free-free emission as the main IR source in \sn\ at those early, photospheric epochs. We surmise that it is because of the rising continuum opacity in the IR, enhanced by the interaction with CSM, that the IR line profiles are such sensitive diagnostics of the ejecta structure. This is, however, most vivid in the second half of the photospheric phase when the photosphere has receded much below the CDS (see Section~\ref{sect_kinks}).


\begin{figure}
\centering
\includegraphics[width=0.9\hsize]{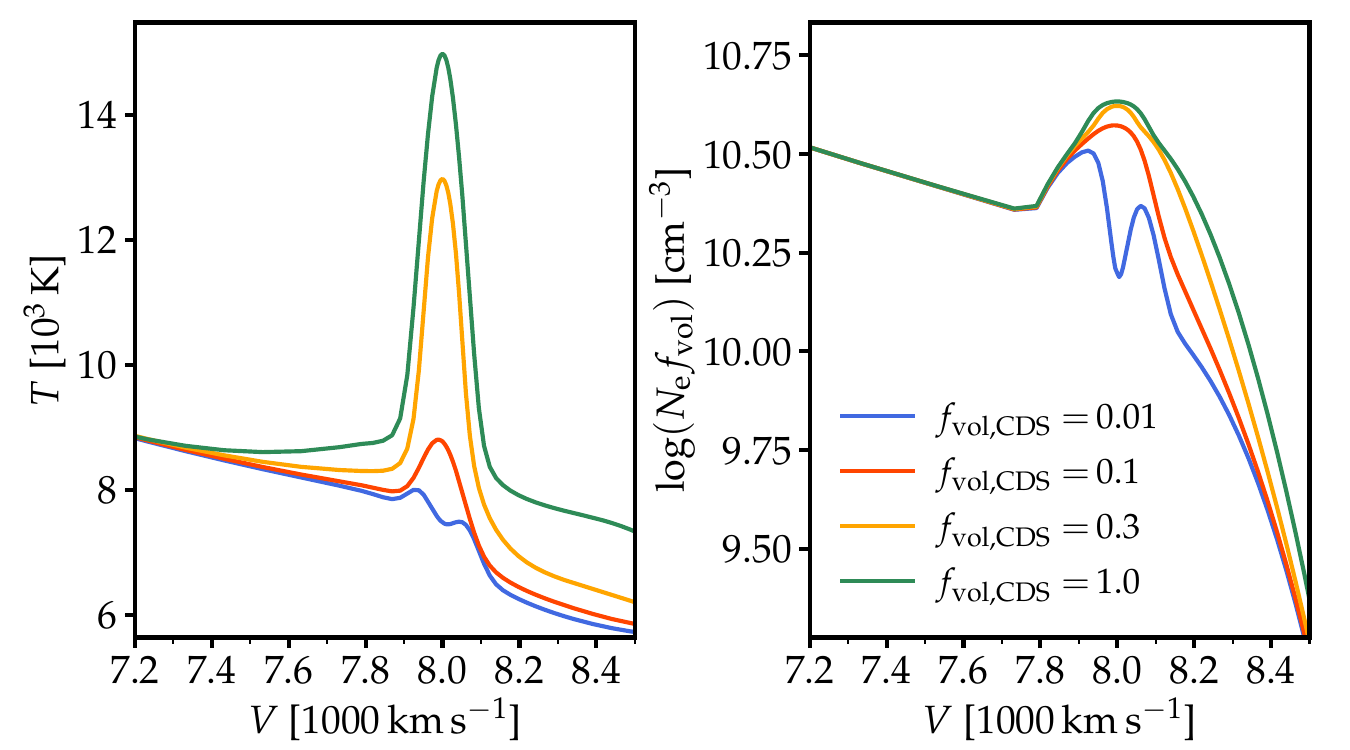}
\includegraphics[width=0.9\hsize]{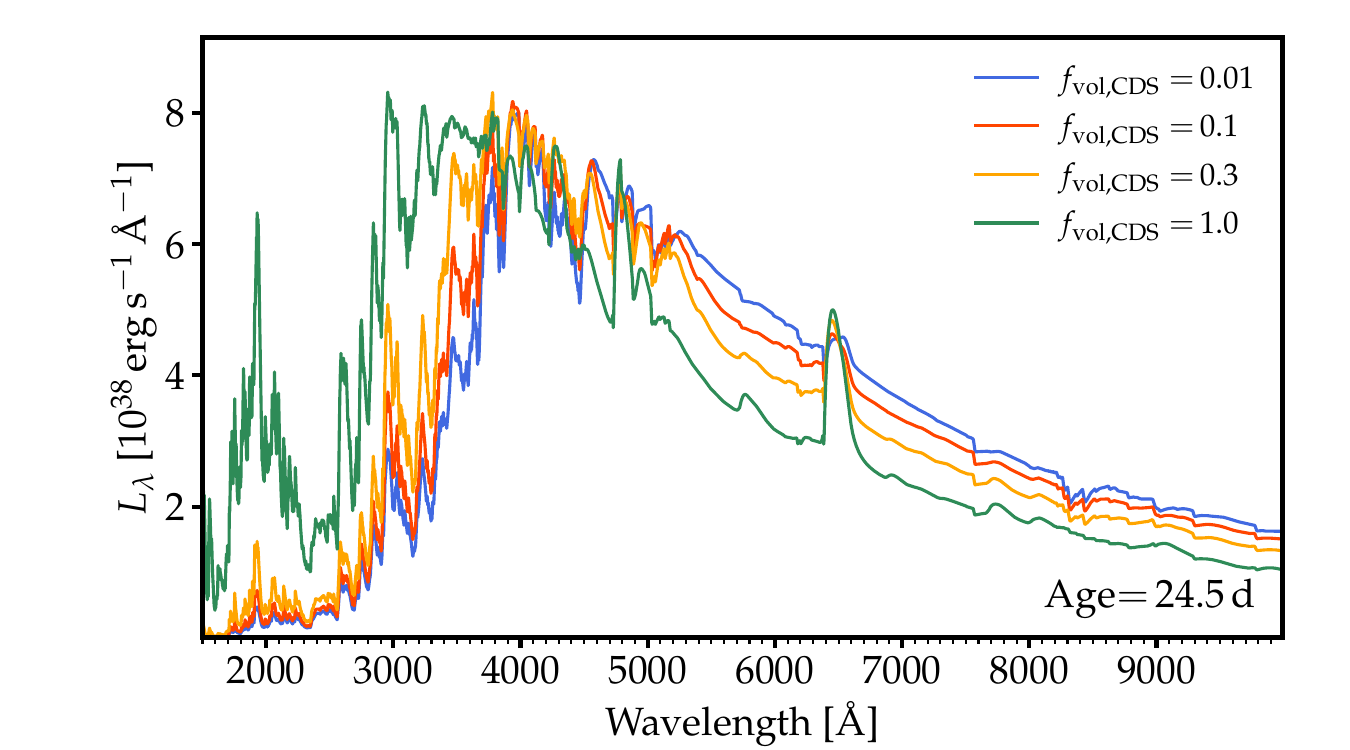}
\includegraphics[width=0.9\hsize]{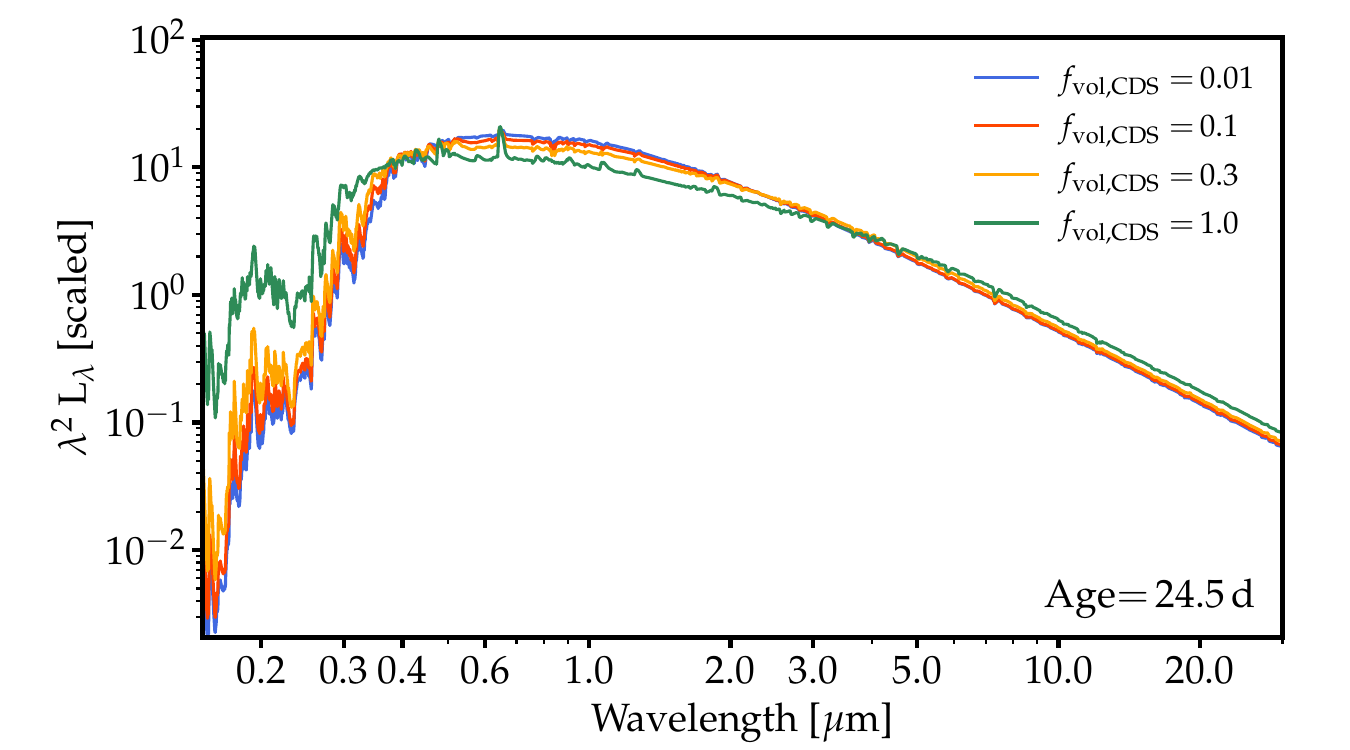}
\caption{Impact of the clumping within the CDS on the gas and radiative properties for model x6p0 with an interaction power of $2 \times 10^{42}$\,\ergs\ at 24.5\,d. From top to bottom, we show the temperature and electron density (corrected for clumping) in the CDS, the optical luminosity, and the (scaled) quantity $\lambda^2 L_\lambda$ from the UV to the IR.
\label{fig_var_fvol_24d}
}
\end{figure}

\section{Impact of clumping within the CDS}
\label{sect_fvol}

All the simulations presented so far in this work use a volume filling factor of the gas of 1\,\% within the CDS (see Fig.~\ref{fig_init} --- the material is smooth outside of the CDS), as also adopted by \citet{dessart_csm_22} and \citet{dessart_late_23}. This corresponds to a maximum density compression of 100. Such rearrangement from a smooth to a clumped ejecta (or CDS) does not change the radial optical depth (at fixed ionization) because it amounts to performing a compression of the material in the radial direction only. In this section, we explore the influence of clumping on the model results and whether modulating this clumping level could reduce or cure some of the model discrepancies relative to observations (see also \citealt{bostroem_23ixf_24}). In this exploration, all other quantities were kept the same, including for example the injected interaction power. These tests were done in a time-dependent calculation over one time step, assuming that in the previous time step the clumping level was that used in the standard model (volume filling factor of the gas of 1\,\%). So, there is some ``inertia'' in the results in the sense that $d/dt$ terms contain results from the previous time step with a different clumping.

Figure~\ref{fig_var_fvol_24d} shows the impact on the CDS properties and emergent spectra when the volume filling factor is varied from 1\,\% to 10\,\%, 30\,\%, and 100\,\%, the last value corresponding to a smooth density structure in the CDS and thus in the entire ejecta. As expected, the temperature and the ionization rise when the volume filling factor is increased (or equivalently when the clumping level is reduced; see also \citealt{d18_fcl}). The impact is not limited to the CDS because the temperature change in the CDS modifies the opacity and emissivity within the CDS. With a higher CDS temperature, a greater ionizing flux is radiated from the CDS and influences the nearby regions where it is absorbed. The rise in ionization for smoother ejecta also implies a weaker metal-line blanketing for optical photons crossing the CDS. There is thus a number of consequences from the change of that quantity alone.

The impact on the spectra varies in a nonlinear way. Raising the volume filling factor leads to a greater optical flux at first, and it is only for smooth ejecta that the UV flux is boosted, following the rise in CDS temperature. This likely arises from a combination of enhanced CDS temperature (emission from the CDS shifts to the blue) and reduced metal-line blanketing from the CDS. More complicated is the impact in the IR (bottom panel of Fig.~\ref{fig_var_fvol_24d}) which reflects the changes in electron density and continuum emission from the CDS. This exploration emphasizes how challenging it is to obtain a good match to the full SED from UV to IR as well as the numerous line profiles therein for SNe interacting with CSM.


\section{Conclusions}
\label{sect_conc}

In this paper, we have presented nonLTE time-dependent radiative-transfer simulations based on a tailored model of the progenitor, its explosion, and its subsequent evolution under the influence of interaction with CSM, and compared with photometric and spectroscopic observations of \sn\ from about 20 to 115\,d after first light. Consistent with a number of independent studies, we find that \sn\ resulted from the explosion of a partially stripped RSG progenitor, compatible with a 15\,\msun\ progenitor star on the main sequence but evolved with enhanced mass loss during the RSG phase (be it wind mass loss or binary mass transfer), and whose explosion produced an ejecta with a mass of 7--8\,\msun, a kinetic energy of $1.2 \times 10^{51}$\,erg and a \nifs\ mass of about 0.05\,\msun. This work is, however, the only existing nonLTE radiative-transfer calculation of the photospheric phase with allowance for interaction power.

\begin{figure}
\centering
\includegraphics[width=\hsize]{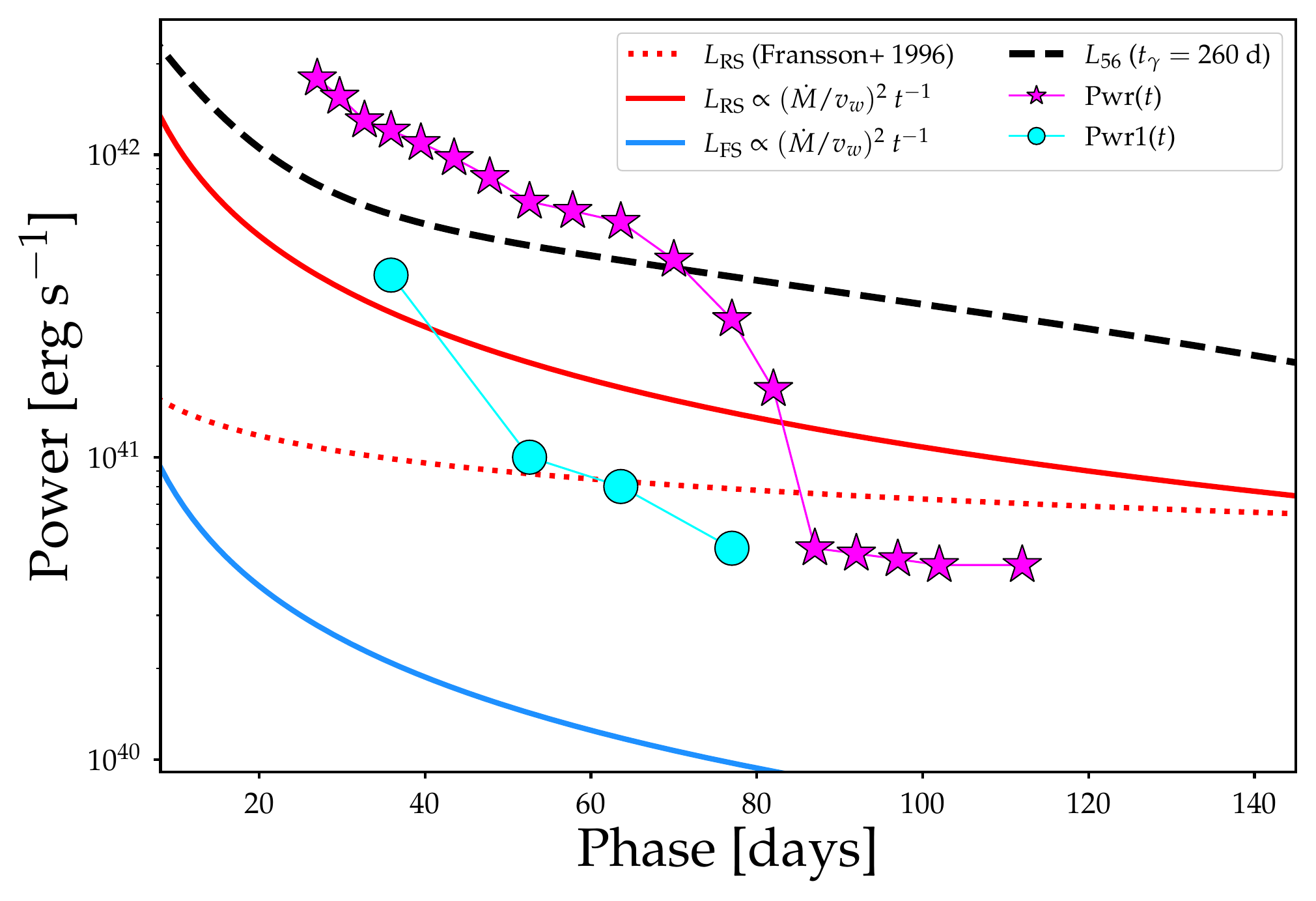}
\caption{Predicted reverse shock (RS) and forward shock (FS) powers for the formalism presented in \citet{fransson_93j_96} (dotted red line) and for thermal bremsstrahlung emission from adiabatic shocks (solid red and blue lines). These shock powers assume a mass loss rate of ${\dot M} = 10^{-4}$ M$_{\odot}$ yr$^{-1}$, a wind velocity $v_w = 20$ km s$^{-1}$, an ejecta density profile index $n = 12$, an ejecta velocity $v_{\rm ej} = 8000$ km s$^{-1}$, a wind-like CSM density profile $s = 2$, and electron-ion equipartition. Radioactive decay power for $M(^{56}{\rm Ni}) = 0.06$~M$_{\odot}$ and $t_{\gamma} = 260$~days shown as dashed black line \citep{wynn_sed_25}. Injected powers from Table~\ref{tab_pwr} are shown as magenta stars and cyan circles.
\label{fig_pwr_rs_fs}
}
\end{figure}

We found that there is interaction with CSM during the entire photospheric phase, as evidenced in part from the excess UV flux but more systematically here from the excess emission affecting the \ha\ profile. This corroborates alternate studies on the X-ray and radio properties of \sn\ at similar epochs \citep{chandra_23ixf_24,nayana_23ixf_25} and fits within the expectations of interaction with CSM (Fig.~\ref{fig_pwr_rs_fs}). This excess emission appears as a mild extension of the profile on the red side (hard to gauge because of the relative proximity between the photosphere and the CDS at early times) but a conspicuous one from the filling-in of the P-Cygni trough. This interaction power also affects the flux level throughout the  optical, leading to a stronger continuum flux. The observed profile morphology suggests a CDS at 8000\,\kms, with a velocity that remains fixed throughout the photospheric phase. This CDS, and in particular the mass budget (fixed at 0.2\,\msun), was built primarily from the sweeping-up of CSM (and decelerated ejecta material) that occurred in the first 1--2 weeks following shock breakout.

We explored with a variety of interaction powers, which have a strong impact on the UV, optical, and IR flux levels. This arises in a direct way from the excess power made available but also indirectly from the rise in ionization and reduction in metal-line blanketing, which impact the optical depth, modulating the UV flux, or varying the magnitude of free-free emission in the IR range. Variations in the adopted CDS clumping also modulate the SED by altering the temperature and ionization in the CDS. Because of this complex and nonlinear behavior, obtaining a model that matches the entire electromagnetic spectrum is a challenge. But our explorations demonstrate that a model with interaction power and a CDS fares considerably better than a model without.

We revisited the peculiar profile morphologies of optical and IR lines at different epochs in the photospheric phase. Guided by our models, we sharpened the analysis of \citet{park_23ixf_25} and \citet{derkacy_23ixf_26}, finding that double absorptions in line profiles are a natural consequence of the presence of a CDS in the outer ejecta, although it may be absent at early times when the photosphere is still within or close to the CDS. Such double absorptions are more easily identifiable in the IR, likely because the continuum optical depth is greater there, pushing out the photosphere to larger radii. All double absorptions are compatible with a CDS at about 8000\,\kms\ and there is no evidence for absorbing material at larger velocities at 20 to 115\,d. Apparent disagreements in this sector arise from overlap with nearby lines.

We also conducted 2D radiative-transfer calculations using a partnership between \cmfgen\ and \longpol, as normally done for the modeling of Type II SN polarization. Constructing 2D ejecta as a combination of interacting and noninteracting models, or interacting and smooth models, we demonstrated that asymmetry may be a fundamental source of profile diversity (e.g., in \ha\ or \naid), primarily in the trough, and appearing in the form of a high-velocity notch, or a notch somewhere within the trough, or as a trough with a flat bottom but without any high-velocity notch. The presence of a high-velocity notch in the P-Cygni trough indicates the presence of a density enhancement in the outer ejecta and located directly along the line of sight --- this feature provides no information on the presence of similar density enhancements along other ejecta-centered radial directions. The P-Cygni trough gives, in that respect, a very biased and restricted view of the ejecta as a whole, essentially limited to one direction. Yet, much interpretation in the literature is based on what appears in those P-Cygni troughs, including speculations on whether there is interaction, or whether there is He (i.e., for a Type Ib SN classification). Obviously, this diagnostic can be very misleading.

In forthcoming studies, we will extend the present work to the nebular-phase evolution as well as the early SN~IIn phase. As done here, we will compare models with the full SED of \sn\ from the UV to the IR ranges.


\begin{acknowledgements}

  L.D. acknowledges support from the ESO Scientific Visitor Program
  for a visit to ESO-Garching during the summer of 2025.
  W.J.-G. is supported by NASA through Hubble Fellowship grant
  HSTHF2-51558.001-A awarded by the Space Telescope Science Institute,
  which is operated for NASA by the Association of Universities for
  Research in Astronomy, Inc., under contract NAS5-26555.
  K.A.B is supported
  by an LSST-DA Catalyst Fellowship; this publication was thus made
  possible through the support of Grant 62192 from the John Templeton
  Foundation to LSST-DA.

  A.V.F.'s research group at UC Berkeley acknowledges
  financial assistance from the Christopher R. Redlich Fund, Gary and
  Cynthia Bengier, Clark and Sharon Winslow, Alan Eustace and Kathy
  Kwan (W.Z. is a Bengier-Winslow-Eustace Specialist in Astronomy),
  Timothy and Melissa Draper, Briggs and Kathleen Wood, Ellen and Alan
  Seelenfreund (T.G.B. is Draper-Wood-Seelenfreund Specialist in
  Astronomy), and numerous other donors.
  Research at Lick Observatory is partially supported by a gift from Google.
  This work was granted access to the HPC resources of TGCC under
  the allocation 2024--A0170410554 and 2025--A0190416871 on Irene-Rome
  made by GENCI, France.
  This work has made use of NASA's Astrophysics Data System
  Bibliographic Services.

\end{acknowledgements}

\bibliographystyle{aa}
\bibliography{./new_sn_library_luc}

\onecolumn
\appendix

\section{Additional comparisons with observations}
\label{sect_more}





\begin{figure*}[h]
\centering
\includegraphics[width=0.45\hsize]{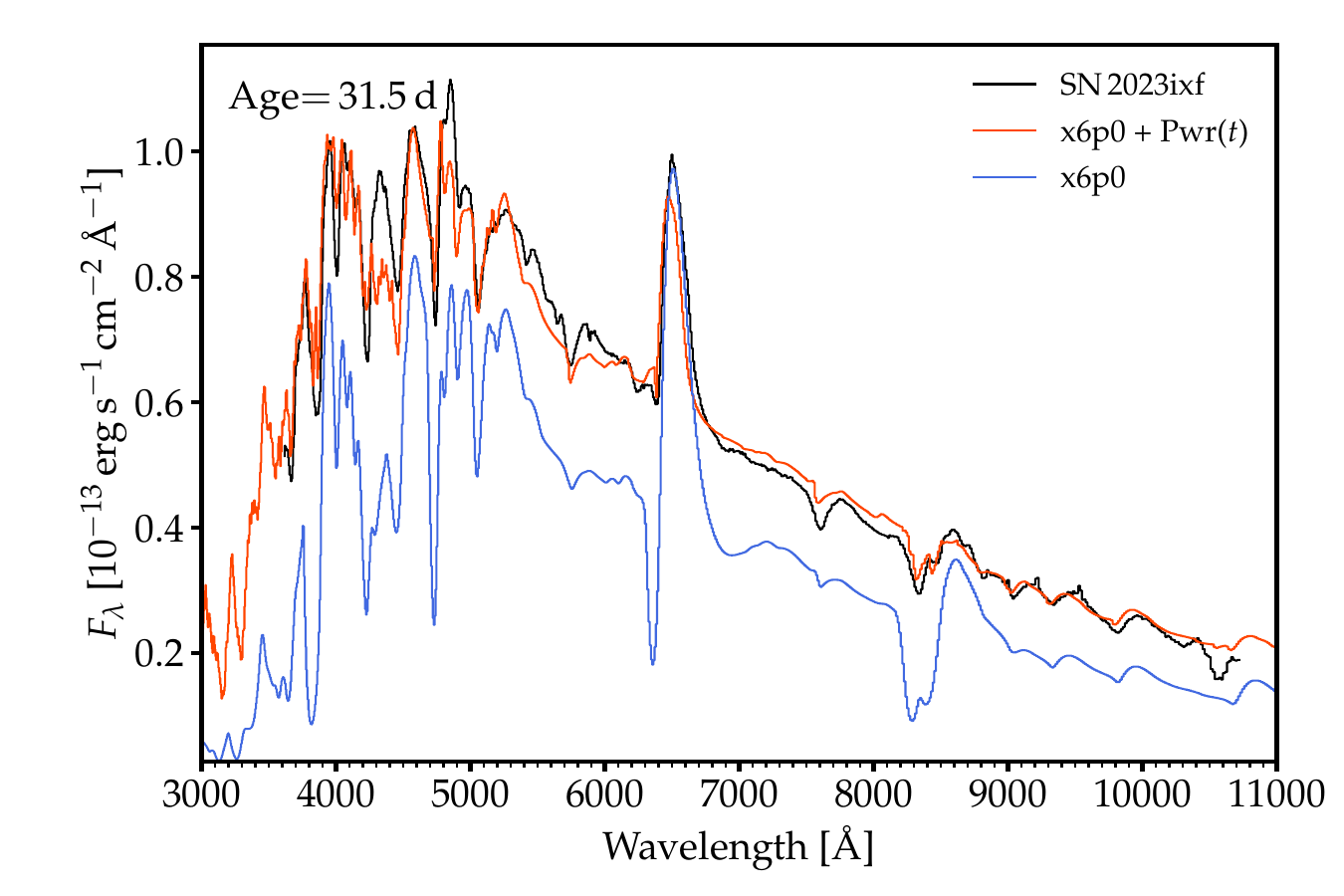}
\includegraphics[width=0.45\hsize]{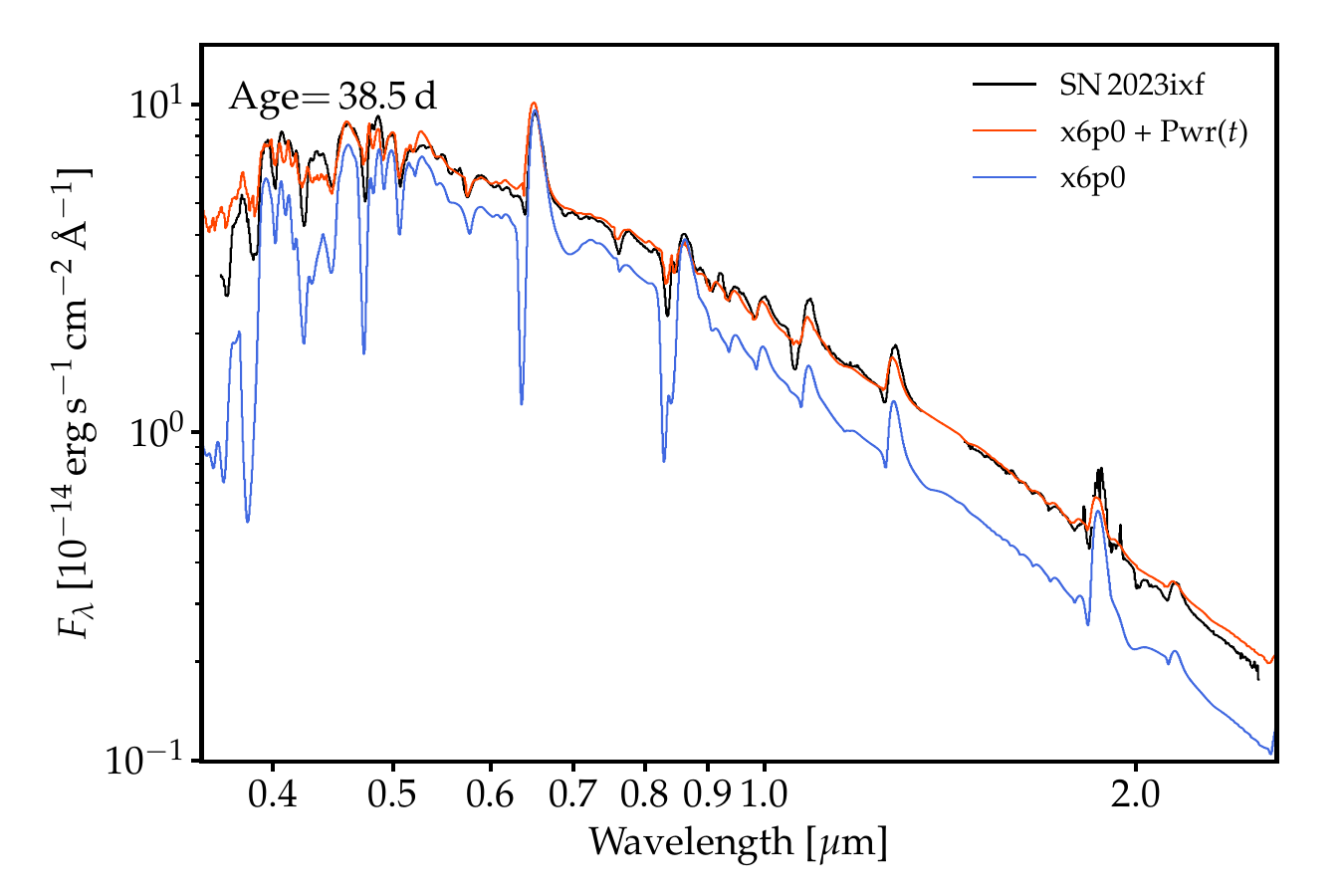}
\includegraphics[width=0.45\hsize]{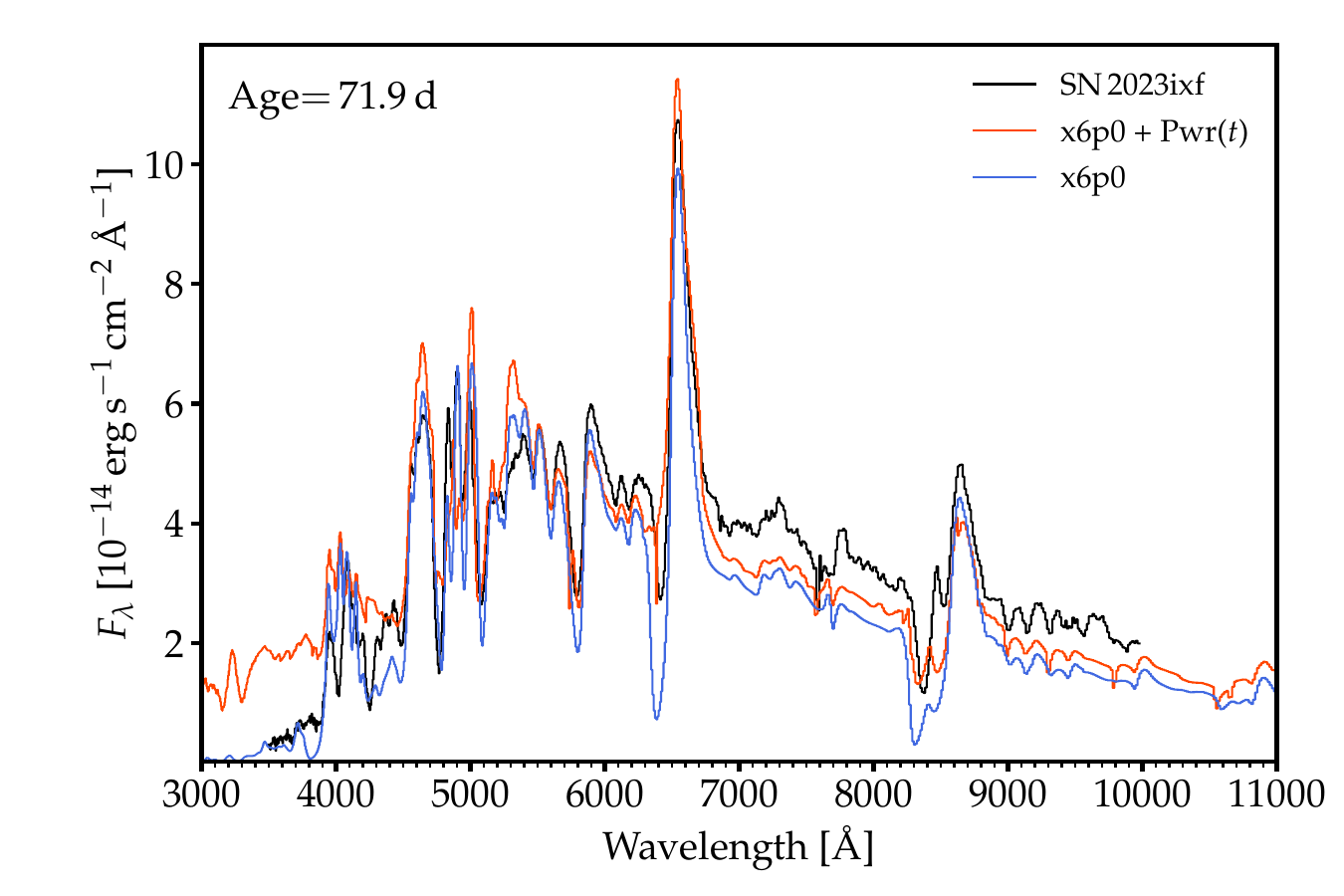}
\includegraphics[width=0.45\hsize]{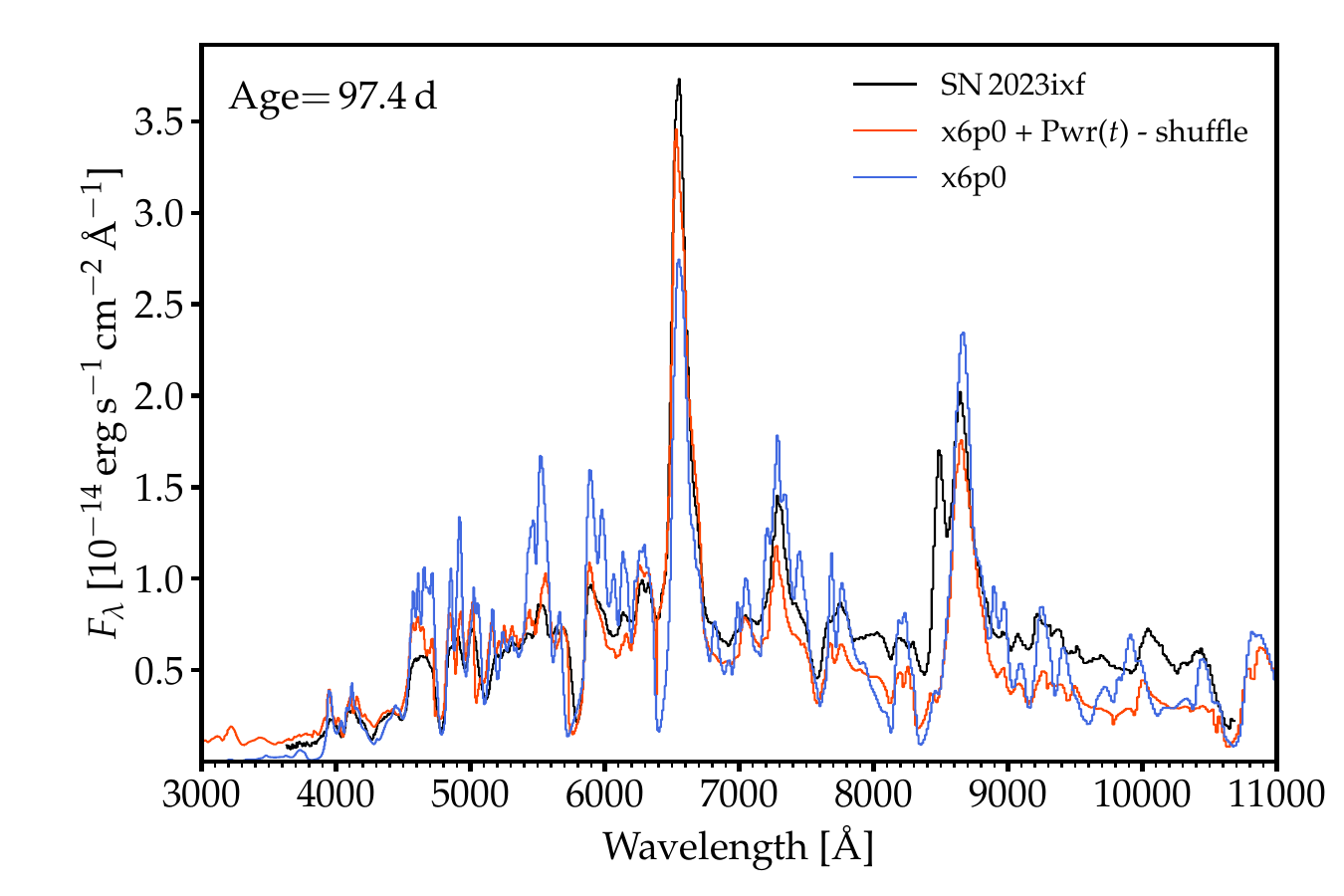}
\caption{Photospheric-phase spectra for models with (or without) shock power and observations of \sn\ at 31.5, 38.5, 71.9, and 97.4\,d after explosion. This figure is a counterpart to Figs.~\ref{fig_spec_22p5d}--\ref{fig_spec_84p5d}.
\label{fig_spec_more}
}
\end{figure*}

\begin{figure*}[h]
\centering
\includegraphics[width=0.9\hsize]{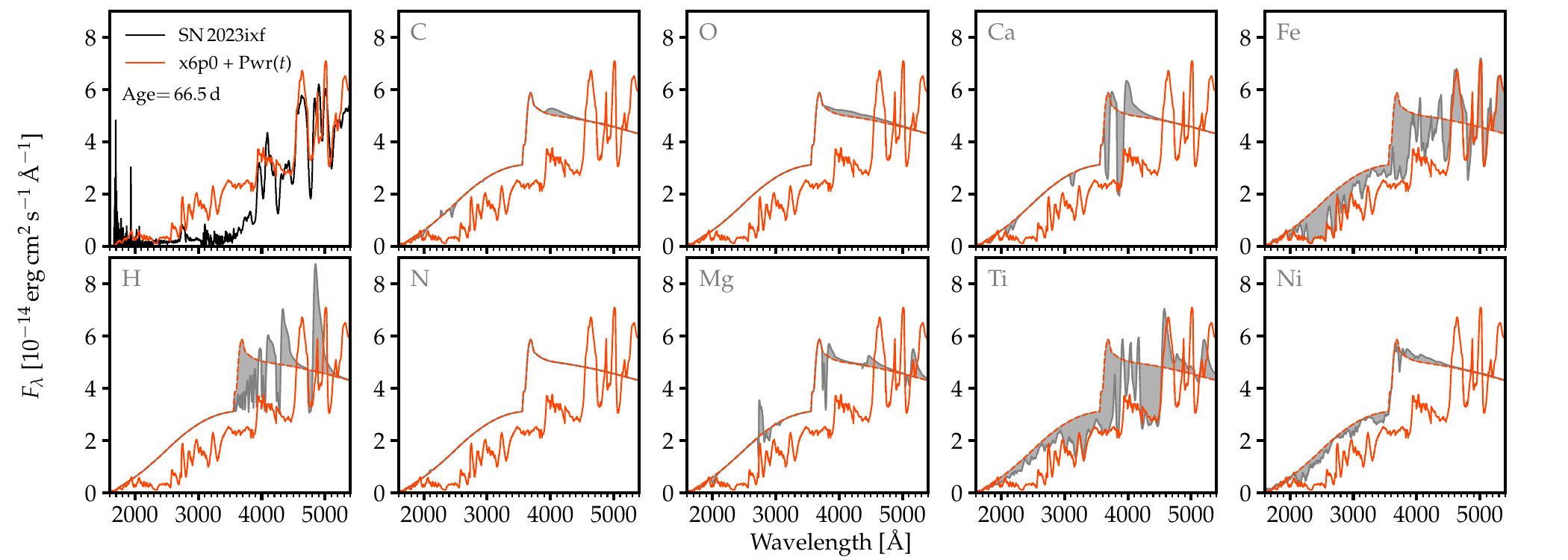}
\caption{Same as Fig.~\ref{fig_UV_24d} but now for model and observations of \sn\ at 66.5\,d.
\label{fig_UV_66d}
}
\end{figure*}

Figure~\ref{fig_spec_more} complements the results presented in Section~\ref{sect_rt} by showing comparisons at additional epochs and specifically at 31.5, 38.5, 71.9, and 97.4\,d (for the last epoch, shuffled-shell ejecta are used; see Section~\ref{sect_84p5d} for details). In this order, the model shown is at 32.65, 37.5, 70.0, and 97.0\,d (hence 1--2\,d offset relative to the observations), and the interacting power injected in the CDS is 130.0, 110.0, 45.0, and $4.6 \times 10^{40}$\,\ergs. The model counterpart without interaction nor a CDS is also shown.
Figure~\ref{fig_UV_66d} complements Fig.~\ref{fig_UV_24d} on the UV properties of \sn\ but now at 66.5\,d after first light. Model x6p0 + Pwr1($t$) yields a better match to the UV flux observed in \sn\ at that time, but the principles are the same and the visible offset between model and observations facilitates the analysis of contributing species and associated lines.

\end{document}